\documentclass{aa}  
\usepackage{natbib}
\bibpunct{(}{)}{;}{a}{}{,}
\usepackage{lscape}
\usepackage{supertabular}
\usepackage{graphicx}
\usepackage{txfonts}
\begin{document} 

   \title{{\em Herschel}-PACS photometry of the five major moons of Uranus
     \thanks{{\em Herschel} is an ESA space observatory with science instruments provided by
     European-led Principal Investigator consortia and with important participation from NASA.} 
          }

%   \subtitle{}

\titlerunning{{\em Herschel}-PACS photometry of Uranus' five major moons}

   \author{Ö.H.~Detre
           \inst{1}
           \and
           T.G.~M\"uller
           \inst{2}
           \and
           U.~Klaas
           \inst{1}
           \and
           G.~Marton
           \inst{3,4}
           \and
           H.~Linz
           \inst{1}
           \and
           Z.~Balog
           \inst{1,5}
          }

   \institute{Max-Planck-Institut f\"ur Astronomie (MPIA),
              K\"onigstuhl 17, 69117 Heidelberg, Germany \\
              \email{detre@mpia.de}
         \and
              Max-Planck-Institut f\"ur extraterrestrische Physik (MPE),
              PO Box 1312, Giessenbachstra{\ss}e, 85741 Garching, Germany
         \and
              Konkoly Observatory, Research Centre for Astronomy and Earth Sciences, 
              Konkoly Thege-Mikl\'os 15-17, 1121 Budapest, Hungary
         \and
              ELTE Eötvös Loránd University, Institute of Physics, 
              Pázmány Péter 1/A, 1171 Budapest, Hungary
         \and
              Astronomisches Recheninstitut des Zentrums f\"ur Astronomie,
              M\"onchhofstra{ss}e 12--14, 69120 Heidelberg, Germany
             }

   \date{Received / Accepted}

% \abstract{}{}{}{}{} 
% 5 {} token are mandatory
 
  \abstract
   {}
   {We aim to determine far-infrared fluxes at 70, 100, and 160\,$\mu$m for the five major 
    Uranus satellites, Titania, Oberon, Umbriel, Ariel, and Miranda. Our study is based on the available calibration 
    observations at wavelengths taken with the  PACS-P photometer aboard 
    the Herschel Space Observatory.
}
   {The bright image of Uranus was subtracted using a scaled Uranus point spread function 
    (PSF) reference established from all maps of each wavelength in an iterative process 
    removing the superimposed moons.
    The photometry of the satellites was performed using PSF photometry.
    Thermophysical models of the icy moons were fitted to the photometry of each measurement 
    epoch and auxiliary data at shorter wavelengths.
}
   {The best-fit thermophysical models provide constraints for important properties 
        of the moons, such as surface roughness and thermal inertia.
    We present the first thermal infrared radiometry longward of 50\,$\mu$m for 
    the four largest Uranian moons, Titania, Oberon, Umbriel, and Ariel, at
    epochs with equator-on illumination. Due to this inclination geometry, 
    heat transport took place to the night side so that thermal inertia played a
    role, allowing us to constrain that parameter. Also, we found some indication for differences 
    in the thermal properties of leading and trailing hemispheres. 
    The total combined flux contribution of the four major moons relative to Uranus 
    is 5.7$\times$10$^{-3}$, 4.8$\times$10$^{-3}$ , and 3.4$\times$10$^{-3}$ at 70, 100, and 
    160\,$\mu$m, respectively. We therefore  precisely specify the systematic error of the 
    Uranus flux by its moons when Uranus is used as a far-infrared prime flux calibrator. 
    Miranda is considerably fainter and always close to Uranus, impeding reliable photometry.
}
   {We successfully demonstrate an image processing technique for PACS photometer data
    that allows us to remove a bright central source and reconstruct point source fluxes
    on the order of 10$^{-3}$ of the central source as close as $\approx$3$\times$ the 
    half width at half maximum (HWHM) of the PSF. We established improved thermophysical 
    models of the five major Uranus satellites. Our derived thermal inertia values resemble those of TNO dwarf planets, Pluto and Haumea, more than those of 
    smaller TNOs and Centaurs at heliocentric distances of about 30\,AU.
}


   \maketitle

\section{Introduction}

The planet Uranus is a well suited primary flux standard at the upper
end of the accessible flux range for a number of contemporary far-infrared 
space and airborne photometers, such as ISOPHOT~\citep{lemke96}, {\it Herschel}-PACS~\citep{poglitsch10}, 
and HAWC$+$~\citep{harper18}. Uranus is also an important 
flux/amplitude calibrator for submm/mm/cm ground-based observatories, 
such as IRAM~\citep{kramer08} or JCMT (SCUBA-2)~\citep{chapin13}.

Uranus was routinely observed during the {\it Herschel} 
mission~\citep{pilbratt10} as part of the PACS photometer 70, 100, and 
160\,$\mu$m filter flux calibration program, in particular for a quantitative 
verification of the flux non-linearity correction for PACS~\citep{mueller16}.

\begin{figure*}[ht!]
   \centering
   \includegraphics[width=0.33\textwidth]{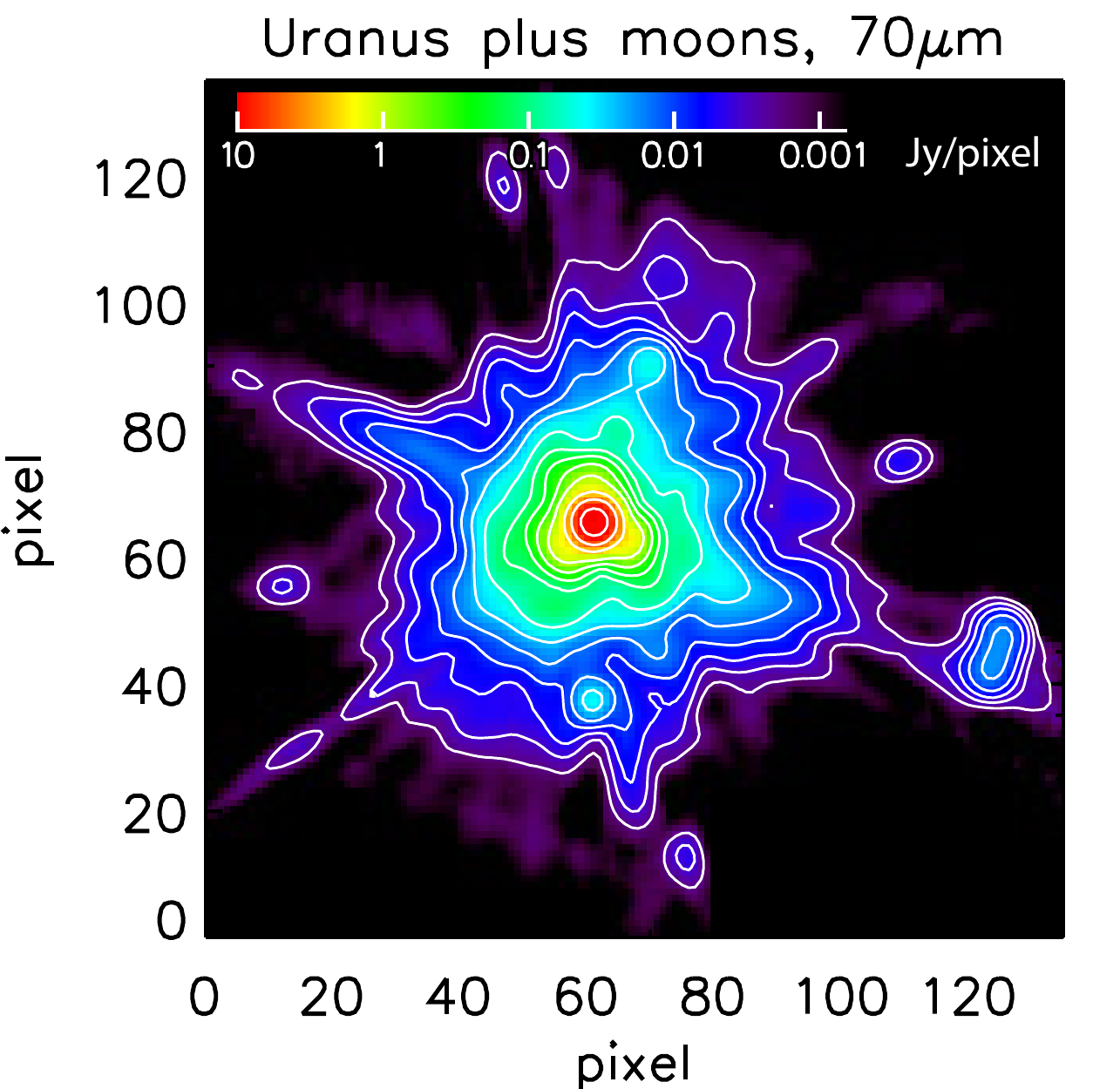}
   \includegraphics[width=0.33\textwidth]{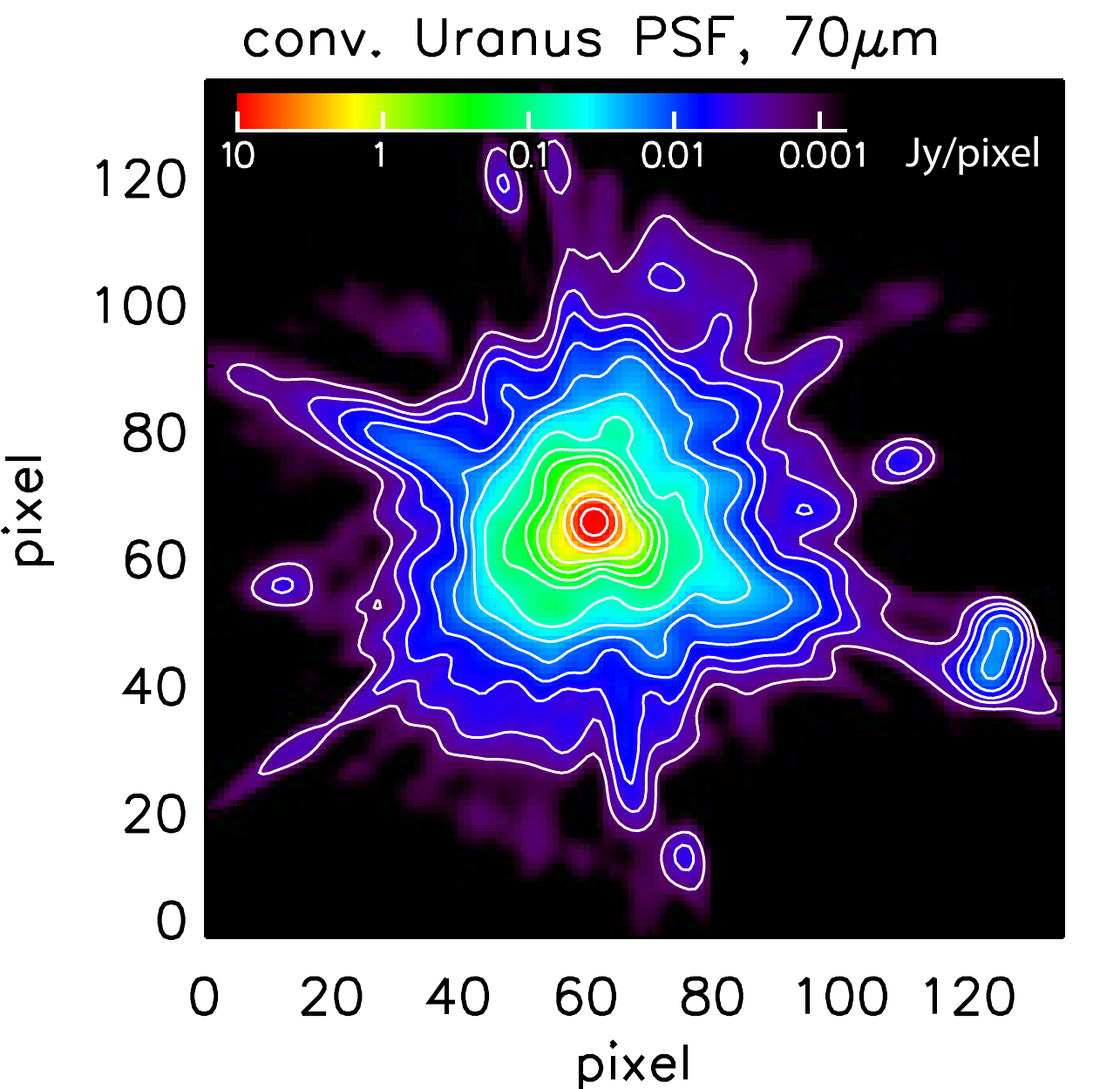}
   \includegraphics[width=0.33\textwidth]{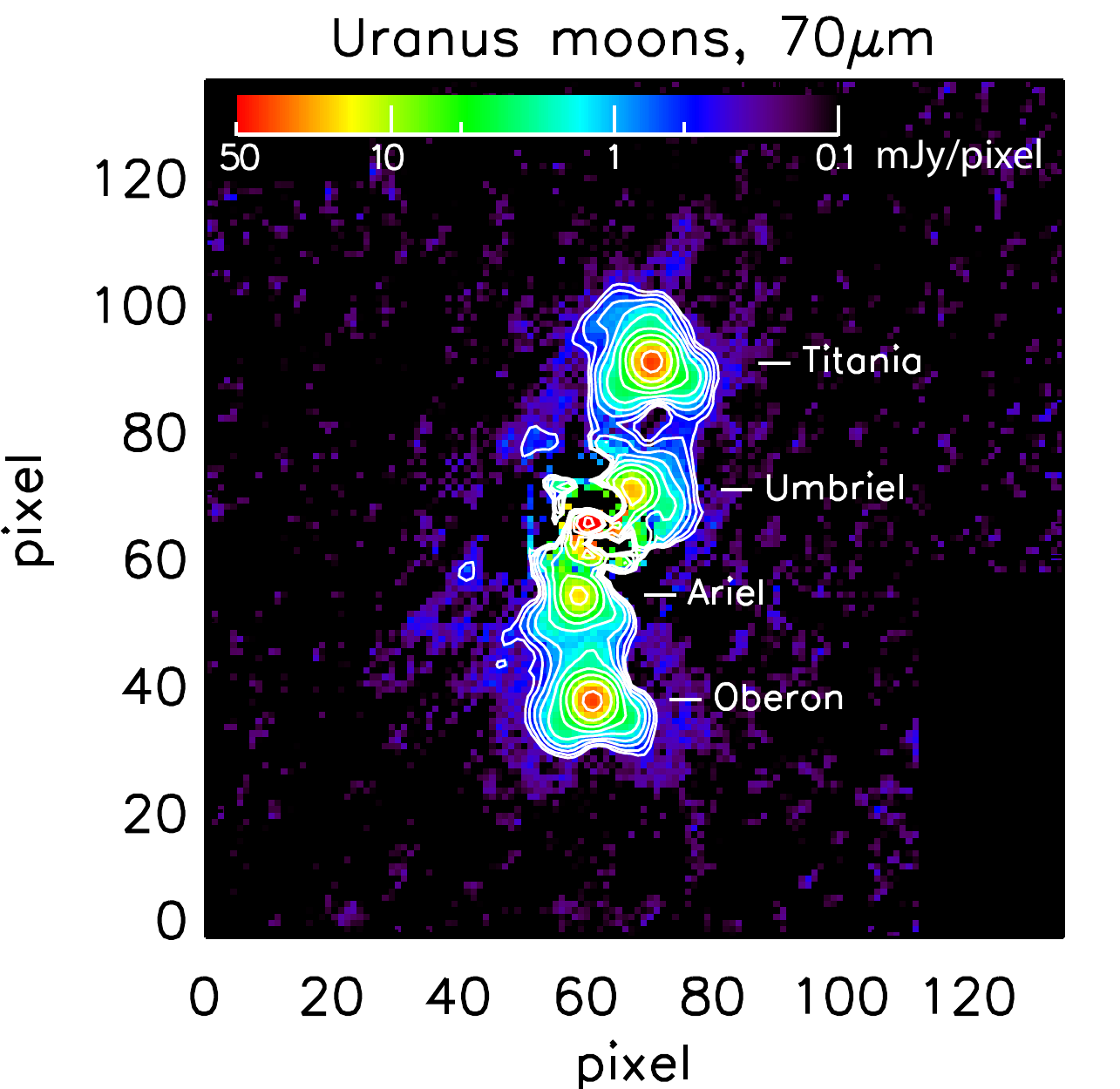}
      \caption{PACS 70\,$\mu$m scan map of Uranus (OBSIDs 1342223982+83) from OD\,789 
              (2011-07-12T01:21:57). Pixel scale is 1\farcs1. Left: Original
              map generated by high-pass filtering and co-addition of scan and 
              cross-scan containing Uranus plus its moons. This map was
              actually generated as an average of the nine different map
              parameter data sets per scan direction 
              Middle: Convolved Uranus point spread function for OD\,789 generated 
              from the Uranus reference PSF by PSF matching, cf.~Sect.~\ref{ssect:psfmatching}. 
              It shows a number of pronounced PSF features, while the absence of Titania and 
              Oberon is clearly visible. Right: Residual map after
              subtraction of the convolved Uranus PSF map from the original
              map. The four Uranus moons Titania, Umbriel, Ariel, and Oberon
              become clearly visible. 
              For better visualisation different flux scales were used for the individual images. 
              }
         \label{fig:uranus1342223982_83psfsub}
   \end{figure*}

Due to its flux density of $>$ 500\,Jy, Uranus exhibits an 
extended intensity profile in the PACS maps which reaches out to radii $>$
1\arcmin, and overwhelms the emission from its moons. An example is 
the Uranus image shown in the left panel of Fig.~\ref{fig:uranus1342223982_83psfsub}. 
Nevertheless, with a detailed comparison of the Uranus image  
with a PACS reference point spread function (PSF; Fig.~\ref{fig:uranus1342223982_83psfsub} middle), it is
possible to trace extra features on top of the Uranus PSF. That is how we 
recognised the two largest and most distant of the five major Uranian moons, 
Titania and Oberon, in the PACS maps.
Titania and Oberon were discovered by the name patron of the {\it Herschel} Space
Observatory, William (Wilhelm) Herschel, himself. In the following sections, we 
describe the method used to generate the Uranus reference PSF and subtract it
from the maps in order to extract FIR fluxes for all five major moons of Uranus. 
This photometry will be compared with the thermophysical modelling of the moons.

%
%                                                Two column figure
%-----------------------------------------------------------

\section{Data reduction}

\subsection{Input maps for Uranus PSF reference}

The key to a good PSF subtraction is to have a good reference PSF.
Since Uranus is a slightly extended source ($\approx$3\farcs5), the standard PACS PSF
references based on maps of the asteroids Ceres and 
Vesta~\citep{lutz15}\footnote{https://www.cosmos.esa.int/documents/12133/996891/PACS+\\photometer+point+spread+function, Fig.~7}
did not provide adequate PSF subtraction results. We therefore decided
to construct a Uranus reference PSF (Ref PSF from now on) out of the individual 
Uranus maps in each PACS filter.

The {\it Herschel} Science Archive contains twenty individual scan map measurements of Uranus,
taken over the entire course of the mission at five distinct epochs (cf.\ Table~\ref{table:psfphotUranus}).
Within each of those five epochs, four scan map observations were taken approximately\ 6\,min apart from 
each other.
The PACS photometer was able to take data simultaneously in the 160\,$\mu$m filter as well as either the 70\,$\mu$m
or 100\,$\mu$m filter.
The starting point of our PSF analysis were, therefore, the ten 70 and 100\,$\mu$m and the twenty 160\,$\mu$m,
high-pass filtered and flux-calibrated level 2 scan maps produced for the 
Uranus photometry as published in~\citet{mueller16}.
The data reduction and calibration performed in HIPE\footnote{HIPE is a joint 
development by the Herschel Science Ground Segment Consortium, consisting of 
ESA, the NASA Herschel Science Center, and the HIFI, PACS and SPIRE 
consortia.}~\citep{ott10} up to this level is described in~\citet{balog14}.
A general description of PACS high-pass filter processing is given in the
PACS Handbook~\citep{exter18}.
In order to determine any dependence of our PSF photometry on the data
reduction, we re-processed the maps with a variety of map parameter
combinations for HPF radius and pixfrac, as listed in Table~\ref{table:scanmapparams}. 
The variation of the results among the nine different created maps
of the same observation identifier (OBSID) is one component in our photometric
uncertainty assessment. The related uncertainty is listed under $\sigma_{\rm red}$ in Tables~\ref{table:psfphotUranus} - \ref{table:psfphotMiranda}.
 
%
%_____________________________________________________________
%                                             One column Table 
%_____________________________________________________________
%
\begin{table}[ht!]
\caption{Applied scan map parameters for the input maps of the PSF fitting 
         step. FWHM$_{PSF}$ is the average FWHM of the
         PSF for a point-like source in the corresponding
         filter. 'Outpix' marks the output pixel size in the final map. This 
         was kept constant, which means a sampling of the PSF FWHM by 5 pixels
         in each filter. 'HPF' is the abbreviation for the high-pass filter, 
         'pixfrac' is the ratio of drop size to input pixel size used for the 
         drizzling algorithm \citep{fruchter02} within the {\sf photProject()} mapper. 
}             
\label{table:scanmapparams}      
\centering
    \begin{tabular}{r c c c c c c c c}
   \hline\hline
            \noalign{\smallskip}
Filter  & FWHM$_{PSF}$ & outpix & HPF radius\tablefootmark{a} & pixfrac \\
($\mu$m)&    (")      &   (")  &                             &         \\
            \noalign{\smallskip}
    \hline
             \noalign{\smallskip}
   70  &  5.6 & 1.1 &  15, 20, 35 & 0.1, 0.5, 1.0 \\
  100  &  6.8 & 1.4 &  15, 20, 35 & 0.1, 0.5, 1.0 \\
  160  & 10.7 & 2.1 &  30, 40, 70 & 0.1, 0.5, 1.0 \\
            \noalign{\smallskip}
\hline
    \end{tabular}
\tablefoot{
\tablefoottext{a}{This parameter determines the elementary section of a scan 
                  over which the high-pass filter algorithm computes a running 
                  median value. Its unit is 'number of read-outs'. The spatial 
                  interval between two readouts is $\alpha_{\rm ro} = 
                  \frac{v_{\rm scan}}{\nu_{\rm ro}}$. For the standard 
                  $\nu_{\rm ro}$ = 10\,Hz read-out scheme in PACS prime mode, 
                  and a scan speed $v_{\rm scan}$ = 20"/s,~the spatial
                  interval $\alpha_{\rm ro}$ between two read-outs corresponds 
                  to 2". The entire width of the HPF window (") = [(2 $\times$ HPF radius) + 1] 
          $\times~\alpha_{\rm ro}$.}
}
\end{table}

\subsection{Establishment of the Uranus reference PSFs}

As a first step the WCS (world coordinate system) astrometries of the images were corrected 
by finding the centre of Uranus. This was crucial to correct the majority of astrometric 
uncertainties of the images. In addition to the standard flux calibration in HIPE a final
flux calibration step was done by removing the dependence of the detector response on the
telescope background, a calibration feature which is described in~\citet{balog14}.
The relation of detector responsivity with telescope background could be established from the 
Uranus observations themselves with a very high signal-to-noise-ratio (S/N). All images were then flux-normalised 
to a mean Uranus-to-{\it Herschel} distance and rotated to the same reference angle. 
The distance correction was on the order of 6\%, while the detector response correction 
with telescope background was on the order of 1\%. Details of these flux corrections are 
detailed in Appendix ~\ref{sect:appb}. Following these corrections, the uncertainty of Uranus flux was within a 
remarkable 0.19\% - 0.27\% depending on the filter, proving the outstanding flux stability of the 
PACS instrument. This was important because flux variation could have a negative effect on the 
creation of the median image for the Ref PSF in the next steps. On the other hand, an arbitrary 
normalisation compensating for the flux differences would render any later photometry unreliable.

\begin{figure}[tp]
        \includegraphics[width=1.0\linewidth]{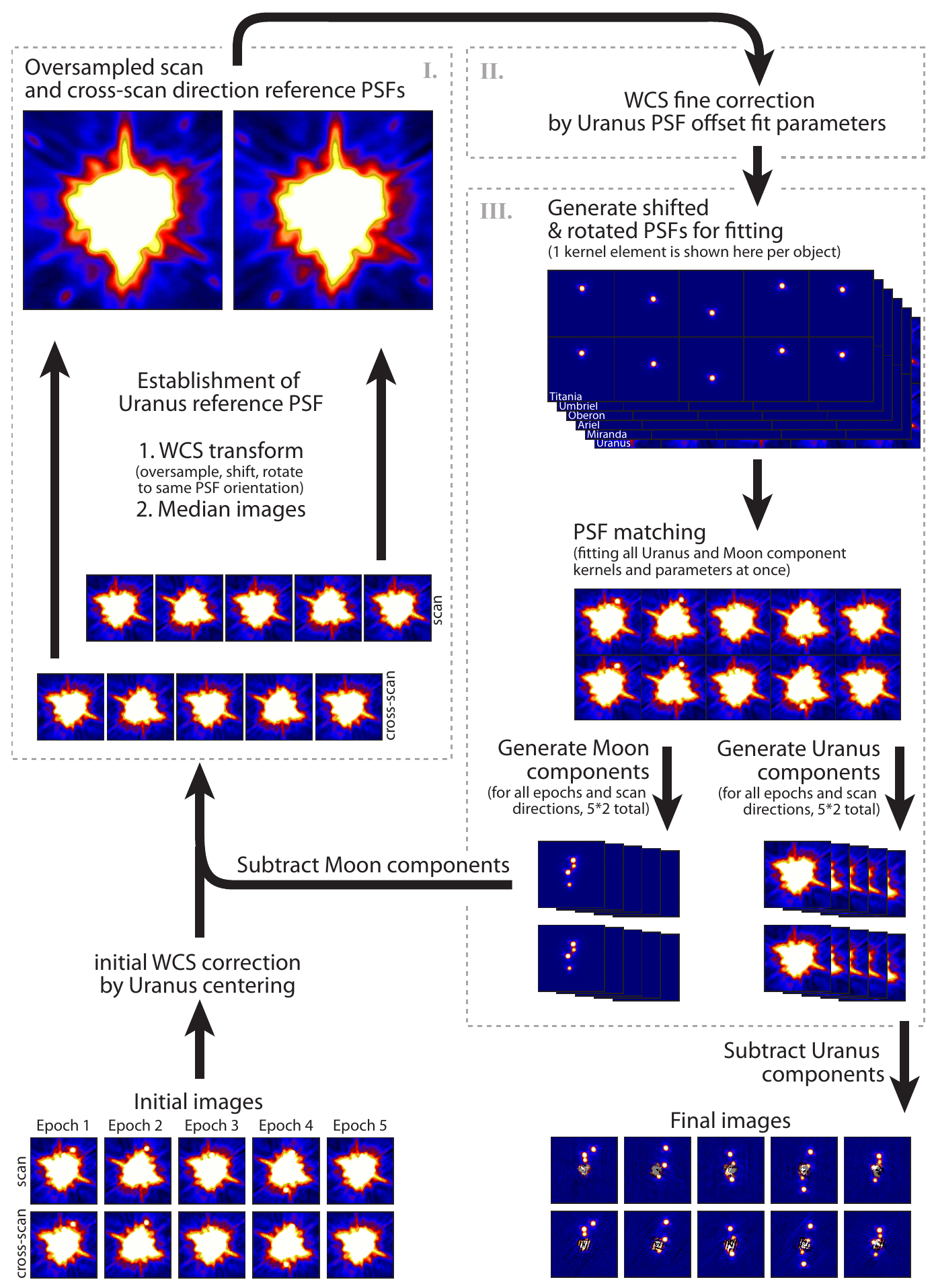}
        \caption[Flowchart of the iteration cycle]{Flowchart for the iteration cycle. 
                The dashed boxes show the three main parts of the iteration loop. The calculation 
                starts at bottom left by initial WCS correction of the raw images. The iteration 
                cycle is stopped when the fit parameters do not change significantly. The 25 iteration 
                cycles were needed for each dataset with different HPF and pixfrac values and, of 
                course, each filter. Finally, Uranus-subtracted images are at the bottom right.}
        \label{fig:uranusmoonsanalysisflowchart}
\end{figure}

Four-time oversampling was used for the Ref PSF (FWHM was sampled by 20 pixels) to mitigate the 
information loss by the re-sampling of the data back and forth. A separate Ref PSF was generated for 
each of the two scan directions due to minor differences between them.
The very first Ref PSF was generated by a simple median over the individual 
images on each pixel. The median removed the orbiting moons for most of the pixels around the PSF centre. 
However, for some areas of the Ref PSF the moons were overlapping multiple times. To remove the remnants 
of the moons at these spots the generation of the Ref PSF was done in an iteration loop. 
The iteration loop also corrected small distortions and flux differences between the images 
(called PSF matching, see Section~\ref{ssect:psfmatching}) and further enhanced the astrometry of the images.
The iteration loop is shown in Fig.~\ref{fig:uranusmoonsanalysisflowchart}. Its three main parts are:
1)  Generating a Ref PSF. The improved Ref PSF was generated from 
moon-cleaned individual images, calculated in the previous loop.
2) Improving the astrometry (RA and Dec) of Uranus and the moons.
3) Decomposing the individual images into matched PSFs at the position of Uranus and 
its five major moons. These are called the Uranus component and the Moon component 
(including all five moons) of a given image.

The iteration loop stopped when no significant change was found for the Ref PSF, nor any 
flux change for the moons.

\subsection{PSF matching}
\label{ssect:psfmatching}

Given the Uranus Ref PSF was generated from the measurements themselves, we obtained already 
good results using the simplest way to generate the Uranus component, namely, by using the Ref PSF from the previous iteration loop, multiplied by a simple relative 
flux parameter. This parameter was fitted for each measurement to take into account 
the flux changes of Uranus. 
Similarly, five flux parameters were used for the Moon component, fitted for each moon to 
take into account the relative flux difference of Uranus and its moons.

The Uranus PSF shape was changing slightly between images. To adjust these individual 
differences, we convolved the Ref PSF with normalised kernel matrices. 
Fitting 5$\times$5 normalised kernel elements to the individual images improved 
the Uranus PSF subtraction near the centre of the PSF, making even 
the inner moons visible in some cases. 

The PSF difference between the Uranus and its moons were clearly visible by leaving doughnut artefacts
at the residual images of the moons. The use of a simple 3$\times$3 sharpening kernel for the moon PSFs 
completely eliminated this issue, which had clearly originated from differences in their PSF size.
The moons as well as Uranus had the same small distortions on the same image, therefore, we applied the 
sharpening kernel to the (already PSF-matched) Uranus component of a given image instead of the Ref PSF.
The Moon component of an image was generated by shifting the Moon PSF to the moon positions at a 
given epoch and multiplied by the relative flux parameter of each moon.

The optimal sizes of the kernels change with wavelength. In order to have the same number 
of free parameters and constraints for all wavelengths, we implemented a spatial scale factor
for the kernels. In this scaled kernel image convolution, the kernel values were used to weight 
the -2d, -1d, 0d, 1d, 2d distance units shifted Ref PSF instances around Uranus in X and Y-direction. 
Where the d units were d$_{\rm 70}$ = 1.5, d$_{\rm 100}$ = 1.25 
and d$_{\rm 160}$ = 1 map pixels for the 70, 100, and 160\,$\mu$m images, respectively. 
Finally, all shifted elements were added together and multiplied by a relative flux parameter. 
We note here that this scaled kernel image convolution becomes a traditional image convolution 
with d = 1 pixel shift distance unit.

\begin{figure}[ht!]
        \centering
        \includegraphics[width=0.50\textwidth]{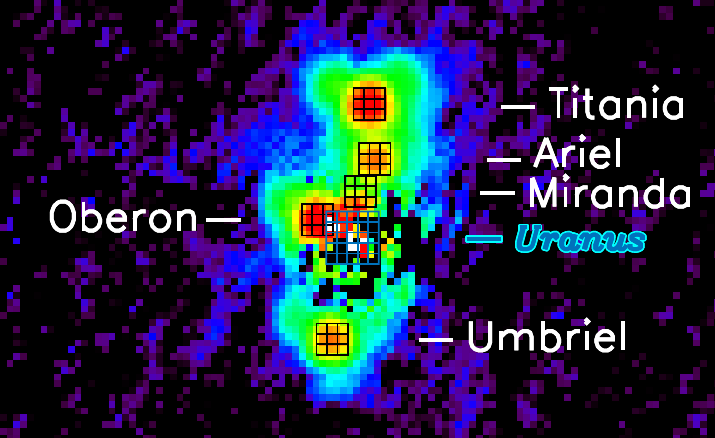}
        \caption{Kernel positions of Uranus (5$\times$5, in blue) and its moons 
                (3$\times$3, in black) shown on the Uranus subtracted product of OBSID 1342211117+18. 
                Overlapping kernel elements (Uranus and Oberon on this given example here) caused redundancy
                in the fit of these kernel elements. Rotational symmetries were introduced into the kernels
                to eliminate this issue. The flux scale of the image is the same as in Fig.~\ref{fig:moonpsfphot}.
        }
        \label{fig:kernelillustration}
\end{figure}

\begin{figure*}[ht!]
   \centering
   \includegraphics[height=0.26\textheight]{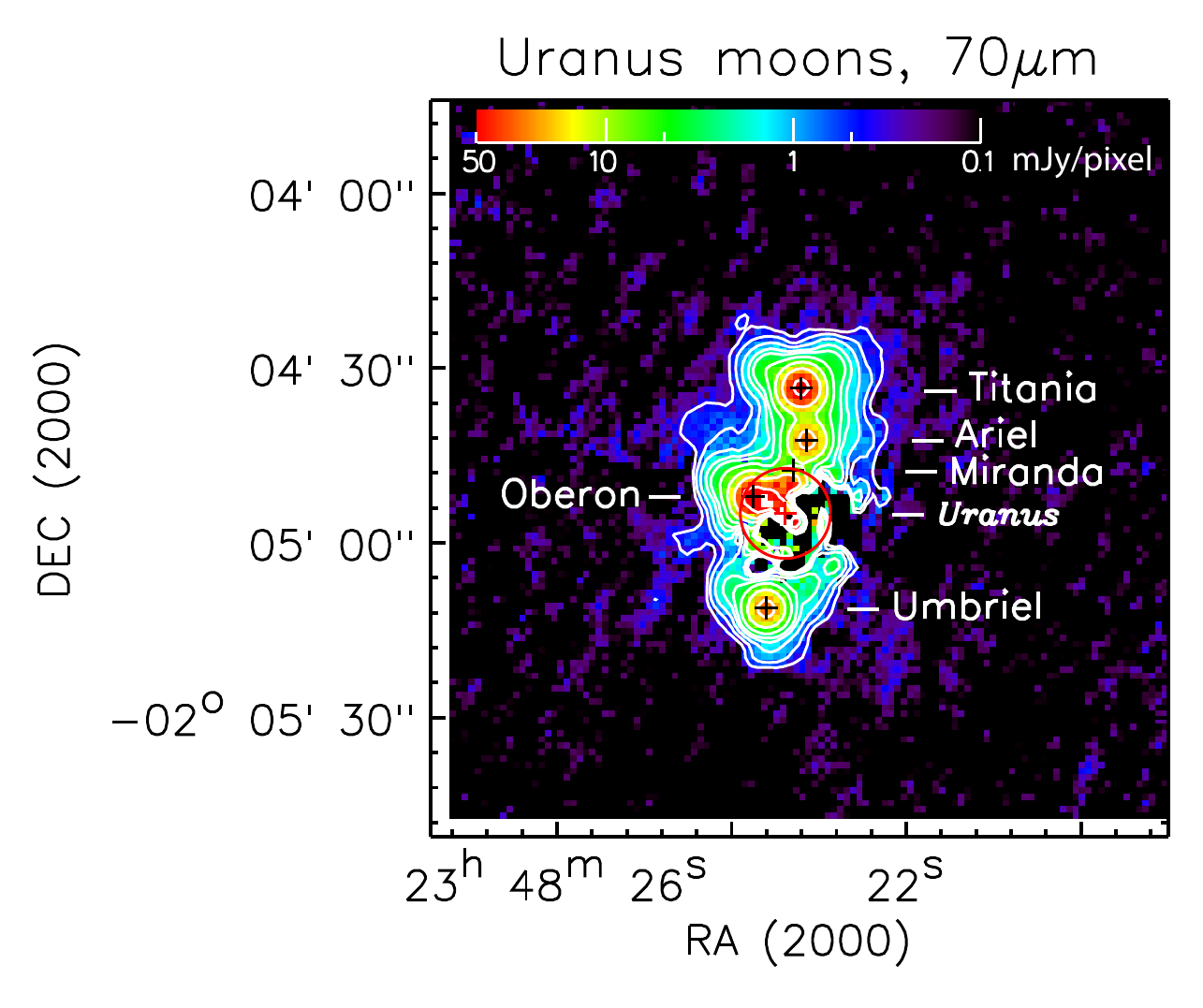}
   \includegraphics[height=0.26\textheight]{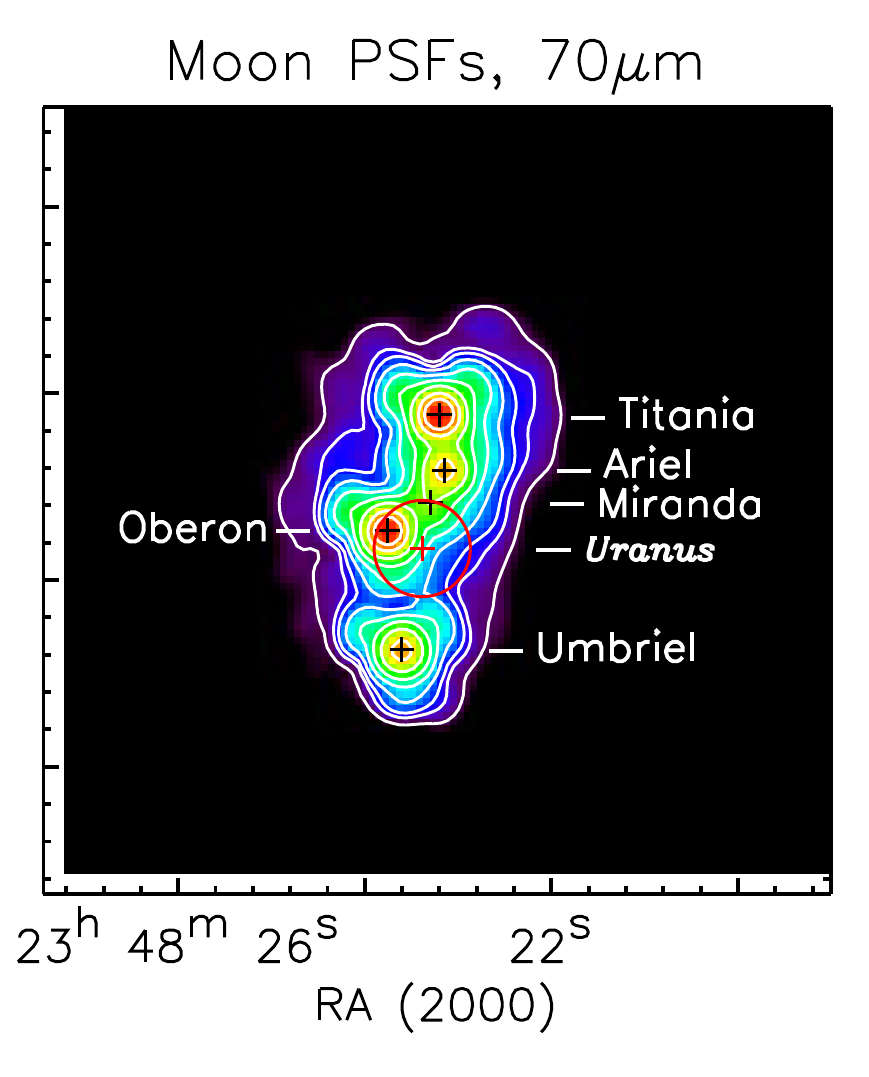}
   \includegraphics[height=0.26\textheight]{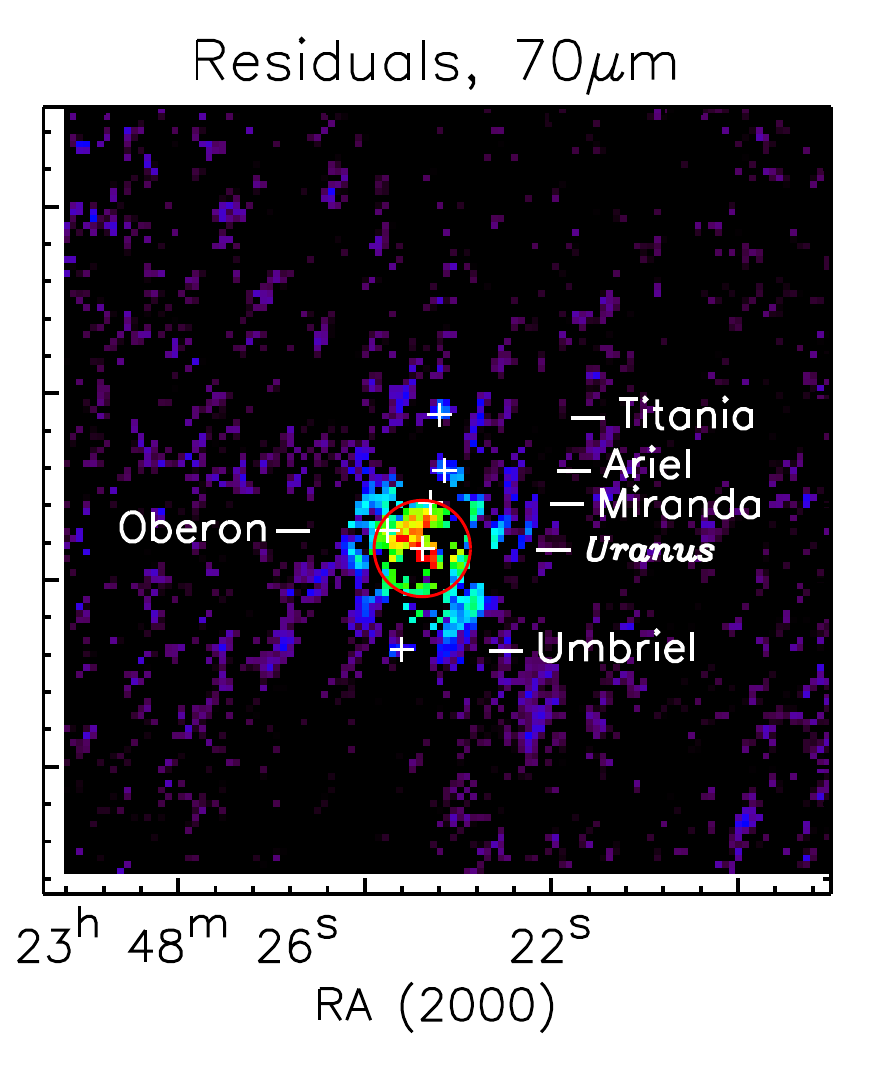}
      \caption{70\,$\mu$m PSF photometry of Uranian moons for OBSIDs 1342211117+18 on OD\,579.
               Left: Actual moon map after subtraction of the convolved Uranus PSF.
               The centres of the five moons are marked by black crosses and are labelled.
               The red cross inside the red circle indicates the approximate centre
               of the Uranus PSF. The circle has a radius of 7\farcs8 and circumscribes
               an area of significant PSF subtraction residuals (see right figure)
               inside which the photometric S/N ratios are degraded (cf.\
               Fig.~\ref{fig:sndepondist}). 
               Middle: Moon component of the image done by fitting a
               sharpening kernel and relative flux parameter for the reference PSF 
               at the position of each moon (indicated by black crosses). This Moon
               component map recovers also intensity inside the circle area where 
               the PSFs are disturbed in the map with the Uranus PSF
               subtracted. The moon PSF map is displayed with a larger dynamic
               range than the moon map. 
               Right: Residuals map (Moon map minus Moon PSF map) providing a judgement
               of the quality of the fit. The centres of the five moons and
               Uranus are marked here by white crosses and are labelled. For comparability,
               we used the same flux scale for all three images.
               }
         \label{fig:moonpsfphot}
   \end{figure*}

Figure~\ref{fig:kernelillustration} shows an example of the kernels. The 5$\times$5 Uranus 
kernel is in blue and the 3$\times$3 moon kernels are in black. An example of a fitted Moon 
component can be seen in the middle and residual image at the right in Fig~\ref{fig:moonpsfphot}.

The major step in the iteration loop was to fit these kernels and flux parameters to each individual image.
For the fitting parameters, the crucial point was to find a good balance between constraints 
and free parameters. The constraints were:
\begin{itemize}
\item[1)] Until the very last iteration loop the flux of each moon was set constant 
           for all observation epochs. This was crucial, because with this constraint the flux of
           a given moon was fitted dominantly to those epochs where it was farther away from 
           the centre of Uranus due to the higher S/N of the image at those pixels.
           The noise estimate was taken from the associated standard deviation 
           map of the image product. 
\item[2.)] Although the optimal kernels were not symmetric for all individual images, it was crucial 
           to impose symmetry on the kernels. The PSFs of the nearby moons were overlapping with some 
           image convolution elements, making the fit redundant for their kernel elements.
           For example, Fig.~\ref{fig:kernelillustration} displays where the kernels of Uranus and Oberon are overlapping.
           This redundancy would incorrectly elevate some of the kernel components of Oberon, reducing the
           Uranus kernel values proportionally. Implementing rotational symmetry for the kernels solved these redundancies. 
           The Uranus kernel therefore was an average of two 5$\times$5 kernels with 180$^{o}$ and 120$^{o}$ 
           rotation symmetric elements. 
\item[3.)] The more point-like Moon PSF was generated by a convolution of the Uranus component 
           with the simplest (two-parameter) 90$^{o}$ rotation-symmetric 3$\times$3 normalised sharpening kernel. 
           These fitted kernel elements were constant for all the epochs and the same for all moons 
           as the relative diameter ratios of Uranus and its moons can be considered as constant.
\item[4.)] The last free parameters to be fitted were the X and Y spatial offsets of the images to improve the relative 
           positions of the individual PSFs. The PSF subtraction is very sensitive to any offset. 
           An uncertainty of $\approx$100\,mas  for the Uranus centre would result in a quite 
           significant residual pattern.
\item[5.)] In the last iteration loop, all previously fitted parameters were fixed, but the 
           constant moon flux constraint was released. This last fit showed the variability of the moon fluxes 
           from their averages for each epoch.
\end{itemize}

%
%                                                One column figure
%-----------------------------------------------------------

%
%                                                Two column figure
%-----------------------------------------------------------

\subsection{PSF subtraction}
After fitting of all parameters to all individual images at the same time, two intermediate outputs were generated.
First is the Uranus component-subtracted images. Second is the Moon component-subtracted images for Ref PSF generation 
at the beginning of the next iteration loop. This ensures that remnants of the moons on the Ref PSF are gradually 
removed with each iteration.

After the last iteration loop, the Uranus and Moon components were saved into the FITS files of the final moon map 
products. Subtracting both the Uranus and Moon components give the residual image. The residual image seen at the right of 
Fig.~\ref{fig:moonpsfphot} clearly proves the correctness of the fit parameters and the correct balance of free fit 
parameters and constraints.

\section{Maps of the Uranian moons}

All data products with the PSF subtracted maps and including the convolved
Uranus PSF and the moon PSFs in additional extensions will be available
in FITS format as {\it Herschel} Highly Processed Data Products (HPDPs)
\footnotemark[3]\footnotetext[3]{
https://www.cosmos.esa.int/web/herschel/highly-processed-data-products} 
in the {\it Herschel} Science Archive.

Figures~\ref{fig:moonmaps70} to~\ref{fig:moonmaps100} show the final actual maps of the Uranian moon 
constellations with the Uranus PSF subtracted for the five observation epochs. The corresponding 
scan and cross-scan maps have been averaged. It is clear that there is an inner area where the PSF 
subtraction does not work perfectly. This area is quantified by the results illustrated in 
Fig.~\ref{fig:sndepondist}.

\section{Photometry of the Uranian moons}

The PSF photometry of the moons is a side product of our PSF subtraction itself, as we have 
to fit and subtract the moons to get a moon-cleared image for the Ref PSF generation.
In comparison with aperture photometry, the constraint of knowing the exact PSF shape provides extra 
information to the PSF photometry, giving better results in crowded fields 
for overlapping sources. To get additional confidence in our PSF photometry, we 
also performed standard aperture photometry whenever any moon was well-separated from Uranus.

\subsection{PSF photometry}

An example of PSF photometry fit results is shown in Fig.~\ref{fig:moonpsfphot} for 
the combined scan and cross-scan map of OBSIDs 1342211117+18, from which
70\,$\mu$m photometry for all five moons can be obtained. The PSF images of Oberon and
Miranda are disturbed in the residual map due to imperfect Uranus PSF
subtraction in this central area, nevertheless a significant fraction of the moon PSF
is available to recover the total flux and reconstruct the intensity
distribution. As already mentioned earlier in this paper, the fitting algorithm weights the 
pixels with their sigma value using the associated standard deviation 
map of the image product. In the case of Miranda, the PSF is fitted dominantly to this outer 
part of the Uranus PSF, where the S/N of the pixels is higher than the ones closer to the Uranus centre. 
Of course the uncertainty of the PSF fit worsens if only part of the PSF is available.

The unitless PSF flux fit parameters were relative fluxes used to weight the Ref PSF. 
To get the flux in Jy from these weights, they have to be multiplied 
with the aperture photometry of the Ref PSF, in other words, the average flux of 
Uranus over the measurements. 
The flux uncertainties were calculated the same way from the unitless 1-sigma parameter
error values of the PSF fit parameters. This is the second component in our photometric uncertainty
assessment. The related value is listed under $\sigma_{\rm par}$.

\subsection{Aperture photometry}

Based on the Uranus subtracted products, we also performed standard aperture photometry, as described 
in the PACS Handbook~\citep{exter18}, Sect.~7.5.2. Subtracting all other moons from the product 
(except the one we were measuring) clearly enhanced the aperture photometry results. Still, it was 
possible when a given moon was well separated from Uranus at a given epoch. This is mainly the case for Oberon and 
Titania, while unfortunately the number of comparison cases for Umbriel and Ariel is quite limited, 
in particular at 160\,$\mu$m (70\,$\mu$m: 8 cases, 100\,$\mu$m: 4 cases, 160\,$\mu$m: 0 cases).

The detailed comparison of PSF photometry with aperture photometry has been compiled in 
Table~\ref{table:aperphotcomp_1}. A statistical overview is given in Table~\ref{table:compPSFaper}. 
From this, it can be seen that the consistency of the two photometric methods is very good 
(within 3--4\%), thus confirming the principal quality of our PSF photometry procedure.
This does not, however, exclude that individual fits may be unreliable or even fail, particularly
in areas with high PSF residuals or confusion from close sources. The uncertainty of the fit gives then
already good advice on the reliability.

The aperture photometry shows on average a systematic 3--4\% negative flux offset with 
regard to the PSF photometry. This flux loss was a result of the small apertures and sky radii to 
achieve good residual rejection.

\begin{table}[h!]
\caption{Comparison of PSF photometry with standard aperture photometry for a number of measurements, n,
         when an Uranian moon was far enough off Uranus and the other moons to allow relatively undisturbed
         aperture measurements.
}             % title of Table
\label{table:compPSFaper}      % is used to refer this table in the text
\centering                          % used for centering table
\begin{tabular}{r c c}        % centered columns (4 columns)
\hline\hline                 % inserts double horizontal lines
            \noalign{\smallskip}
Filter & n & $\frac{f^{\rm PSF}}{f^{\rm aper}}$ \\
            \noalign{\smallskip}
\hline
            \noalign{\smallskip}
  70   & 26 & 1.030$\pm$0.003 \\
 100   & 22 & 1.032$\pm$0.005 \\
 160   & 28 & 1.036$\pm$0.004 \\
            \noalign{\smallskip}
\hline\hline                 % inserts double horizontal lines
\end{tabular}
\end{table}

%
%                                                Two column figure
%-----------------------------------------------------------
   \begin{figure*}[ht!]
   \centering
   \includegraphics[width=0.33\textwidth]{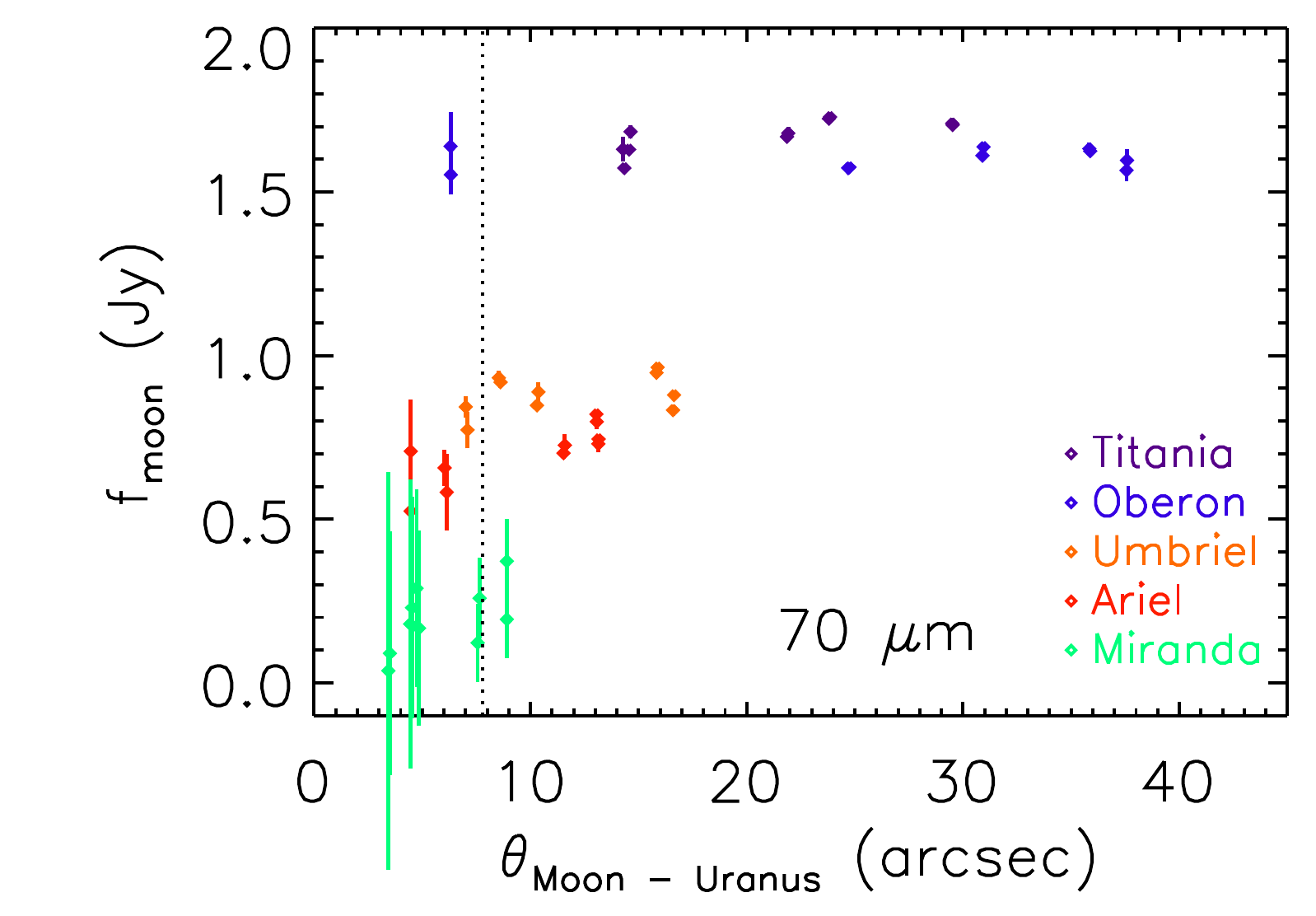}
   \includegraphics[width=0.33\textwidth]{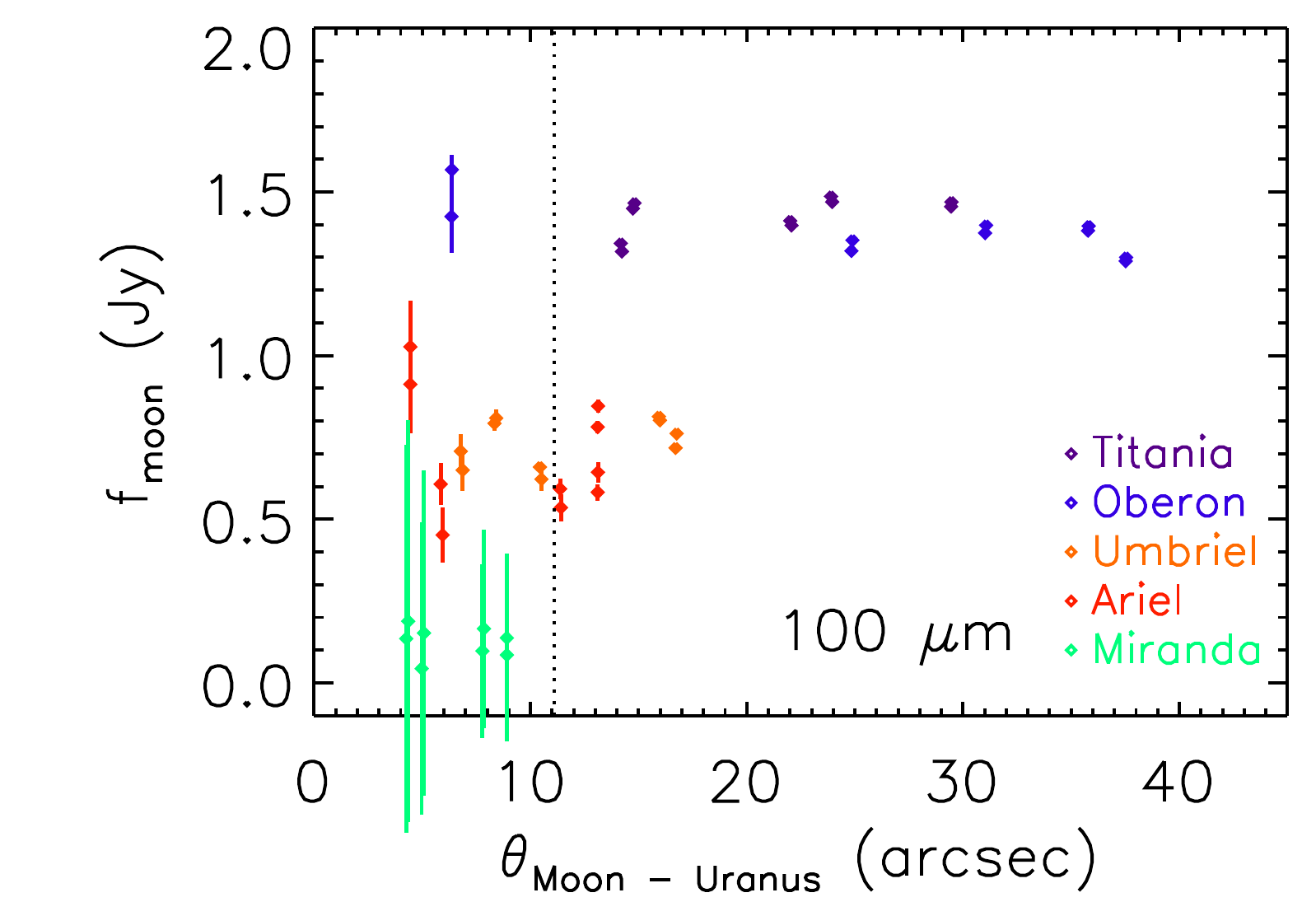}
   \includegraphics[width=0.33\textwidth]{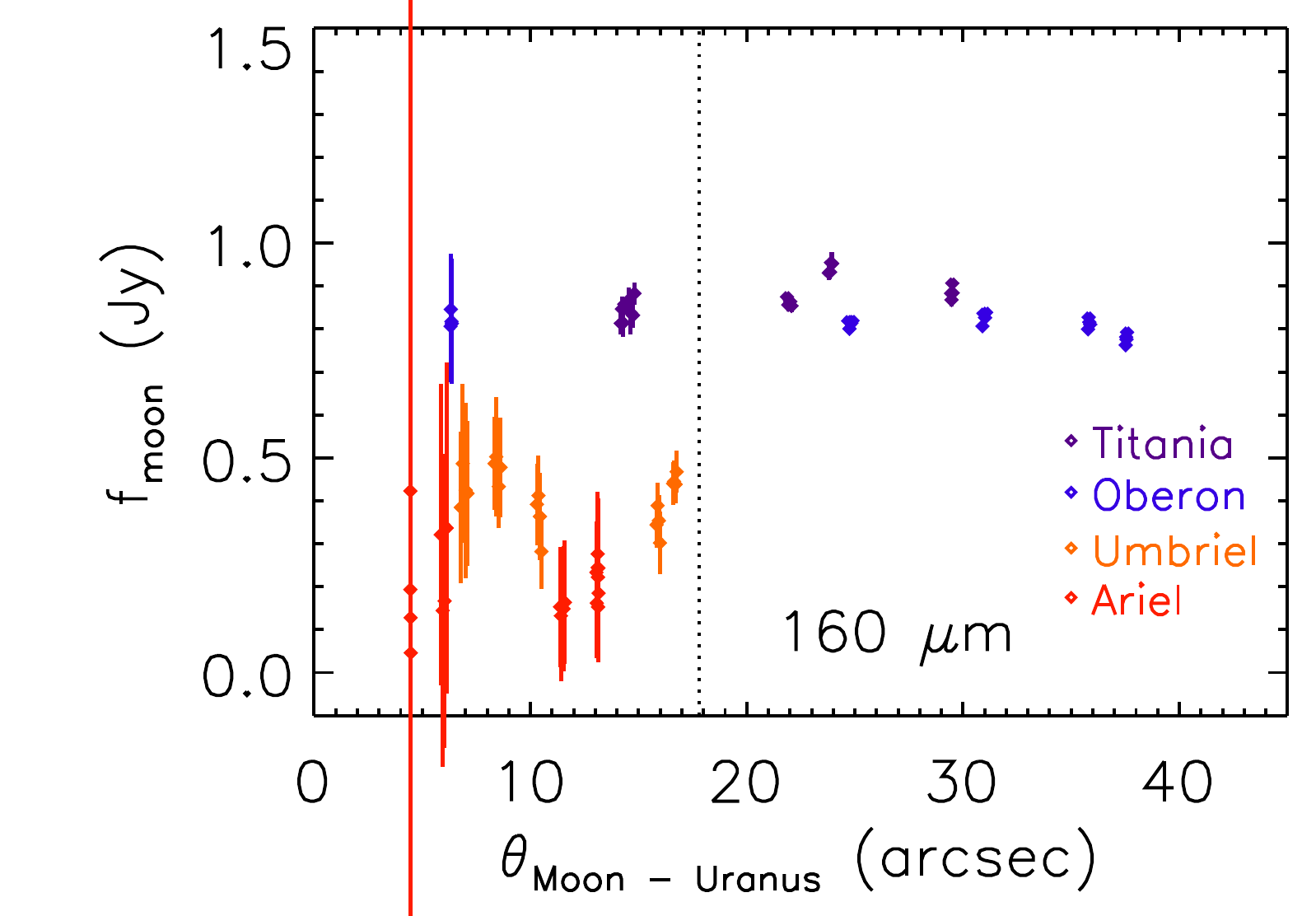}
   \includegraphics[width=0.33\textwidth]{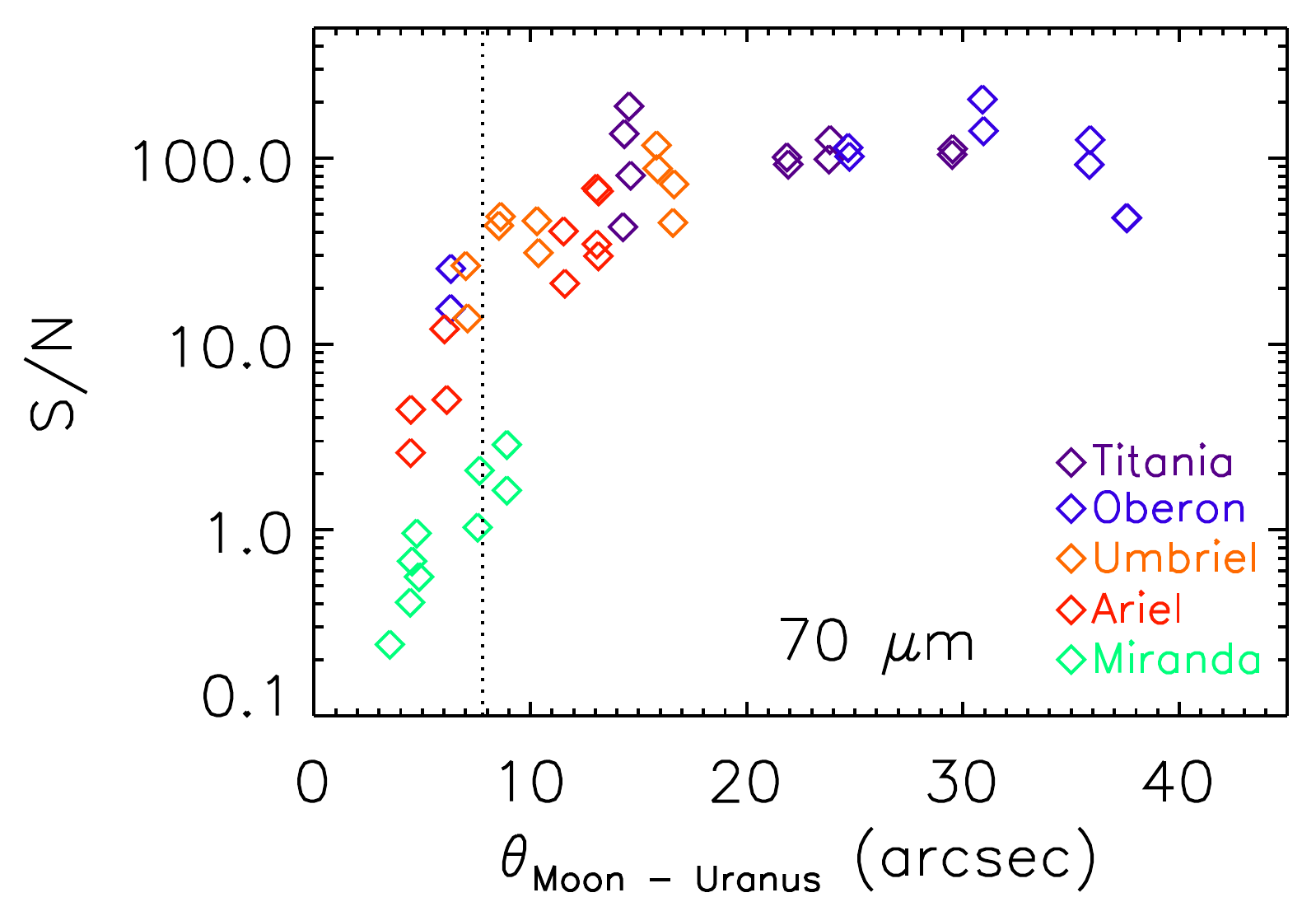}
   \includegraphics[width=0.33\textwidth]{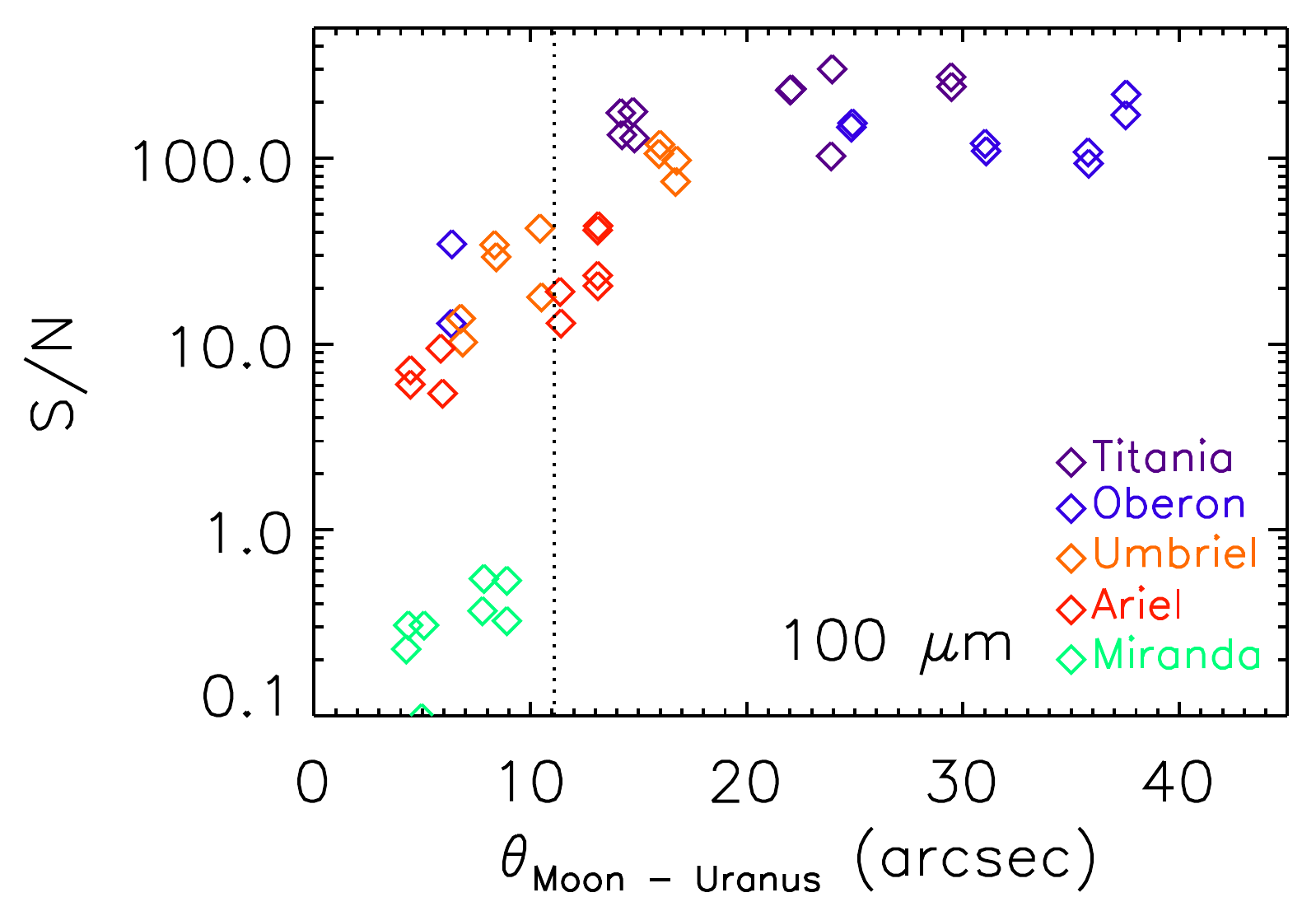}
   \includegraphics[width=0.33\textwidth]{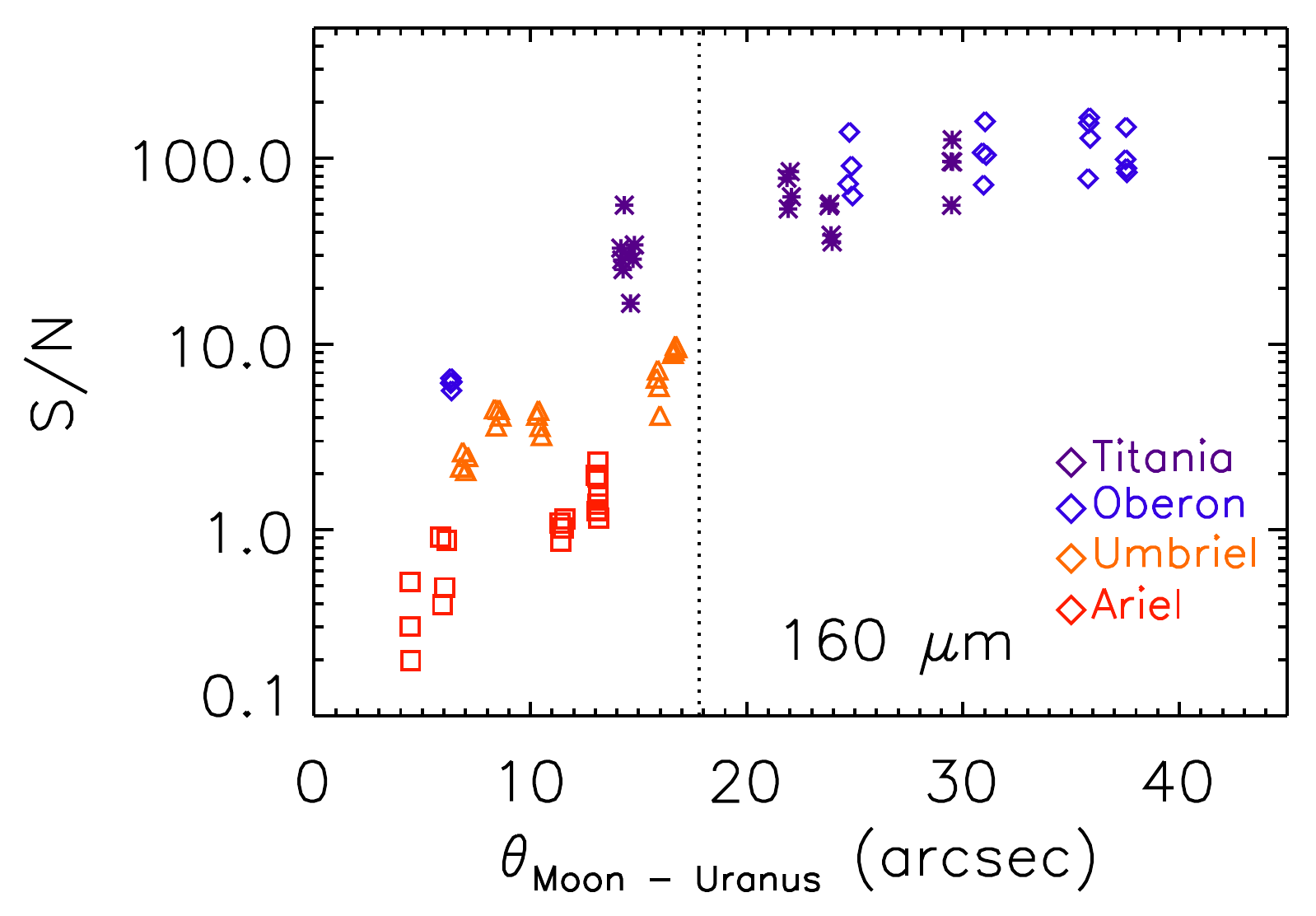}
      \caption{     Upper panel: Derived fluxes $f_{\rm moon}$ from PSF photometry 
                    and their uncertainties $\sigma_{tot}$ depending on the distance 
                    of the Uranian moon from the Uranus position for the 
                    70, 100, and 160\,$\mu$m filter, respectively.
                    Lower panel: Corresponding signal-to-noise ratios 
                    (S/N = $\frac{f_{\rm moon}}{\sigma_{tot}}$). The dashed 
                    vertical line at $\approx$ 7\farcs8, 11\farcs1, and
                    17\farcs8, respectively (scaling with $\lambda_{\rm c}$ 
                    of the filter) indicates a radius inside which
                    the uncertainty increases and the S/N degrades noticeably 
                    due to PSF residuals.
               }
         \label{fig:sndepondist}
   \end{figure*}

\subsection{Photometry results}

In Fig.~\ref{fig:sndepondist}, we plotted the PSF photometry fluxes and their uncertainties 
and the corresponding S/N of the individual measurements depending on the distance 
of the Uranian moon from the Uranus position for each filter. As a general feature, we note that the 
uncertainties increase and, hence, the S/N degrades noticeably inside a certain radius, which is
$\approx$ 7\farcs8, 11\farcs1, and 17\farcs8 for 70, 100, and 160$\mu$m, respectively (these
radii scale with $\lambda_{\rm c}$ of the filter). This is due to PSF residuals as seen in 
Figs.~\ref{fig:moonmaps70} to~\ref{fig:moonmaps160}. It should be noted that negative fluxes and, hence, the
negative S/N ratios do not occur since the PSF fit produces either positive fluxes or fails. For the
photometry of the individual moons, the following can be concluded:
\begin{itemize}
\item[$\bullet$] The S/N ratios of all Titania measurements are $>$10, so that all measurements
                 should be very reliable.
\item[$\bullet$] The S/N ratios of the Oberon measurements for epochs 2 -- 5 are all $>$10,
                 so that all these measurements should be very reliable. Regarding the measurements
                 of the first epoch, the moon is inside the critical radius. Nevertheless S/N
                 at 70 and 100\,$\mu$m are still $\gtrsim$10, so that their quality should be medium.
                 At 160$\mu$m the S/N ratios are $<$10, so that this photometry is less reliable.
\item[$\bullet$] For Umbriel, the S/N ratios of the 70 and 100\,$\mu$m measurements of epochs 1 and 5,
                 which are outside the critical radius, are of very high quality. The corresponding 
                 160\,$\mu$m fluxes have S/N ratios
                 $\lesssim$10, so that they are less reliable. The S/N ratios for the 70 and 100\,$\mu$m
                 measurements of epochs 2 -- 4 are between 10 -- 50, so that their quality should be
                 still medium. However, the corresponding 160\,$\mu$m fluxes have S/N ratios between 2 -- 5,
                 so that this photometry is less reliable.
\item[$\bullet$] For Ariel, the S/N ratios of the 70 and 100\,$\mu$m measurements of epochs 1 to 3
                 have medium to high quality ($\gtrsim$10 -- $<$100). The S/N ratios of the 70 and 
                 100\,$\mu$m measurements of epochs 4 and 5 are $\lesssim$10, so that they are less reliable.
                 The S/N ratios of all 160\,$\mu$m measurements are $\lesssim$3, so that they are likely
                 to be quite inaccurate.
\item[$\bullet$] For Miranda, which is considerably fainter than the other four moons and always close to Uranus, 
                 the S/N ratios of the 70\,$\mu$m measurements of epochs 1 and 3 are
                 in the range between 1 -- 3. These measurements indicate the order of flux, but they are
                 not very reliable. All other measurements at 70 and 100$\mu$m have S/N ratios $\lesssim$1,
                 so that individual measurements are not reliable at all. At 160\,$\mu$m S/N ratios are $<<$1. 
\end{itemize} 

Small S/N ratios indicate that there are some restriction in the subsequent analysis. It is important to bear in mind that this is not
a deficiency of the observational design since the original design with just one modular mini scan map was 
meant to observe Uranus, so that the S/N for the moons is, naturally, not optimal. 

The results of photometry from the individual scan maps are given in 
Tables~\ref{table:psfphotTitania} to ~\ref{table:psfphotMiranda} in Appendix~\ref{sect:tabmoonphot}.
For completeness we compile the Uranus photometry in Table~\ref{table:psfphotUranus} of 
Appendix~\ref{sect:taburanusphot}. This table gives both the actually measured flux $f_{\rm i, Uranus}^{\rm measured}$ 
and a flux normalised to a reference distance $f_{\rm i,Uranus}^{\rm distance corrected}$ which is needed in generating 
the PSF reference. The determination of the distance corrected Uranus flux is described below. 

Table~\ref{table:photUranianmoons} provides an overview of the Uranian moon photometry with mean fluxes.
For it a weighted mean moon-to-Uranus flux ratio was calculated from the individual photometry results 
listed in Tables~\ref{table:psfphotUranus} to~\ref{table:psfphotMiranda},
\begin{equation}
\label{eqn:meanMoonToUranusFluxRatio} 
\left( \frac{f_{\rm moon}}{f_{\rm Uranus}} \right)_{\rm \lambda}^{\rm mean} =
\frac{\sum_{i = 1}^{n} \left( \frac{f_{\rm i,moon}^{\rm measured}}{f_{\rm i,Uranus}^{\rm measured}} \right)_{\rm \lambda}~ \left( \frac{1}{\sigma_{\rm i,tot}} \right) ^2}{\sum_{i = 1}^{n} \left( \frac{1}{\sigma_{\rm i,tot}} \right)^2} , 
\end{equation}
using the $\sigma_{\rm i,tot}$ of the moon photometry as weights. For the calculation of 
the mean moon fluxes, a weighted mean Uranus flux at a mean distance of all {\it Herschel} observations is used. The 
mean Uranus distance is derived from the $\Delta_{\rm obs,i}$ of the 20 individual observations ($\Delta_{\rm obs,mean}$ 
= $\frac{\sum_{i = 1}^{20} \Delta_{\rm obs,i}}{20}$ = 20.024\,AU; for the $\Delta_{\rm obs,i}$ cf.\ 
Table~\ref{table:psfphotUranus}). 
Individual distance corrected Uranus fluxes $f_{\rm i,Uranus}^{\rm distance corrected}$ are determined by scaling the 
measured Uranus flux $f_{\rm i, Uranus}^{\rm measured}$ with the correction factor $c_{\rm dist} = 
\left( \frac{\Delta_{\rm obs,i}}{\Delta_{\rm obs,mean}} \right)^{2}$ (see also Table~\ref{table:corrtelbg} for values of 
$c_{\rm dist}$ per observation epoch). The weighted mean Uranus flux is then calculated as
\begin{equation}
\label{eqn:meanUranusFlux} 
f_{\rm Uranus,\lambda}^{\rm mean}$ = $\frac{\sum_{i = 1}^{n}\frac{f_{\rm i,Uranus}^{\rm distance corrected}}{\sigma_{\rm i,tot}^2}}{\sum_{i = 1}^{n} \frac{1}{\sigma_{\rm i,tot}}^2}, 
\end{equation}
using the $\sigma_{\rm i,tot}$ of the individual Uranus measurements as weights. These mean distance corrected fluxes of Uranus are listed in Table~\ref{table:photUranianmoons}, too. The mean moon flux is then calculated as \begin{equation} 
\label{eqn:meanMoonFlux}
f_{\rm moon,\lambda}^{\rm mean} = \left( \frac{f_{\rm moon}}{f_{\rm Uranus}} \right)_{\rm \lambda}^{\rm mean} 
\times f_{\rm Uranus,\lambda}^{\rm mean}.
\end{equation}

\begin{table*}[ht!]
\caption{Mean fluxes of the Uranian moons (Eq.~\ref{eqn:meanMoonFlux}) calculated from a weighted 
mean moon-to-Uranus flux ratio (Eq.~\ref{eqn:meanMoonToUranusFluxRatio}) and and a mean Uranus flux 
(Eq.~\ref{eqn:meanUranusFlux}) over the {\it Herschel} observation campaign. $\sigma_{tot}$ of the 
individual moon photometry was used as weight. The applied mean distance (20.024\,AU) normalised Uranus flux is given in the last line. 
n$_{\rm 70}$, n$_{\rm 100}$ and n$_{\rm 160}$ give the number of reliable measurements used in the determination 
of $\frac{f_{\rm moon}}{f_{\rm Uranus}}$. 
}             % title of Table
\label{table:photUranianmoons}      % is used to refer this table in the text
\centering                          % used for centering table
\begin{tabular}{l r c c r c c r c c}        % centered columns (4 columns)
\hline\hline                 % inserts double horizontal lines
            \noalign{\smallskip}
Object & n$_{\rm 70}$ & $\left( \frac{f_{\rm moon}}{f_{\rm Uranus}} \right)_{\rm 70}^{\rm mean}$ & $f_{\rm object,70}^{\rm mean}$& n$_{\rm 100}$ & $\left( \frac{f_{\rm moon}}{f_{\rm Uranus}} \right)_{\rm 100}^{\rm mean}$& $f_{\rm object,100}^{\rm mean}$& n$_{\rm 160}$ & $\left( \frac{f_{\rm moon}}{f_{\rm Uranus}} \right)_{\rm 160}^{\rm mean}$& $f_{\rm object,160}^{\rm mean}$ \\
            \noalign{\smallskip}
       &     & (10$^{-3}$)   &  (Jy) &    & (10$^{-3}$) &  (Jy) &   & (10$^{-3}$) &  (Jy) \\
            \noalign{\smallskip}
\hline
            \noalign{\smallskip}
Titania & 10 & 1.931$\pm$0.0095 & 1.663$\pm$0.008 & 10 & 1.619$\pm$0.0104 & 1.423$\pm$0.009 & 20 & 1.317$\pm$0.0074 & 0.873$\pm$0.005 \\
Oberon  & 10 & 1.847$\pm$0.0067 & 1.591$\pm$0.006 & 10 & 1.537$\pm$0.0074 & 1.351$\pm$0.007 & 20 & 1.217$\pm$0.0038 & 0.807$\pm$0.003 \\
Umbriel & 10 & 1.055$\pm$0.0140 & 0.909$\pm$0.012 & 10 & 0.869$\pm$0.0138 & 0.764$\pm$0.012 & 20 & 0.613$\pm$0.0187 & 0.406$\pm$0.012 \\
Ariel   & 10 & 0.876$\pm$0.0150 & 0.754$\pm$0.013 & 10 & 0.799$\pm$0.0428 & 0.702$\pm$0.038 & 20 & 0.294$\pm$0.0171 & 0.195$\pm$0.011 \\
Miranda & 10 & 0.261$\pm$0.0349 & 0.225$\pm$0.030 &  8 & 0.135$\pm$0.0164 & 0.119$\pm$0.016 & -- &        --        &        --       \\          
            \noalign{\smallskip}
\hline
            \noalign{\smallskip}
Uranus  & 10 &                 & 861.287$\pm$0.535 & 10 &                 & 879.061$\pm$0.488 & 20 &                 & 663.011$\pm$0.403 \\
            \noalign{\smallskip}
\hline\hline                 % inserts double horizontal lines
\end{tabular}
\end{table*}

The combined flux contribution of the four largest moons relative to the Uranus flux is 5.7$\times$10$^{-3}$, 
4.8$\times$10$^{-3}$ , and 3.4$\times$10$^{-3}$ at 70, 100, and 160\,$\mu$m, respectively. Hence, earlier 
published photometry of Uranus~\citep{mueller16} not subtracting the moon contribution is not invalidated 
by our new results. Rather, we specify more precisely the systematic error of the Uranus flux by its moons 
when using Uranus as a far-infrared prime flux calibrator. The fluxes in column f$_{\rm total}$ of 
Table~\ref{table:psfphotUranus} are very consistent with those in Table~B.1, column 'Flux' in~\citet{mueller16} 
($\frac{f_{\rm tot}^{\rm this paper}}{f_{\rm Uranus}^{\rm Mueller2016}}$ = 1.005$\pm$0.005, 1.009$\pm$0.003, and 
1.014$\pm$0.004 at 70, 100, and 160\,$\mu$m, respectively).

No dependence of the moon fluxes on the distance to Uranus is expected since all the moons have orbits
with small eccentricity. The variation of the angular separation of the moons to Uranus stands as a pure
projection effect that is due to the inclination of the Uranian system. 

\begin{figure}[ht!]
   \centering
   \includegraphics[width=0.50\textwidth]{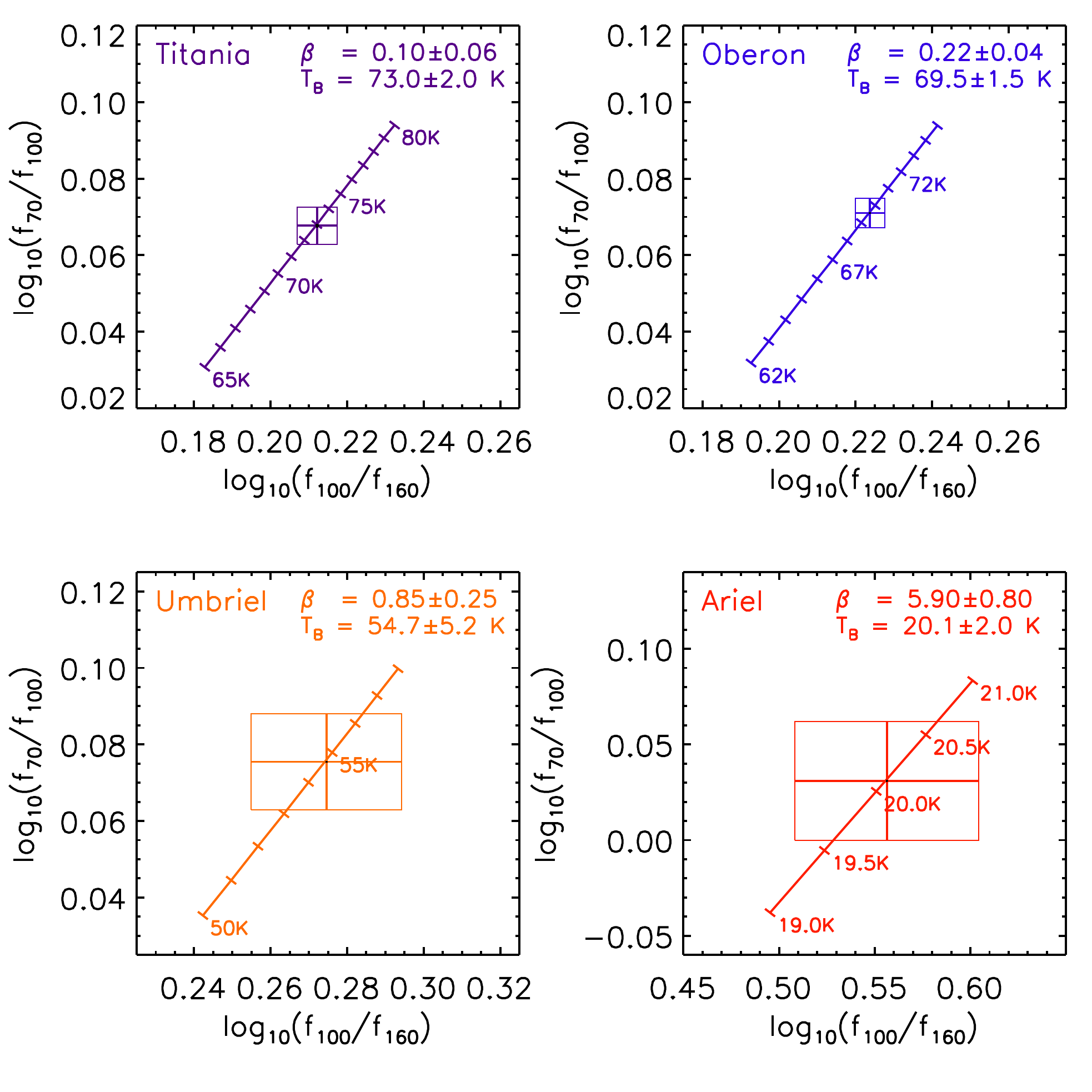}
      \caption{PACS two-colour diagrams for the Uranian moons.  
               The PACS fluxes are not colour corrected and refer to the PACS standard
               photometric reference SED $\nu \times f_{\nu}$ = const. The 
               boxes with the central cross indicate the uncertainty range of
               the measured colours (determined by 
               log$_{10} \left( \frac{f_{\lambda_{1}}+\sigma_{\lambda_{1}}}{f_{\lambda_{2}}-\sigma_{\lambda_{2}}} \right) $
               and log$_{10} \left( \frac{f_{\lambda_{1}}-\sigma_{\lambda_{1}}}{f_{\lambda_{2}}+\sigma_{\lambda_{2}}} \right)$, 
               respectively.)
               The straight lines with the temperature tick marks represent
               the colours of modified blackbodies calculated according to Eq.~(\ref{eqn:modbbcolour})
               for the displayed temperature range and with $\beta$ as indicated in the upper right
               corner of each panel. 
               Derived $\beta$ and T$_{\rm b}$ are used for a plausibility check of the PACS photometry.
              }
         \label{fig:pacstwocolourmap}
   \end{figure}

Another plausibility check of the PACS photometry can be obtained from FIR two-colour diagrams.
In Fig.~\ref{fig:pacstwocolourmap}, we show the individual two-colour diagrams for the Uranian moons.
The PACS fluxes are not colour corrected and refer to the PACS standard photometric reference SED 
$\nu \times f_{\nu}$ = const. Modified blackbody functions 
$\frac{\nu^{\beta}}{\nu_{0}^{\beta}}$\,B$_{\nu}$(T$_{\rm b}$) are good first order approximations for dust 
emission. Emission from the surface regolith of satellites is usually well approximated by pure blackbody 
emission, that is,\ $\beta$ should be zero or small. We calculated the two PACS colours of modified blackbody 
emission as: 
\begin{equation}
\label{eqn:modbbcolour}
 {\rm log}_{10} \left( \frac{\lambda_{1}^{(2-\beta)}\,\times\, 
               B_{\lambda}(\lambda_{1},T)\,\times\,cc_{\lambda_{1}}}{\lambda_{2}^{(2-\beta)}\,\times\,B_{\lambda}(\lambda_{2},T) 
               \,\times\,cc_{\lambda_{2}}} \right) 
.\end{equation}
The modified blackbody fluxes have been colour corrected (cc$_{\lambda}$) to the PACS photometric 
reference SED for a homogeneous comparison with the moon colours (cf.\ PACS Handbook~\citep{exter18}, 
formula 7.20 for the calculation for any SED shape). We checked which combination of 
$\beta$ and T$_{\rm b}$ best match the measured colours. Fig.~\ref{fig:pacstwocolourmap} shows the line
for the best matching $\beta$ value and a range of T$_{\rm b}$ which crosses the measured combination
of colours. For Titania and Oberon, the approximation of the measured colours by pure blackbodies
is quite good since  the match yields $\beta$ = 0.10$\pm$0.06, T$_{\rm b}$ = 73.0\,K$\pm$2.0\,K and 
$\beta$ = 0.22$\pm$0.04, T$_{\rm b}$ = 69.5\,K$\pm$1.5\,K, respectively.
For Umbriel, we find $\beta$ = 0.85$\pm$0.25, T$_{\rm b}$ = 54.7\,K$\pm$5.2\,K,
which shows that the 160\,$\mu$m flux is somewhat too low, so that the 
log($\frac{f_{\rm 100}}{f_{\rm 160}}$) value is too high, thus requiring higher $\beta$ values.
For Ariel, the fit gives $\beta$ = 5.9$\pm$0.8, T$_{\rm b}$ = 20.1\,K$\pm$2.0\,K, which
is a completely unphysical spectral energy distribution solution for this moon. We conclude 
that the mean 160\,$\mu$m flux is far too low (about a factor of 2) and unreliable, as suggested 
by the S/N analysis above.  On the other hand, the mean 70 and 100$\mu$m photometry appears to be 
adequate for all four moons, since the log($\frac{f_{\rm 70}}{f_{\rm 100}}$) values are all similar.
Because of the partial deficiency or incompleteness of the measured SEDs, we derived
colour correction factors for the PACS photometry from the best fitting models (see 
Table~\ref{table:derivedmodelparams}). 

%
%                                                One column figure
%-----------------------------------------------------------

\section{Auxiliary thermal data}

In addition to the new PACS measurements, we searched in the literature
to find more thermal data for the Uranian satellites. \citet{brown82}
presented standard broad-band Q filter measurements taken by the 3-m
IRTF\footnote{Infrared Telescope Facility on Mauna Kea, Hawaii} telescope.
We re-calibrated the Q-band magnitudes (after applying the listed monochromatic
correction factors and taking the specified -3.32\,mag for $\alpha$Boo) with the
template flux of 185.611\,Jy at 20.0\,$\mu$m~\citep{cohen96}. The resulting
flux densities are given in Table~\ref{tbl:irtf}.

\begin{table*}[ht!]
        \caption{Flux densities and uncertainties at 20.0\,$\mu$m based on measured Q-band magnitudes
                from~\citet{brown82} and re-calibrated via the reference standard star $\alpha$Boo.
                Data were taken in May 1982 with the IRTF (Miranda was not part of the study).
                r$_{\rm helio}$ is the light-time corrected heliocentric range, $\Delta_{\rm obs}$ 
                is the range of target centre wrt.\ the observer, i.e.\ IRTF, $\alpha$ is the phase 
                angle and "ang-sep" is the apparent angular separation from Uranus. The aspect angle
                during the measurements was around 163.5$^{\circ}$ which means that IRTF saw
                mainly the South-pole region of Uranus and the four satellites.
                % The corresponding "deldot" values are very small (-1.2, +1.0, +1.2, +0.4\,km/s for
                % Ariel, Umbriel, Titania, and Oberon, respectively) and the data contain therefore
                % no clean information about the leading or trailing hemispheres.
        }
        \label{tbl:irtf}
\centering                          % used for centering table
\begin{tabular}{l c c c c c c}
\hline\hline
            \noalign{\smallskip}
%Object & JD  & ang-sep [$^{\prime \prime}$] & FD [Jy] \\
Object & MJD  & r$_{\rm helio}$ & $\Delta_{\rm obs}$ & $\alpha$ & ang-sep  & $f_{\rm 20}$   \\
%Object & JD  & ang-sep  & FD  \\
       &     & [AU]      &    [AU]       & [deg] & [\arcsec]& [Jy] \\
\noalign{\smallskip} \hline \noalign{\smallskip}
Ariel (UI) [701]     & 45111.50 & 18.879 & 17.867 & 0.09 & 14.65 & 0.142 $\pm$ 0.026 \\ % 1982-May-22 12:00:00 APmag: 14.09 deldot=-1.2
Umbriel (UII) [702]  & 45109.33 & 18.879 & 17.869 & 0.21 & 19.86 & 0.131 $\pm$ 0.018 \\ % 1982-May-20 07:55:12 APmag: 14.74 deldot=+1.0
Titania (UIII) [703] & 45109.50 & 18.879 & 17.868 & 0.20 & 33.31 & 0.250 $\pm$ 0.024 \\ % 1982-May-20 12:00:00 APmag: 13.66 deldot=+1.2
Oberon (UIV) [704]   & 45108.40 & 18.880 & 17.871 & 0.26 & 43.23 & 0.280 $\pm$ 0.038 \\ % 1982-May-19 09:36:00 APmag: 13.87 deldot=+0.4
\noalign{\smallskip} \hline \hline \noalign{\smallskip}
\end{tabular}
\end{table*}

An important set of measurements was taken by Spitzer-IRS (14-37\,$\mu$m)~\citep{houck04}. 
We took the reduced and calibrated low-resolution spectra~\citep{lebouteiller11} and
high-resolution spectra~\citep{lebouteiller15} from
CASSIS\footnote{\url{https://cassis.sirtf.com/atlas/welcome.shtml}}, all related to
the Spitzer Program ID 71\footnote{ID 71: Observations of Outer Solar System Satellites and 
Planets; PI: J.\ R.\ Houck}.
The programme includes thermal emission spectroscopy between 10 and 40\,$\mu$m for
Uranus' synchronous satellites (among other objects), observing the leading and trailing
hemispheres at large separations from the planet. These observations were taken under
aspect angles between 96.8$^{\circ}$ and 104.6$^{\circ}$, that is, close to an equator-on view
of Uranus and its four satellites. An overview of these observations is given in 
Table~\ref{tbl:irs_cassis}.

\begin{table*}[ht!]
        \caption{        Overview of the {\it Spitzer}-IRS CASSIS spectra. r$_{\rm helio}$ is the light-time corrected 
                     heliocentric range, $\Delta_{\rm obs}$ is the range of target centre wrt.\ the observer, i.e.\ 
                     {\it Spitzer}, $\alpha$ is the phase angle and "ang-sep" is the apparent angular separation 
                     from Uranus. The observations were designed to observe either the leading or the trailing
                     hemisphere, which is indicated in column 'hemisp'. 'UsefulSpectrum' indicates the wavelength 
                     range not affected by Uranus stray-light; in the case of Ariel the whole spectrum is affected.
        }
        \label{tbl:irs_cassis}
\centering                          % used for centering table
\begin{tabular}{l c c c c c c c c}
\hline\hline
            \noalign{\smallskip}
Object         & MJD  & r$_{\rm helio}$ & $\Delta_{\rm obs}$ & $\alpha$ &  ang-sep & AORkey  & hemisp. & UsefulSpectrum\\
                     &            &  [AU]  &  [AU] & [deg] & [\arcsec] &    &    &  [$\mu$m] \\
\noalign{\smallskip} \hline \noalign{\smallskip}
Ariel (UI) [701]     & 53563.98055 &    20.067  &   19.547   &     2.54    &   13.45  & 4521984 &  L  &      --     \\
                     & 53323.25416 &    20.056  &   19.724   &     2.76    &   13.33  & 4522240 &  T  &      --     \\
Umbriel (UII) [702]  & 53695.76250 &    20.073  &   19.673   &     2.69    &   18.63  & 4522496 &  L  & 14.1 -- 22.0\\
                     & 53183.96250 &    20.049  &   19.582   &     2.62    &   18.72  & 4522752 &  T  & 14.1 -- 22.0\\
Titania (UIII) [703] & 53181.21458 &    20.048  &   19.624   &     2.67    &   30.61  & 4523008 &  L  & 14.1 -- 37.3\\
                     & 53716.59166 &    20.074  &   20.017   &     2.89    &   30.08  & 4523264 &  T  & 14.1 -- 37.3\\
Oberon (UIV) [704]   & 53184.28680 &    20.049  &   19.577   &     2.61    &   41.09  & 4523520 &  L  & 14.1 -- 30.0\\
                     & 53325.65486 &    20.056  &   19.763   &     2.80    &   40.70  & 4523776 &  T  & 14.1 -- 30.0\\
\noalign{\smallskip} \hline \hline \noalign{\smallskip}
\end{tabular}
\end{table*}

The 'optimal' extraction of the spectra from the CASSIS data assumes a perfect
point-like source which is certainly the case for the Uranian satellites at 18-20\,AU 
distance from the 0.85\,m Spitzer Space Telescope. We also looked into the high-resolution 
scans, but they only cover the longer wavelength range (19.5-36.9\,$\mu$m) and from the comparison
with the low-resolution spectra we concluded that they do not add any new information.
At longer wavelengths ($>$22\,$\mu$m for Umbriel and $>$30\,$\mu$m for Oberon),
the CASSIS spectra (both, low- and high-resolution ones) show significant additional
fluxes, which probably originate from the Uranus PSF (cf.\ Figs.~\ref{fig:tpmodel_umbriel} 
and~\ref{fig:tpmodel_oberon}, respectively). The Ariel spectra have fluxes which are
at least a factor of 2--3 too high and it seems the data are still affected by the
influence of Uranus (cf.\ Fig.~\ref{fig:tpmodel_ariel}). We eliminated those 
parts which show a strong deviation from a typical satellite thermal emission spectrum. We 
rebinned the spectra down to 10-15 wavelength points and added 10\% to the measurement 
errors to account for absolute flux calibration uncertainties in the close proximity 
of a very bright source.

\citet{cartwright15} presented IRTF/SpeX ($\sim$0.81 - 2.42\,$\mu$m) and
Spitzer/IRAC (3.6, 4.5, 5.8, and 8.0\,$\mu$m) measurements. But even at 8\,$\mu$m,
the measured fluxes are dominated by reflected sunlight. In the most favourable
case, the thermal contribution was still well below 10\%. We therefore excluded
these measurements from our radiometric studies.

\citet{hanel86}  studied the Uranian system with infrared observations obtained
by the infrared interferometer spectrometer (IRIS) on Voyager 2. The measurements
were taken for Miranda and Ariel and cover the range between 200 and 500\,cm$^{-1}$ 
(20-50\,$\mu$m). The South polar region was seen for both targets (under phase angles 
of 38$^{\circ}$ for Miranda and 31$^{\circ}$ for Ariel). They measured a maximum brightness 
temperature near the subsolar point, T$_{SS}$, of 86$\pm$1\,K and 84$\pm$1\,K for Miranda 
and Ariel, respectively. We tested our final model solutions against these two brightness 
temperatures.

%
%_____________________________________________________________
%                                             Two column Table 
%_____________________________________________________________
%
\begin{table*}[ht!]
        \caption{TPM input parameters for the radiometric calculations and the 
                interpretation of the obtained/available mid-/far-IR flux densities.
                The numbers are taken from~\citet{karkoschka01}:
                H$_V$ are the mean values between V$_{max}$ and V$_{min}$ with an
                uncertainty of 0.04\,mag, D$_{eff}$ was calculated from the specified
                radii (both from Table IV). The phase integral $q$ and the Bond albedo
                $A$ are from Table VII ($q$ and $qI_0/F$, respectively). The geometric
                albedo $p_V$ was calculated via $A = p\,q$, with $p/p_V \approx 1.0$
                \citep{morrison79}.
        }
        \label{tbl:tpm_input}
\centering
\begin{tabular}{l c c c c c c}
   \hline\hline
            \noalign{\smallskip}
Object                  & H$_{V}$ & A                & q             & p$_V$ & D$_{eff}$ & Orbital period \\
                        & [mag]  &                  &               &       & [km]      & [days] \\
\noalign{\smallskip} \hline \noalign{\smallskip}                                                  % G=(q-0.29)/0.684
Ariel (UI) [701]     &  0.99  & 0.230$\pm$0.025 & 0.43$\pm$0.05 & 0.53  &  1159.0   &   2.520 \\  % 0.205
Umbriel (UII) [702]  &  1.76  & 0.100$\pm$0.010 & 0.39$\pm$0.04 & 0.26  &  1170.0   &   4.144 \\  % 0.146
Titania (UIII) [703] &  0.78  & 0.170$\pm$0.015 & 0.46$\pm$0.05 & 0.37  &  1578.0   &   8.706 \\  % 0.249
Oberon (UIV) [704]   &  0.99  & 0.140$\pm$0.015 & 0.44$\pm$0.05 & 0.31  &  1522.0   &  13.463 \\  % 0.219
Miranda (UV) [705]   &  3.08  & 0.200$\pm$0.030 & 0.44$\pm$0.07 & 0.45  &   474.0   &   1.413 \\  % 0.219
\noalign{\smallskip} \hline \hline \noalign{\smallskip}
\end{tabular}
\end{table*}

\section{Thermophysical modelling of the Uranian moons}

For the interpretation of the available thermal IR fluxes, we used the thermophysical model
(TPM) by~\citet{lagerros96, lagerros97, lagerros98} and~\citet{mueller98, mueller02}. The
calculations are based on the true observer-centric illumination and observing geometry
for each data point (topocentric for IRTF, Herschel-/Spitzer-centric). The model considers
a one-dimensional heat conduction into the surface, controlled by the thermal inertia.
The surface roughness is implemented via segmented hemispherical craters where the
effective r.m.s.\ of the surface slopes is controlled by the crater depth-to-radius ratio
and the surface coverage of the craters~\citep{lagerros98}. Additional input parameters
are the object's thermal mid-/far-IR emissivity (assumed to be 0.9), the absolute
V-band magnitudes H$_V$ of the Uranian satellites, the phase integrals $q$, the measured
sizes $D_{eff}$ and albedos p$_V$. The H$_{\rm V}$ is only relevant in cases where we solve 
for radiometric size-albedo solutions. In cases where we keep the size fixed, H$_{\rm V}$ is 
not used. Table~\ref{tbl:tpm_input} summarises these values.
For the satellites' rotation properties, we assume a spin-axis orientation perpendicular
to Uranus' equator (orbital inclinations are below 0.5$^{\circ}$, only for Miranda it
is 4.2$^{\circ}$), and a (presumed) synchronous rotation.

Using the above properties (and their uncertainties) allows us now to determine
the moons' thermal properties. We vary the surface roughness from very smooth
(r.m.s.\ of surface slopes $<$0.1) up to very rough surfaces (r.m.s.\ of surface
slopes $>$0.7). In addition, low-conductivity surfaces can have very small
thermal inertias (here, we use a lower limit of 0.1\,Jm$^{-2}$s$^{-0.5}$K$^{-1}$)
and compact solid surfaces have high conductivities (we consider thermal inertias
up to 100\,Jm$^{-2}$s$^{-0.5}$K$^{-1}$).

One problem of radiometric studies in general is related to objects seen pole-on, 
or very close to pole-on (especially for distant objects where the Sun and observer face 
the same part of the surface). In these cases, there is no significant heat transfer 
to the night side and it is much more difficult to constrain the object's thermal 
properties. A pole-on geometry is connected to an aspect angle of 0$^{\circ}$ 
(north pole) or 180$^{\circ}$ (south pole), while an equator-on geometry has 90$^{\circ}$. 
During the 1980s (including the IRTF measurements, but also the time of the Voyager 2 
flyby) mainly the south pole region (of Uranus and also the synchronous satellites) 
was visible, the 2004/2005 Spitzer measurements were taken at aspect angles between 
about 97$^{\circ}$ and 105$^{\circ}$, the 2010-2012 Herschel observations saw the 
Uranus system under aspect angles between about 70$^{\circ}$ and 81$^{\circ}$,
meaning that\ both were close to an equator-on view. The phase angles are typically small 
(below 3$^{\circ}$) and the measured signals are in all cases related to almost 
fully illuminated objects.

Representative examples of our thermophysical models for the epoch 2011-07-12 (period 2) 
covering the wavelength range between 5 -- 300\,$\mu$m are shown for Oberon, Titania, Umbriel,
and Ariel in Figs.~\ref{fig:tpmodel_oberon} to~\ref{fig:tpmodel_ariel}, including 
the corresponding surface temperature maps.
All derived (and approved) flux densities for the 5 satellites, as well
as the corresponding best TPM SEDs will be made available by the
Herschel Science Centre through 'User Provided Data Products' in the
Herschel Science Archive\footnote{\url{http://archives.esac.esa.int/hsa/whsa/}}.
Our {\it Herschel} flux densities and the auxiliary photometry shall also be
imported into the 'Small Bodies: Near and Far' (SBNAF) data 
base\footnote{\url{https://ird.konkoly.hu/}} for thermal infrared observations
of Solar System's small bodies~\citep{szakats20}.

At the time of Voyager flyby, when the south pole of the moons was facing the Sun, 
maximum surface temperatures reached or exceeded 85\,K, but nighttime polar 
temperatures are predicted to drop to 20 or 30\,K, because each pole spends 
about 40\,yr in darkness~\citep{veverka91}. This means that under an 
illumination geometry close to pole-on the satellite surface is hotter than under an
illumination geometry close to equator-on when heat transport to the night side
results in a colder surface temperature. Therefore, a simple scaling of photometric
measurements taken under different illumination geometry by just correcting for 
different ranges of target centre with regard to the observer will not allow a direct
comparison. This has to be kept in mind for Figs.~\ref{fig:tpmodel_oberon} 
to~\ref{fig:tpmodel_ariel}, where there seems to be some flux inconsistency between
the IRTF fluxes and the thermophysical model fluxes matching the PACS observations.  
When taking the illumination geometry at the time of the IRTF measurements into account 
for the TPM, the consistency is very good, that is, for Titania 
$\frac{f_{\rm IRTF}}{f_{\rm TPM(t_{\rm IRTF})}}$ = $\frac{0.250\,Jy}{0.233\,Jy}$.

As part of the analysis we also looked into differences between the leading (LH)
and trailing hemispheres (TH) of the satellites. The tidally-locked and large
satellites display stronger H$_{2}$O ice bands on the leading hemispheres, but this
effect decreases with distance from Uranus. In addition, Titania and Oberon show
spectrally red material on their leading hemisphere. \citet{cartwright18}
discuss the possible origin of the hemispherical differences and speculate
that inward-migrating dust from the irregular satellites might be the cause of
the observed H$_{2}$O ice bands and red material differences in the two hemispheres.
Since the IRTF measurements viewed only the south pole regions, the measured fluxes 
did not allow such a separation. The Spitzer-IRS measurements were aiming for epochs 
were either the leading or trailing faces were seen. The measurements were timed 
for maximum elongation from Uranus, which are close to the epochs of the minimum
and maximum heliocentric range-rate values. This was possible since the Uranus 
system was seen almost equator-on. The Herschel measurements were not timed to catch
the objects at their range-rate maxima. Therefore, we consider {\it Herschel} 
observations as leading and trailing cases, if the apparent (heliocentric) 
range-rates were larger than 2/3 of the maximum possible. In all other cases, 
the observed signals are attributed to both hemispheres (labelled LH, TH, or BH 
in column $\frac{\dot{r}_{\rm helio}}{|{\dot{r}_{\rm helio}^{\rm max}}|}$ of 
Tables~\ref{table:psfphotTitania} to~\ref{table:psfphotMiranda}).

\subsection{Oberon}

A standard radiometric analysis of the combined Herschel-PACS, IRTF, and Spitzer-IRS
measurements leads to a range of size-albedo-thermal solutions with reduced $\chi^2$-values
close to or below 1.0. However, the optimum solutions resulted in an effective diameter
which is about 3-5\% above the object's true size (and a geometric albedo of 0.29),
connected to a thermal inertia in the range 20 -- 40\,Jm$^{-2}$s$^{-0.5}$K$^{-1}$.

When we keep the diameter fixed to 1522\,km (with p$_V$=0.31) we can still find acceptable 
solutions (with reduced $\chi^2$-values close to 1): for an intermediate level of surface 
roughness (r.m.s.\ of surface slopes between 0.3 and 0.7) and thermal inertias between 
9 and 33\,Jm$^{-2}$s$^{-0.5}$K$^{-1}$ (higher thermal inertias are connected to higher levels 
of surface roughness and vice versa).

The best solution is found for a thermal inertia of around 20\,Jm$^{-2}$s$^{-0.5}$K$^{-1}$
and an intermediate level of surface roughness (r.m.s.\ = 0.5). We confirmed
the solution by using a modified input data set where the close-proximity PACS data
(at only 6\arcsec apparent separation from Uranus) were eliminated.

%
%                                                One column figure
%-----------------------------------------------------------
   \begin{figure}[ht!]
   \centering
   \includegraphics[width=0.47\textwidth]{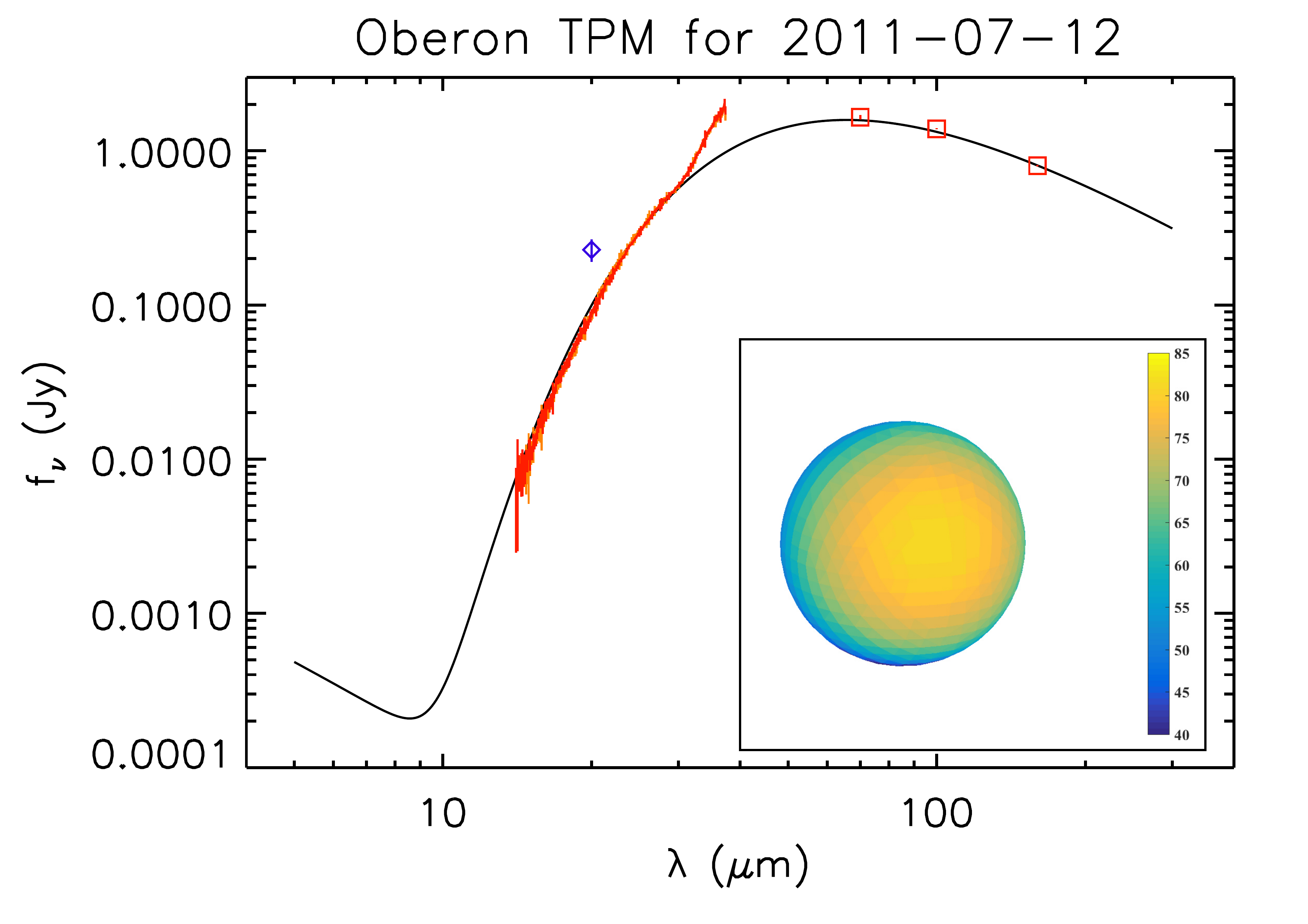}
      \caption{Thermophysical model of Oberon between 5 and 300\,$\mu$m (black line) for 
               the second epoch (2011-07-12). Photometric measurements are PACS observations (red
               boxes), the IRTF observation (blue diamond), and the {\it Spitzer}-IRS CASSIS data
               (orange spectrum: leading hemisphere observation; red spectrum: trailing hemisphere
               data). The CASSIS spectra suffer from Uranus stray light longward of 30\,$\mu$m.
               The IRTF and IRS data were re-scaled to the model epoch with 
               $\left( \frac{\Delta_{\rm obs,IRS,IRTF}}{\Delta_{\rm obs,2011-07-12}} \right)^{2}$.
               Nevertheless, the IRTF flux appears to be too high with regard to the model, because
               the IRTF measurement was done under close to pole-on illumination, when the moon was
               hotter, while the model reflects more a viewing geometry close to equator-on, when
               the moon was colder due to heat transport to the night side.    
               The insert shows the resulting TPM surface temperature map of Oberon for the range 
               40 -- 85\,K.
                }
         \label{fig:tpmodel_oberon}
   \end{figure}

The leading (PACS 2$^{nd}$ epoch, IRS-1) / trailing (PACS 4$^{th}$ and 5$^{th}$ epoch, IRS-2)
analysis did not show any clear differences: both data subsets led to the same thermal
properties (thermal inertia of 20\,Jm$^{-2}$s$^{-0.5}$K$^{-1}$) with very similar
reduced $\chi^2$ values. From the available measurements, we cannot distinguish the leading
and trailing hemispheres. The IRS spectra confirm this finding: both spectra (in comparison
with the corresponding optimum TPM prediction) agree within 5\%, except at the shortest end
below 16\,$\mu$m where the difference is about 10\%.

The overall consistency ($\frac{f_{\rm moon,cc}}{f_{\rm model}}$) of the models with colour 
corrected PACS fluxes for all five periods is 0.95$\pm$0.02, 0.94$\pm$0.06, and 1.00$\pm$0.02 
at 70, 100, and 160\,$\mu$m, respectively.
Excluding the first epoch photometry, where Oberon was at an only 6\arcsec apparent separation 
from Uranus, gives ratios of 0.95$\pm$0.01, 0.96$\pm$0.01, and 1.00$\pm$0.02. An illustrated 
comparison for epoch 2 is shown in Fig.~\ref{fig:tpmodel_oberon}.

% measured flux/model comparison; possibly exclude first measurement set
%  70um: 1.056+-0.018    1.055+-0.013 without 1st period
% 100um: 1.067+-0.058    1.043+-0.012 without 1st period
% 160um: 1.001+-0.021    0.997+-0.019 without 1st period

\subsection{Titania}

The standard thermal analysis (PACS, IRTF, IRS) led to reduced $\chi^2$ values close to 1
for a radiometric size which is again 2--4\% larger than the true value, and a thermal inertia
of 9-31\,Jm$^{-2}$s$^{-0.5}$K$^{-1}$. All data are taken at sufficient separation from Uranus
($>$14\arcsec), but the IRS spectra still seem to be contaminated around 30\,$\mu$m.
Adding the constraints from Titania's known size and albedo (see Table~\ref{tbl:tpm_input}), 
leads to thermal inertia values of 5-15\,Jm$^{-2}$s$^{-0.5}$K$^{-1}$, with optimum values of
7-11\,Jm$^{-2}$s$^{-0.5}$K$^{-1}$, again for an intermediate level of surface roughness
(r.m.s.\ = 0.4). We explicitly tested also other solutions for the thermal inertia, but
a value of 20\,Jm$^{-2}$s$^{-0.5}$K$^{-1}$, as found for Oberon, caused already severe
problems in fitting our thermal measurements. The TPM predictions for the PACS measurements
would decrease by 5-15\% and the match to the observations would not be acceptable (outside
3-$\sigma$). The higher thermal inertia predictions would fit the IRS spectrum at short
wavelength below 22\,$\mu$m and beyond 35\,$\mu$m, but not in between. Overall, we can exclude
a thermal inertia larger than about 15\,Jm$^{-2}$s$^{-0.5}$K$^{-1}$ and smaller than about
5\,Jm$^{-2}$s$^{-0.5}$K$^{-1}$ for Titania.

%
%                                                One column figure
%-----------------------------------------------------------
   \begin{figure}[ht!]
   \centering
   \includegraphics[width=0.47\textwidth]{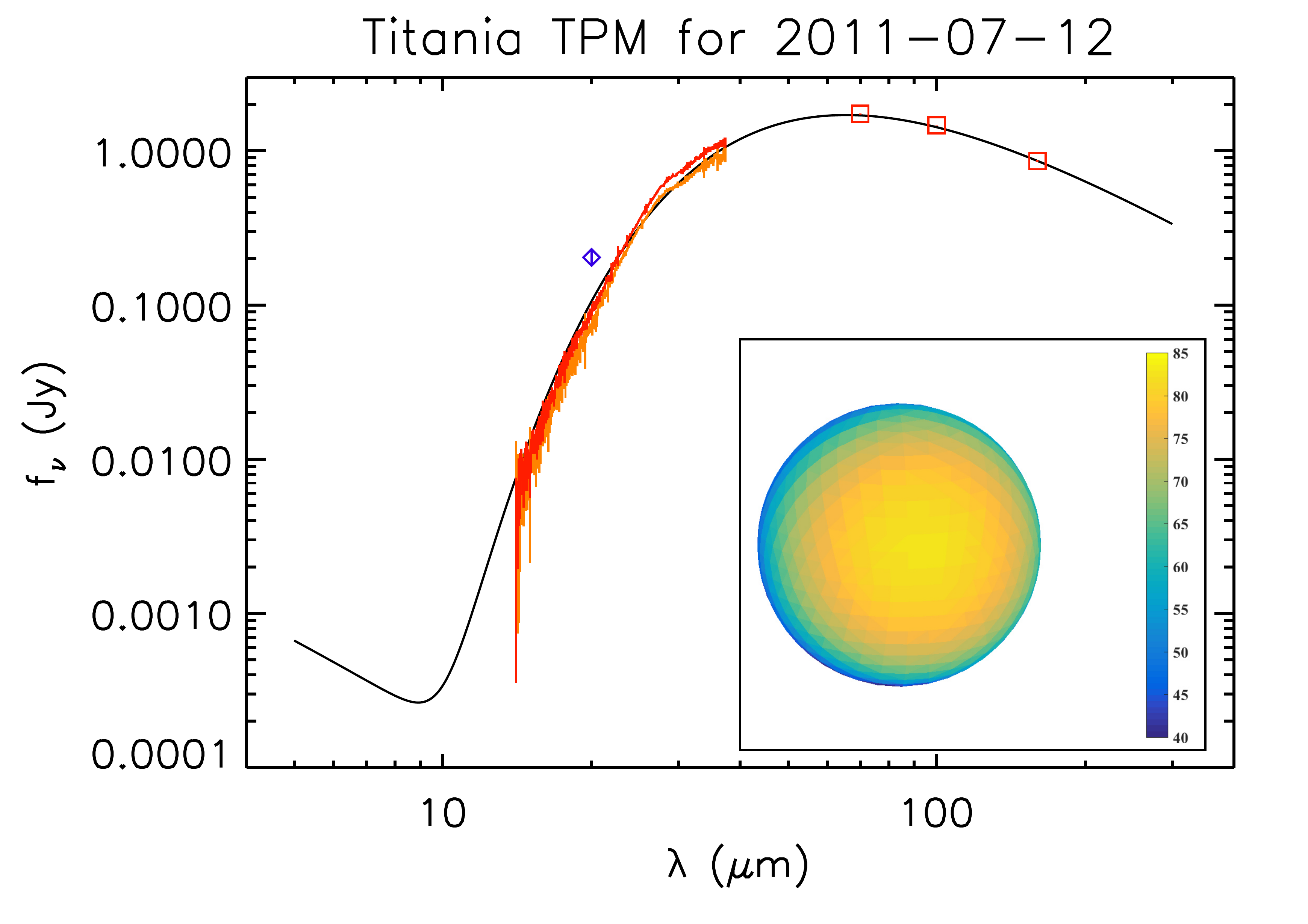}
      \caption{Thermophysical model of Titania between 5 and 300\,$\mu$m (black line) for 
               the second epoch (2011-07-12). Photometric measurements are PACS observations (red
               boxes), the IRTF observation (blue diamond) and the {\it Spitzer}-IRS CASSIS data
               (orange spectrum: leading hemisphere observation; red spectrum: trailing hemisphere
               data). For an explanation of the IRTF and IRS data re-scale, see text and 
               caption of Fig.~\ref{fig:tpmodel_oberon}.
               The insert shows the resulting TPM surface temperature map of Titania for the 
               range 40 -- 85\,K.
                }
         \label{fig:tpmodel_titania}
   \end{figure}

The Herschel-PACS measurements of Titania cover mainly the trailing hemisphere (1$^{st}$, 
2$^{nd}$, and 5$^{th}$ epoch) and a clean leading or trailing analysis is not possible. However, 
we ran our analysis on these trailing hemisphere measurements (three PACS epochs and IRS-2)
and compared the results with the leading hemisphere IRS-1 measurement. The trailing data
give a very consistent (reduced $\chi^2$ of 0.7) solution with a thermal inertia between
5 and 9\,Jm$^{-2}$s$^{-0.5}$K$^{-1}$. But this solution overestimates the fluxes from the 
IRS-1 spectrum. A higher thermal inertia of 9 -- 15\,Jm$^{-2}$s$^{-0.5}$K$^{-1}$
is needed to explain the leading hemisphere data. There are no PACS data to confirm this
finding and due to the reduction and stray light residuals in the IRS spectra so close to 
Uranus; thus, this can only be considered an indication of differences between both hemispheres.

The overall consistency ($\frac{f_{\rm moon,cc}}{f_{\rm model}}$) of the models with colour 
corrected PACS fluxes for all five epochs are 0.97$\pm$0.02, 0.97$\pm$0.03, and 0.99$\pm$0.03 
at 70, 100, and 160\,$\mu$m, respectively. An illustrated comparison for epoch 2 is shown 
in Fig.~\ref{fig:tpmodel_titania}.

% measured flux/model comparison; 
%  70um: 1.026+-0.015
% 100um: 1.028+-0.027
% 160um: 1.010+-0.034

\subsection{Umbriel}

The standard radiometric search for the object's best size, albedo, and thermal properties
led to an unrealistically small thermal inertia below 5\,Jm$^{-2}$s$^{-0.5}$K$^{-1}$ and
a diameter of just below 1100\,km (p$_V$$\approx$0.30), with reduced $\chi^2$ values close
to 1.0. The size and albedo values are in clear contradiction to the published values
of 1170\,km and p$_V$=0.26~\citep{karkoschka01}. Taking the larger size requires a higher
thermal inertia to fit all observed fluxes. Intermediate levels of surface roughness,
combined with thermal inertias between 5 and 15\,Jm$^{-2}$s$^{-0.5}$K$^{-1}$ seem to fit
best (reduced $\chi^2$ values just below the 1.7 threshold). 

However, if we look at the observation-to-model ratios we can identify a few observations 
which suffer from low signal-to-noise ratios (all five PACS measurements at 160\,$\mu$m 
and both IRS 15\,$\mu$m spectral parts have S/N$\le$3), but our radiometric weighted 
solutions handle correctly the proper flux errors. More problematic are the long-wavelengths 
fluxes when Umbriel had only a small apparent separation (below 7\arcsec) from Uranus:
the PACS 100 and 160\,$\mu$m measurements from 26-Dec-2011 and also the
long-wavelength parts of both IRS spectra beyond about 22\,$\mu$m seem to be affected
by residual Uranus PSF features. Excluding these problematic measurements, we obtained
reduced $\chi^2$-values close to 1.0 (for the fixed size of 1170\,km), with a preference 
for a lower surface roughness (around 0.3) than for Oberon and Titania, and a thermal 
inertia in the range between 5 and 12\,Jm$^{-2}$s$^{-0.5}$K$^{-1}$.

%
%                                                One column figure
%-----------------------------------------------------------
   \begin{figure}[ht!]
   \centering
   \includegraphics[width=0.47\textwidth]{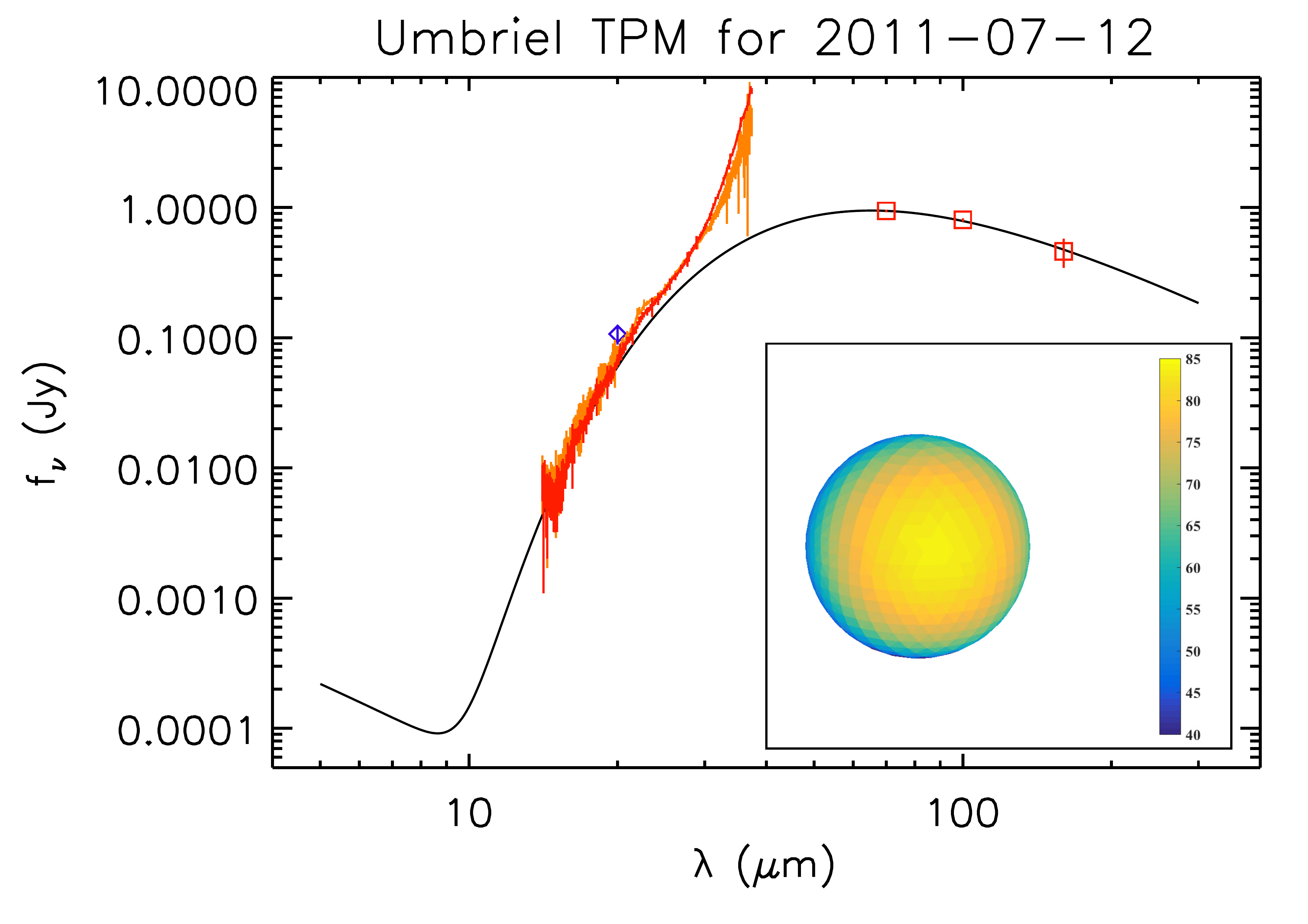}
      \caption{Thermophysical model of Umbriel between 5 and 300\,$\mu$m (black line) for 
               the second epoch (2011-07-12). Photometric measurements are PACS observations (red
               boxes), the IRTF observation (blue diamond), and the {\it Spitzer}-IRS CASSIS data
               (orange spectrum: leading hemisphere observation; red spectrum: trailing hemisphere
               data). The CASSIS spectra suffer from Uranus stray light longward of 22\,$\mu$m.
               For an explanation of the IRTF and IRS data re-scale, see text and 
               caption of Fig.~\ref{fig:tpmodel_oberon}. 
               The insert shows the resulting TPM surface temperature map of Umbriel for the range 
               40 -- 85\,K.
                }
         \label{fig:tpmodel_umbriel}
   \end{figure}

The Umbriel data have a well-balanced coverage of the leading (PACS 1$^{st}$ epoch, IRS-1) and
trailing (PACS 5$^{th}$ epoch, IRS-2) hemispheres. Separate  fits to the data for the two 
hemispheres led to the following results: the fits to the trailing hemisphere data are excellent 
(reduced $\chi^2$ well below 1.0) with a thermal inertia at the lower end (around 
5\,Jm$^{-2}$s$^{-0.5}$K$^{-1}$). The leading hemisphere data show an indication for a slightly 
higher thermal inertia closer to 10\,Jm$^{-2}$s$^{-0.5}$K$^{-1}$. However, within the error bars, 
both sets can be fit with an intermediate solution.

The overall consistency ($\frac{f_{\rm moon,cc}}{f_{\rm model}}$) of the models with colour corrected 
PACS fluxes for all five epochs is 1.02$\pm$0.05, 1.05$\pm$0.08, and 1.16$\pm$0.12 
at 70, 100, and 160\,$\mu$m, respectively. Excluding the second and third epoch, where 
Umbriel is at less than 9\arcsec~apparent separation from Uranus, slightly improves 
the 70\,$\mu$m ratio (1.01$\pm$0.04), but not the 100 and 160\,$\mu$m ratios (1.05$\pm$0.08 
and 1.23$\pm$0.13, respectively). An illustrated comparison for epoch 2 is shown in 
Fig.~\ref{fig:tpmodel_umbriel}.

% measured flux/model comparison; 
%  70um: 0.976+-0.053     0.991+-0.043 excluding 2nd and 3rd period
% 100um: 0.955+-0.075     0.953+-0.080 excluding 2nd and 3rd period 
% 160um: 0.862+-0.124     0.811+-0.125 excluding 2nd and 3rd period

%
%_____________________________________________________________
%                                             Two column Table   
%_____________________________________________________________
%
\begin{table*}[ht!]
	\caption{Overview of the best thermophysical model parameter ranges and the resulting PACS filter
		colour correction factors cc.
	}             
	\label{table:derivedmodelparams}      
	\centering
	\begin{tabular}{l c c c c c}
		\hline\hline
		\noalign{\smallskip}
		Object              &           $\Gamma$        & Surface Roughness & cc$_{\rm 70}$ & cc$_{\rm 100}$ & cc$_{\rm 160}$ \\
		& [Jm$^{-2}$s$^{-0.5}$K$^{-1}$]&  [r.m.s.\ ]       &             &               &              \\
		\noalign{\smallskip}
		\hline
		\noalign{\smallskip}
		Titania (UIII) [703]  &     7 -- 11            &       0.4           & 0.984 & 0.999 & 1.032 \\
		Oberon (UIV) [704]    &       20               &       0.5           & 0.984 & 0.999 & 1.032 \\
		Umbriel (UII) [702]   &     5 -- 12            &       0.3           & 0.984 & 0.999 & 1.032 \\
		Ariel (UI) [701]      & 5--13 (LH), 13--40 (TH)&       0.5           & 0.983 & 0.997 & 1.030 \\
		Miranda (UV) [705]    &      $<$ 20            &       0.5           & 0.983 & 0.998 & 1.030 \\
		\noalign{\smallskip}
		\hline
	\end{tabular}
\end{table*}

\subsection{Ariel}
\begin{figure}[ht!]
   \centering
   \includegraphics[width=0.47\textwidth]{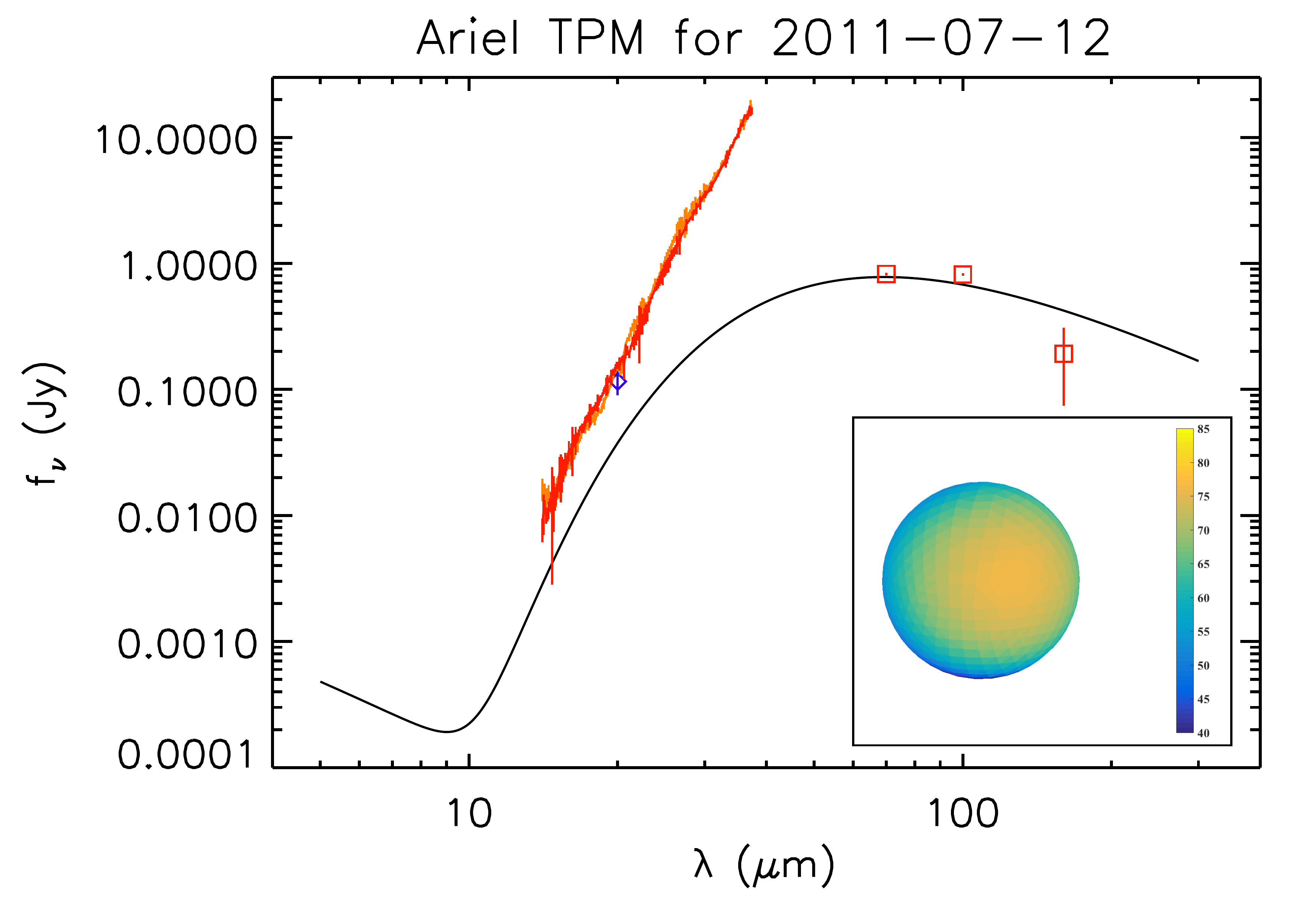}
      \caption{Thermophysical model of Ariel between 5 and 300\,$\mu$m (black line) for 
               the second epoch (2011-07-12). Photometric measurements are PACS observations (red
               boxes), the IRTF observation (blue diamond). and the {\it Spitzer}-IRS CASSIS data
               (orange spectrum: leading hemisphere observation; red spectrum: trailing hemisphere
               data). The CASSIS spectra suffer from Uranus stray light over their full wavelength 
               range. For an explanation of the IRTF and IRS data re-scale, see the text and 
               caption of Fig.~\ref{fig:tpmodel_oberon}.
%               Nevertheless, the IRTF flux appears to be too high with regard to the model, because
%               the IRTF measurement was done under close to pole-on illumination, when Ariel was
%               hotter, while the model reflects more a viewing geometry close to equator-on, when
%               Ariel was colder due to heat transport to the night side.    
               The insert shows the resulting TPM surface temperature map of Ariel for the range 
               between 40 -- 85\,K.
                }
         \label{fig:tpmodel_ariel}
   \end{figure}

Ariel was also seen by IRTF, Spitzer-IRS (leading and trailing), and Herschel-PACS.
However, the thermal IR fluxes are even lower than for Umbriel and the apparent
distances to Uranus are smaller. Two PACS measurement sequences (08-Jun-2012 and 14-Dec-2012)
were taken with Ariel below 6\arcsec separation and had to be skipped. None of the
IRS spectra are usable: the fluxes are too high by factors of 3\,--\,45 (cf.\ Fig.~\ref{fig:tpmodel_ariel}).

%
%                                                One column figure
%-----------------------------------------------------------

A first radiometric analysis (just PACS 70/100\,$\mu$m fluxes and the IRTF flux) produced
sizes between about 1100 and 1400\,km and only a very weak constraint on the thermal
inertia (values below 100\,Jm$^{-2}$s$^{-0.5}$K$^{-1}$).
Using the size constraint of 1159\,km ($a \times b$: 581\,km $\times$ 578\,km~\citep{karkoschka01})
requires a thermal inertia between 6 and 25\,Jm$^{-2}$s$^{-0.5}$K$^{-1}$ for an intermediate
surface roughness. However, the reduced $\chi^2$ is larger than 2.0 and a closer inspection shows
a clear separation in the fits to the leading and trailing hemispheres. Taking the PACS
measurements for the leading hemisphere (2011-Jul-12) and the trailing hemisphere
(2010-Dec-13, and 2011-Dec-26) separately gives much better fits (reduced $\chi^2$
close to 1.0), indicating a lower thermal inertia (5 -- 13\,Jm$^{-2}$s$^{-0.5}$K$^{-1}$) for
the leading hemisphere, and a higher thermal inertia (13 -- 40\,Jm$^{-2}$s$^{-0.5}$K$^{-1}$)
for the trailing hemisphere. Although the IRS spectra cannot be used for the radiometric
studies, the flux levels for the leading hemisphere are about 5-10\% higher. This also
points to a lower thermal inertia for the leading side compared to the trailing side.
With our final solution, we calculated a maximum brightness temperature of about 
86\,K for the South-pole viewing geometry in early 1986. This compares very well with 
the maximum brightness temperature of 84$\pm$1\,K seen by Voyager-2/IRIS~\citep{hanel86}.

The overall consistency ($\frac{f_{\rm moon,cc}}{f_{\rm model}}$) of the models with colour 
corrected PACS fluxes for all five epochs is 1.07$\pm$0.12, 0.94$\pm$0.29, and 2.09$\pm$0.48 
at 70, 100, and 160\,$\mu$m, respectively. The high 160\,$\mu$m ratio of $\gtrsim$2 is due 
to the fact that the measured values are all, except for one, far too low. Excluding the fourth 
and fifth epoch, where Ariel is at less than 6\arcsec apparent separation from Uranus, improves 
the consistency at 70 and 100\,$\mu$m considerably with ratios of 0.99$\pm$0.04, 1.00$\pm$0.17, 
respectively. However, due to the generally low 160\,$\mu$m fluxes, this ratio (2.20$\pm$0.11) 
does not improve. An illustrated comparison for epoch 2 is shown in Fig.~\ref{fig:tpmodel_ariel}.

% measured flux/model comparison; 
%  70um: 0.935+-0.119     1.008+-0.042 excluding 4th and 5th period
% 100um: 1.060+-0.289     1.003+-0.166 excluding 4th and 5th period
% 160um: 0.479+-0.204     0.454+-0.111 excluding 4th and 5th period

\subsection{Miranda}

For Miranda we have only the PACS measurements, but neither IRTF nor Spitzer-IRS data. 
The object was always within 10\arcsec~from Uranus and the contamination problems 
are severe. We eliminated all 160\,$\mu$m fluxes which are clearly completely off.
In addition, we skipped the second- and fourth-epoch data, when Miranda was only at
3\farcs4 and 4\farcs4 apparent distance, respectively. For the last
epoch, we also had to take out the 100\,$\mu$m data. 

In the end, only very few data points remained and the required coverage (in aspect angles, 
wavelengths, leading and trailing geometries, etc.) is missing for a robust 
radiometric analysis. With the size (474\,km) and albedo (p$_V$ = 0.45) we only obtained 
an upper limit of about 50\,Jm$^{-2}$s$^{-0.5}$K$^{-1}$ for Miranda's thermal inertia. Larger
values would force the TPM calculations to smaller fluxes which are not compatible
with the highest S/N detections by PACS (the upper limit goes down to 
20\,Jm$^{-2}$s$^{-0.5}$K$^{-1}$, if we consider only the best 70\,$\mu$m fluxes). The 
corresponding TPM calculations (with a thermal inertia below 20\,Jm$^{-2}$s$^{-0.5}$K$^{-1}$)
for the Voyager-2/IRIS measurements in January 1986 produce a maximum temperature
of about 87\,K, in excellent agreement with the 86$\pm$1\,K by~\citet{hanel86}.

% measured flux/model comparison; 
%  70um: 1.520+-0.782

\subsection{Discussion}

Table~\ref{table:derivedmodelparams} provides an overview of the derived model parameters.
Using these model SEDs, we also calculated the colour correction factors to be applied to the
measured PACS fluxes (cf.\ Tables~\ref{table:psfphotTitania} to~\ref{table:psfphotMiranda}).

How do the derived properties for the Uranian satellites compare with
thermal inertias of other satellites and distant TNOs?
\citet{lellouch13} analysed a large sample of TNOs and found a
$\Gamma$ = 2.5$\pm$0.5\,J m$^{-2}$ s$^{-1/2}$ K$^{-1}$ for objects
at heliocentric distances of r$_{helio}$ = 20 -- 50\,AU (decreasing values
for increasing heliocentric distance). The Uranian system is at about 
20\,AU and therefore one would expect (under the assumption of TNO-like
surfaces) to find low values, maybe up to 5\,J m$^{-2}$ s$^{-1/2}$ K$^{-1}$.

However, looking at dwarf planets, these general TNO-derived values are usually
exceeded: Haumea is at r$_{helio}$ = $\sim$51\,AU and it was found to have
a thermal inertia of around 10\,J m$^{-2}$ s$^{-1/2}$ K$^{-1}$~\citep{mueller18}.
The thermal inertias of Pluto and Charon (at r$_{helio}$ $>$ 30\,AU) are even larger:
   $\Gamma_{Pluto}$ = 16-26\,J m$^{-2}$ s$^{-1/2}$ K$^{-1}$ and
   $\Gamma_{Charon}$ = 9-14\,J m$^{-2}$ s$^{-1/2}$ K$^{-1}$ \citep{lellouch11,lellouch16}.
And putting the Pluto-Charon system closer to the Sun would increase the
values significantly (assuming that the T$^3$ term dominates in the thermal
conductivity, then the thermal inertia scales with $\propto$ r$^{-3/4}$;
see e.g. \citealt{delbo15}). In case of Pluto-Charon, the high $\Gamma$-values
are attributed to a large diurnal skin depth due to their slow rotation
($\sim$ P$^{1/2}$ dependence; see also discussion in~\citet{kiss19}).
In summary, the Uranian satellites Oberon, Titania, Umbriel, Ariel, and 
Miranda have thermal inertias which are higher than the very low values found 
for TNOs and Centaurs at 30\,AU heliocentric distance. It seems that the 
thermal properties of the icy satellite surfaces are closer to the properties 
found for the TNO dwarf planets Pluto and Haumea.

\section{Conclusions}

In this study, we successfully demonstrate an image processing technique for PACS photometer data,
allowing us to remove the bright central point spread function of Uranus and reconstructing 
source fluxes of its five major satellites on the order of 10$^{-3}$ of Uranus. We 
obtained reliable moon fluxes outside radii of 7\farcs8, 11\farcs1, and 17\farcs8 at
70, 100, and 160\,$\mu$m, respectively, which corresponds to $\approx$3$\times$ the HWHM 
of the standard PSF (FWHM$_{\rm PSF}$ = 5\farcs6, 6\farcs8 and 10\farcs7, respectively). 
For Titania and Oberon we have established full sets of 70, 100, and 160\,$\mu$m PSF photometry 
for all five observing epochs. For Umbriel, there are two epochs (1 \& 5) with high quality 
70 and 100\,$\mu$m photometry and for Ariel, there are three epochs (1 -- 3).
The 160\,$\mu$m photometry of these two moons is either of low quality (Umbriel) or
unreliable (Ariel). For Miranda, 70\,$\mu$m flux estimates could be obtained for two
epochs (1 \& 3).
This new FIR photometry and auxiliary photometry at shorter wavelengths compiled from
the literature and retrieved from data archives has allowed for improved 
thermophysical models of the five major Uranus satellites to be established, particularly with regard to the thermal inertia
and surface roughness.

\begin{acknowledgements}
      PACS has been developed by a consortium of institutes led by 
      MPE (Germany) and including UVIE (Austria); KUL, CSL, IMEC (Belgium); 
      CEA, OAMP (France); MPIA (Germany); IFSI, OAP/AOT, OAA/CAISMI, LENS, 
      SISSA (Italy); IAC (Spain). This development has been supported by the 
      funding agencies BMVIT (Austria), ESA-PRODEX (Belgium), 
      CEA/CNES (France), DLR (Germany), ASI (Italy), and CICYT/MCYT (Spain).
      ZB acknowledges funding by DLR for this work. 
      TM has received funding from the European Union’s Horizon 2020 Research and
      Innovation Programme, under Grant Agreement no.\ 687378, as part of the project
      "Small Bodies Near and Far" (SBNAF).
      We have made use of the JPL Horizons On-Line Ephemeris System to derive
      orbital parameters of the Uranian moons. 
      The Combined Atlas of Sources with Spitzer IRS Spectra (CASSIS) is a
      product of the IRS instrument team, supported by NASA and JPL. CASSIS is
      supported by the "Programme National de Physique Stellaire" (PNPS) of
      CNRS/INSU co-funded by CEA and CNES and through the "Programme National
      Physique et Chimie du Milieu Interstellaire" (PCMI) of CNRS/INSU with
      INC/INP co-funded by CEA and CNES.
      GM was supported by the Hungarian National Research, Development and 
      Innovation Office (NKFIH) grant PD-128360. GM acknowledges partial support 
      from the EC Horizon 2020 project OPTICON (730890) and the ESA PRODEX 
      contract nr. 4000129910.
      We thank the referee for constructive comments.
\end{acknowledgements}

\bibliographystyle{aa}
\bibliography{detre_uranusmoons.bib}

\clearpage

\begin{appendix}

\begin{onecolumn}

\section{PSF subtracted maps and photometry of individual maps}
\label{sect:appa}

\subsection{70\,$\mu$m maps of Uranian moons}
\label{sect:maps70obsid}

%
%                                                Two column figure
%-----------------------------------------------------------
%   \begin{figure*}[H!]
   \begin{figure*}[ht!]
   \centering
   \includegraphics[width=0.46\textwidth]{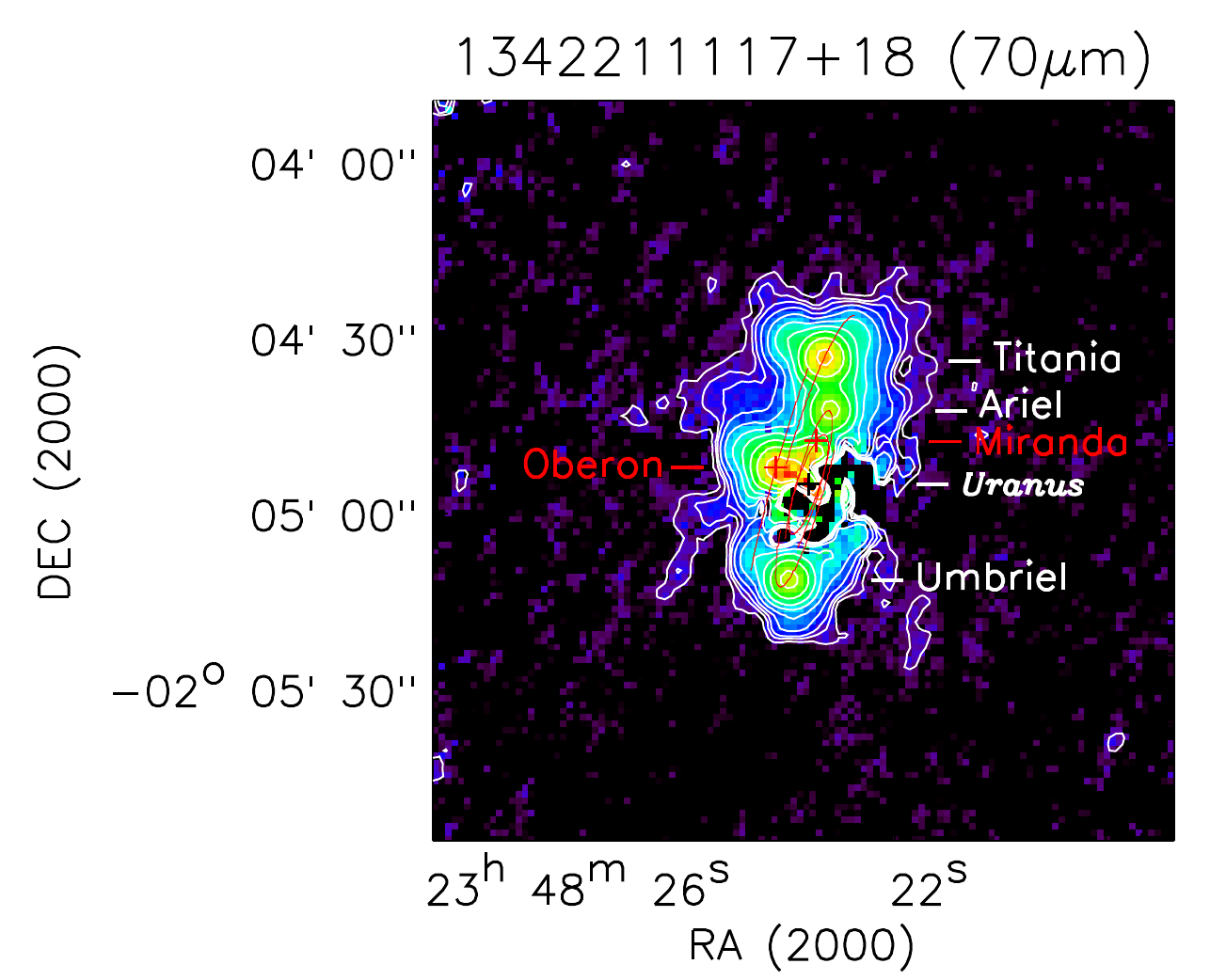}
   \includegraphics[width=0.46\textwidth]{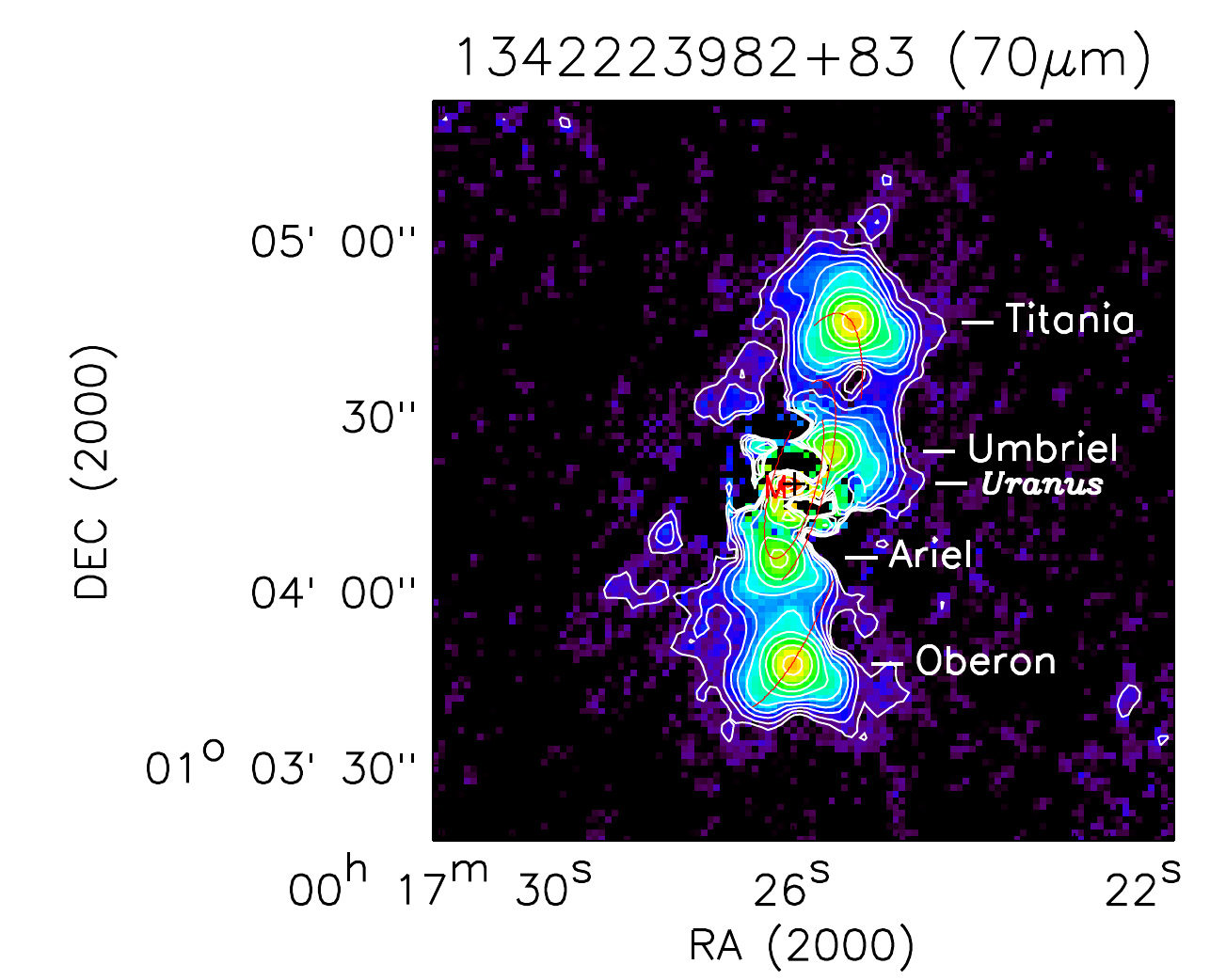}
   \includegraphics[width=0.46\textwidth]{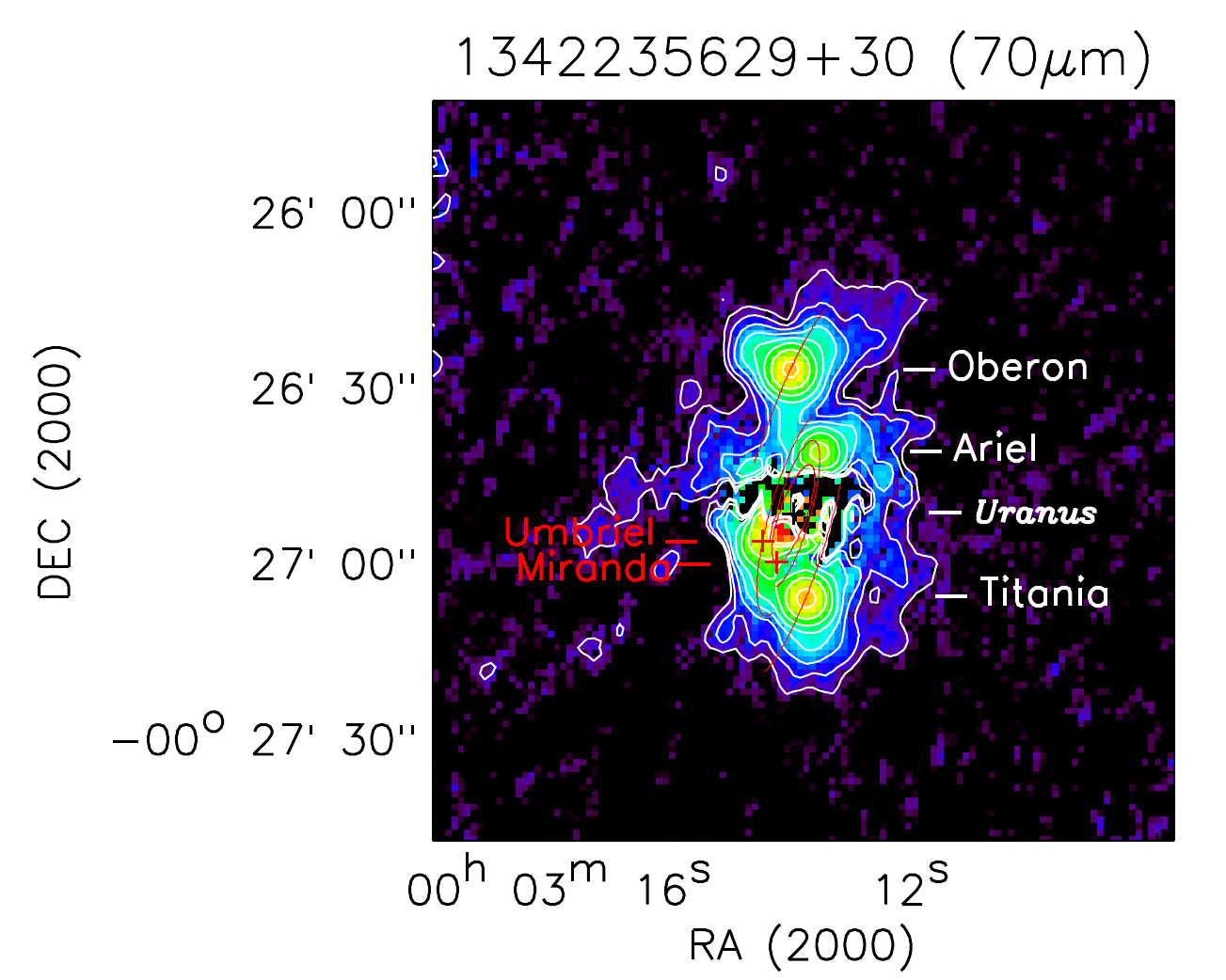}
   \includegraphics[width=0.46\textwidth]{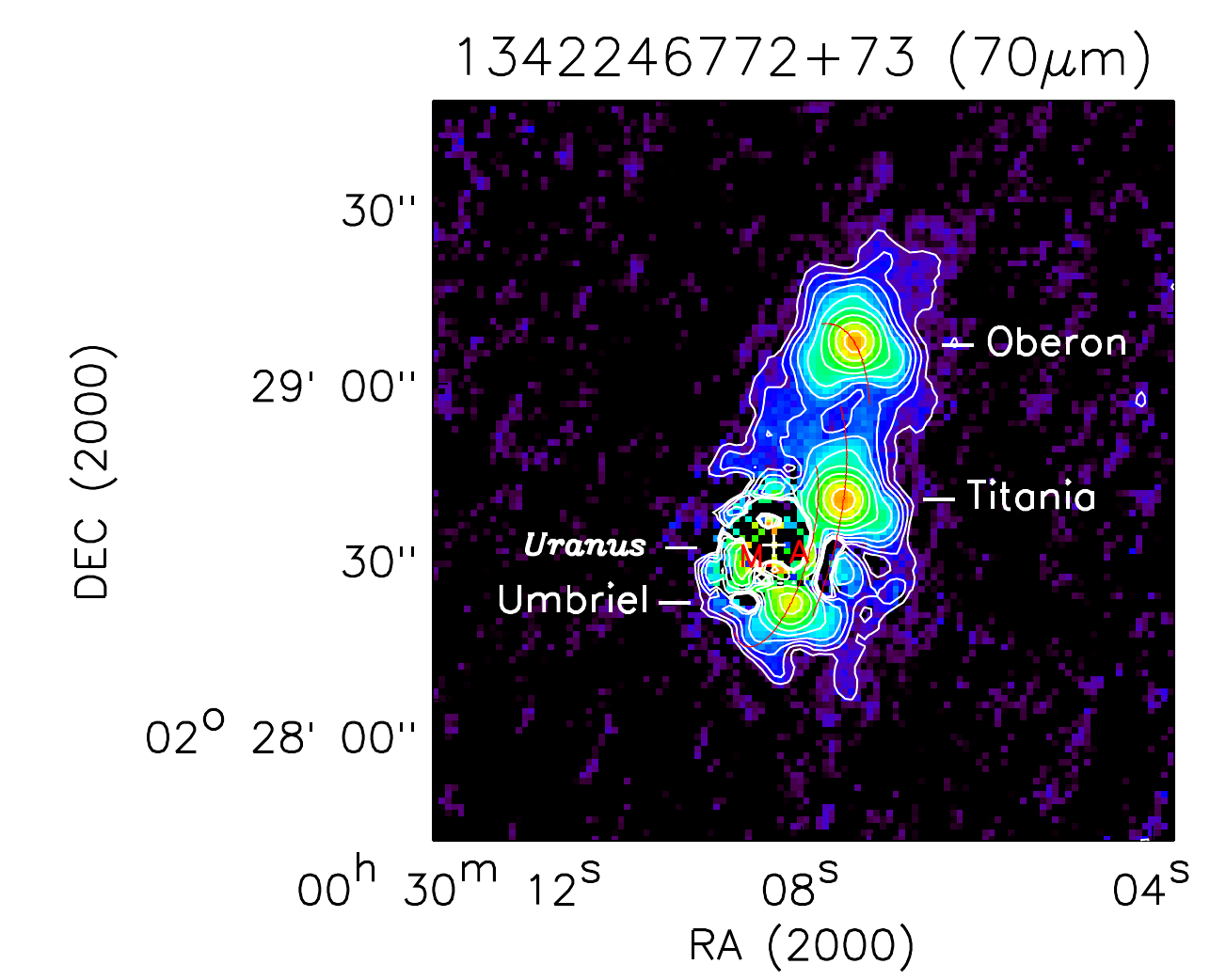}
   \includegraphics[width=0.46\textwidth]{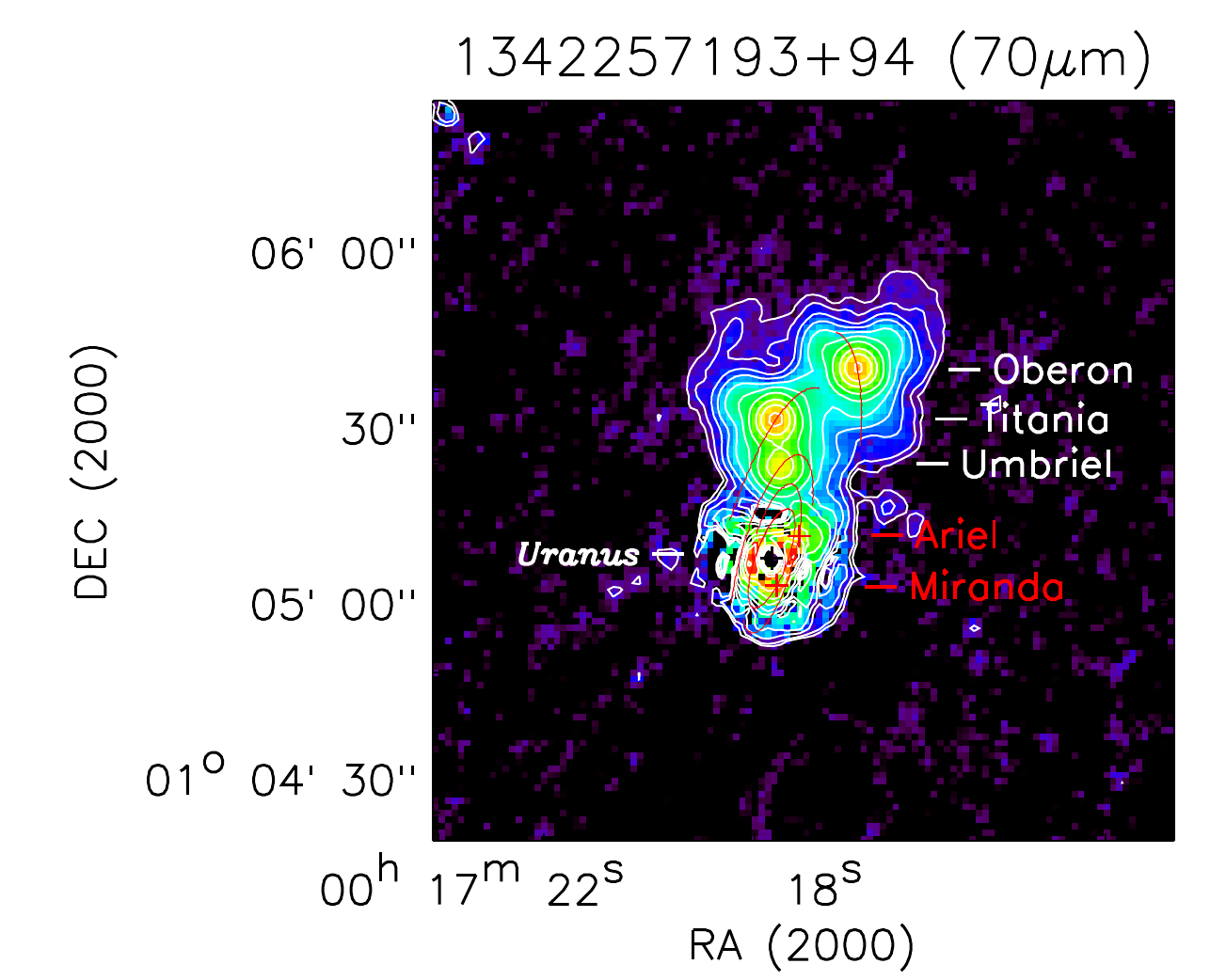}
   \includegraphics[height=0.37\linewidth]{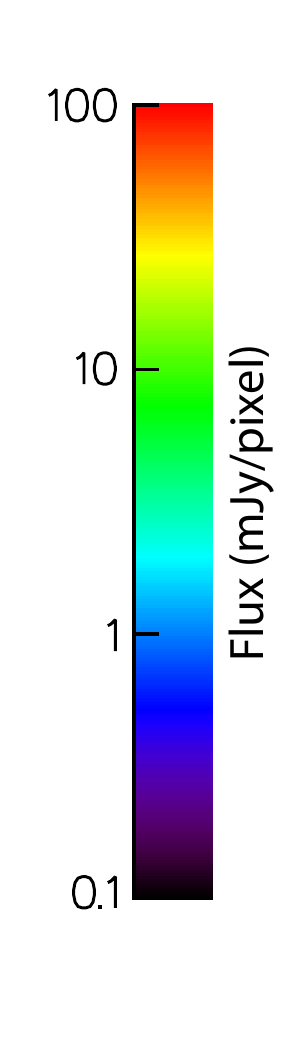}
      \caption{70\,$\mu$m maps of the Uranus moons after subtraction of the Uranus 
               PSF reference. Pixel scale is 1\farcs1. Moons for which the PSF is not or 
               only slightly affected by PSF subtraction residuals are labelled in white. 
               The central positions of moons for which the PSF is more significantly
               affected by PSF subtraction residuals or which are located closely together
               are marked by a red cross.  If only the initial of a moon is
               labelled, then its PSF peak is located inside the critical
               residual area. The central position of Uranus is marked by a
               black or white cross and labelled in italics. The positions of the moons
               relative to Uranus within plus minus one day of the observation are
               indicated by small red lines (we note that these refer to the Uranus position
               at the time of observation). Since Miranda, has a orbital period P = 1.413\,d,
               a closed orbit line is seen.
              }
         \label{fig:moonmaps70}
   \end{figure*}

\clearpage

\subsection{100\,$\mu$m maps of Uranian moons}
\label{sect:maps100obsid}

%
%                                                Two column figure
%-----------------------------------------------------------
   \begin{figure*}[ht!]
   \centering
   \includegraphics[width=0.47\textwidth]{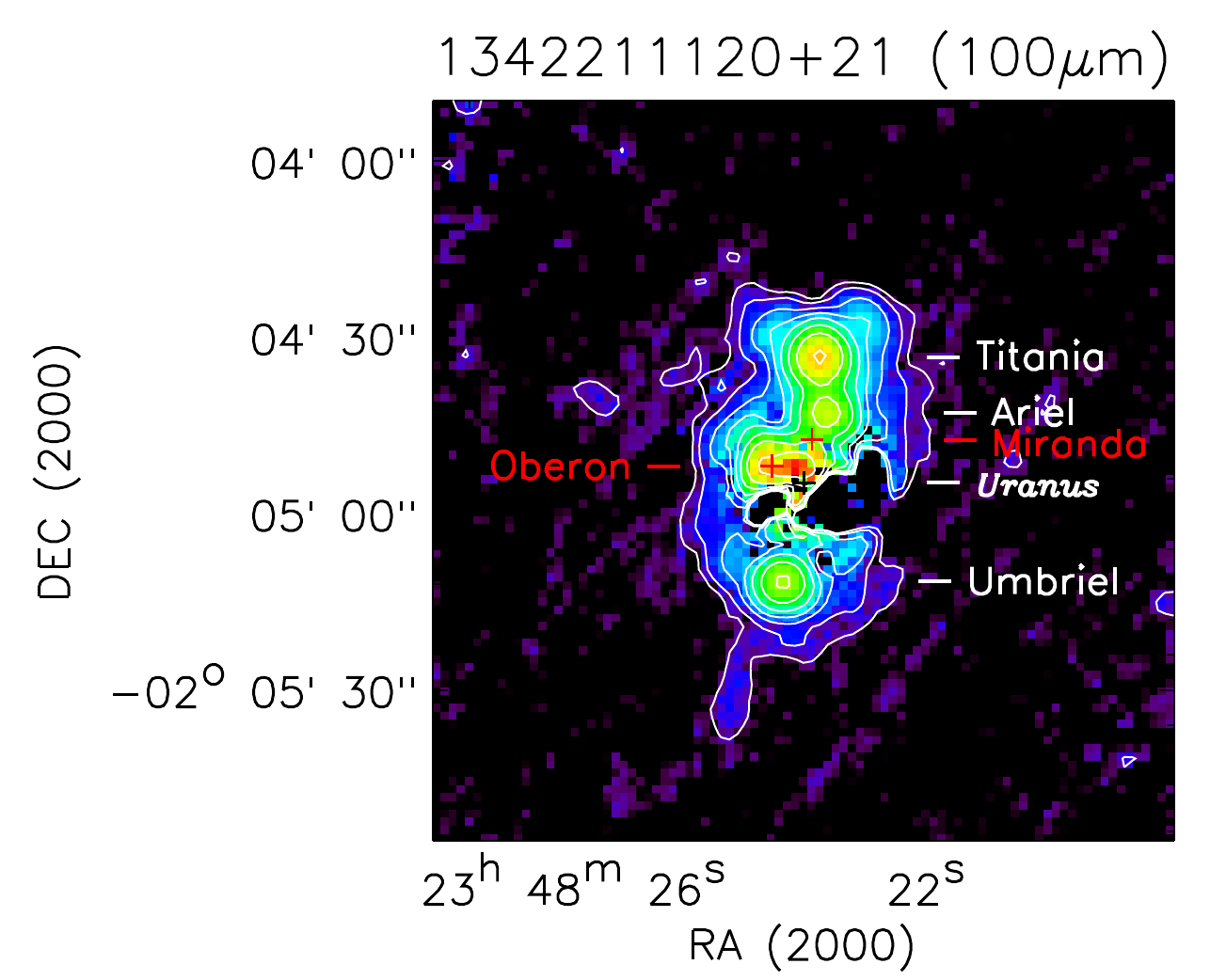}
   \includegraphics[width=0.47\textwidth]{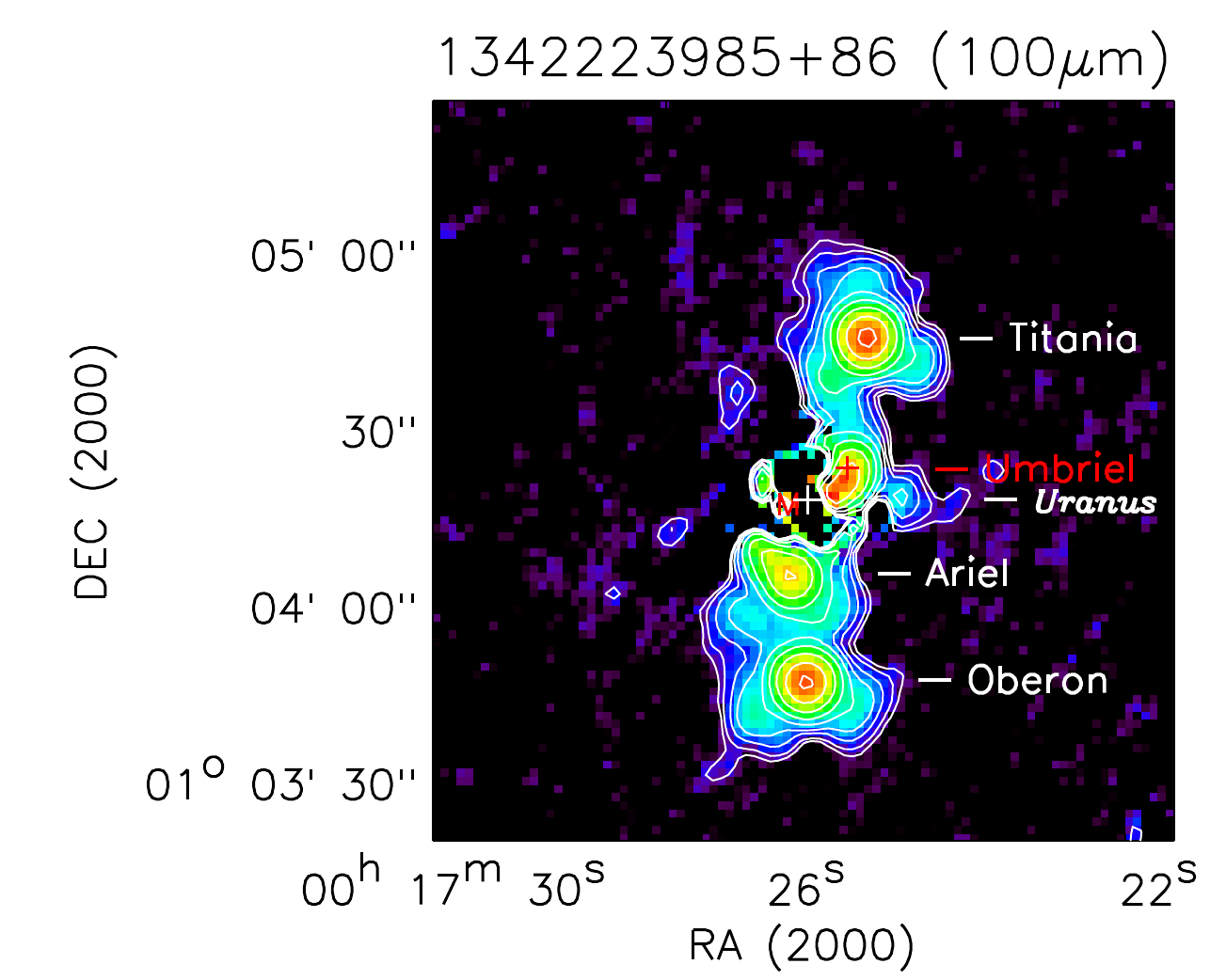}
   \includegraphics[width=0.47\textwidth]{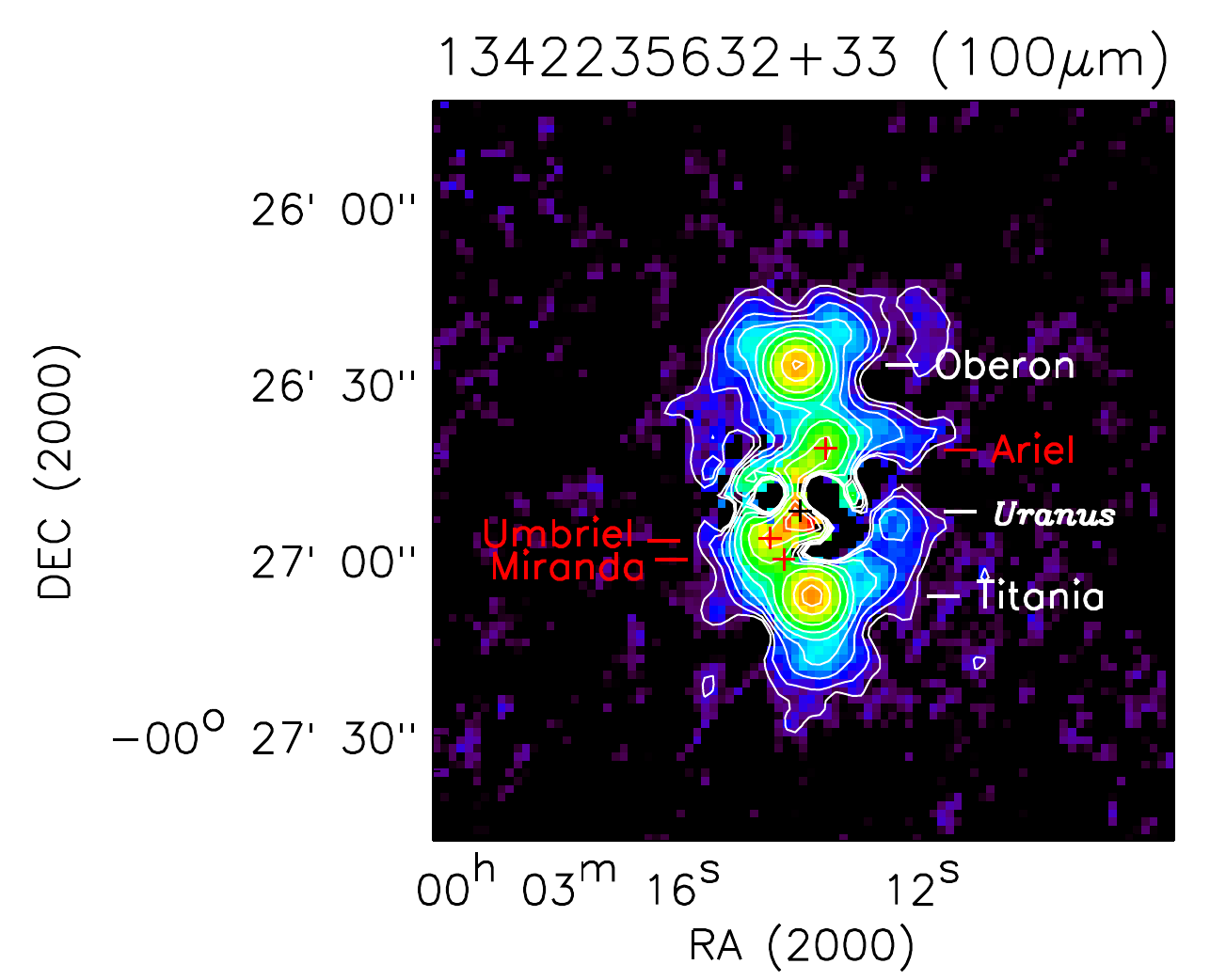}
   \includegraphics[width=0.47\textwidth]{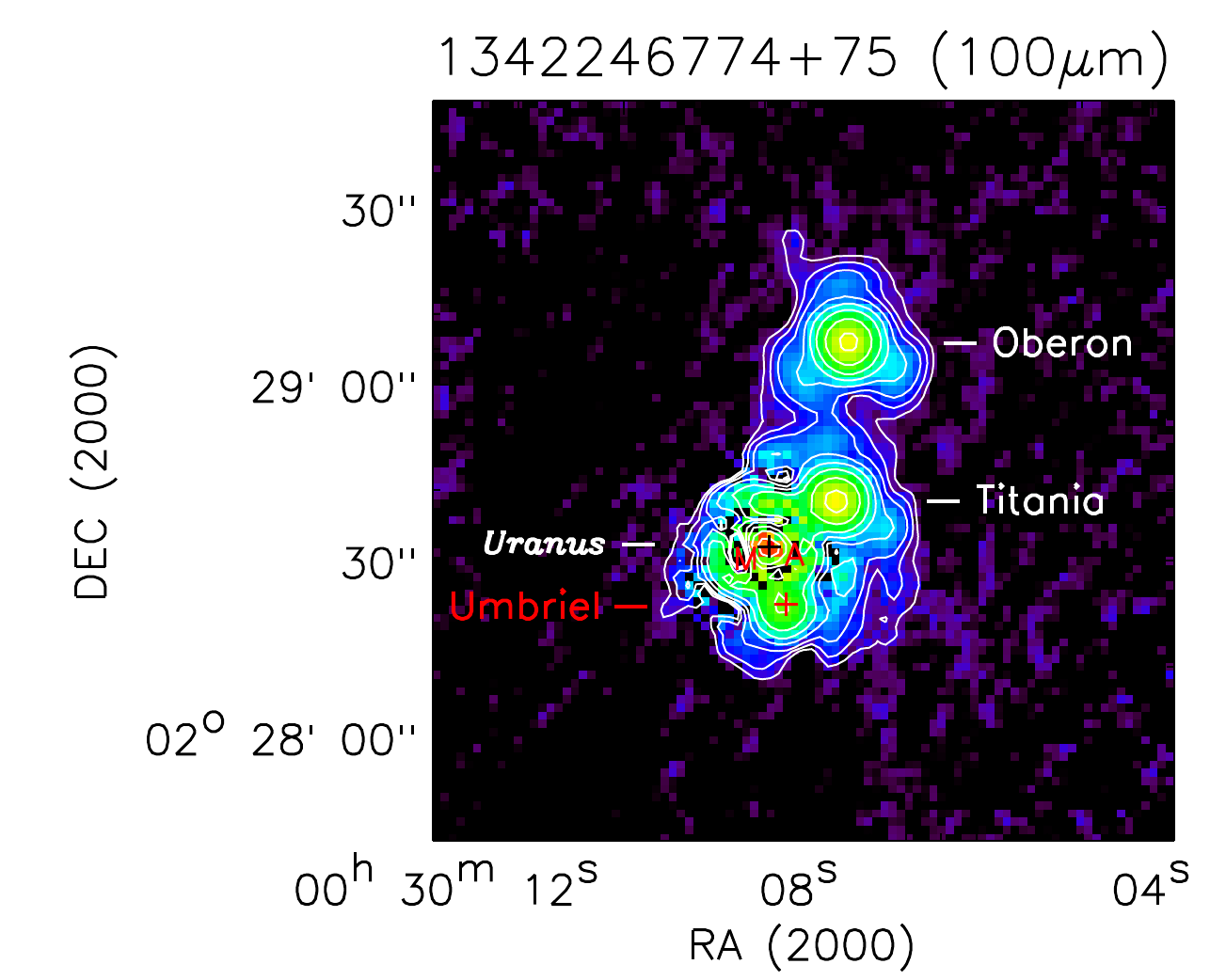}
   \includegraphics[width=0.47\textwidth]{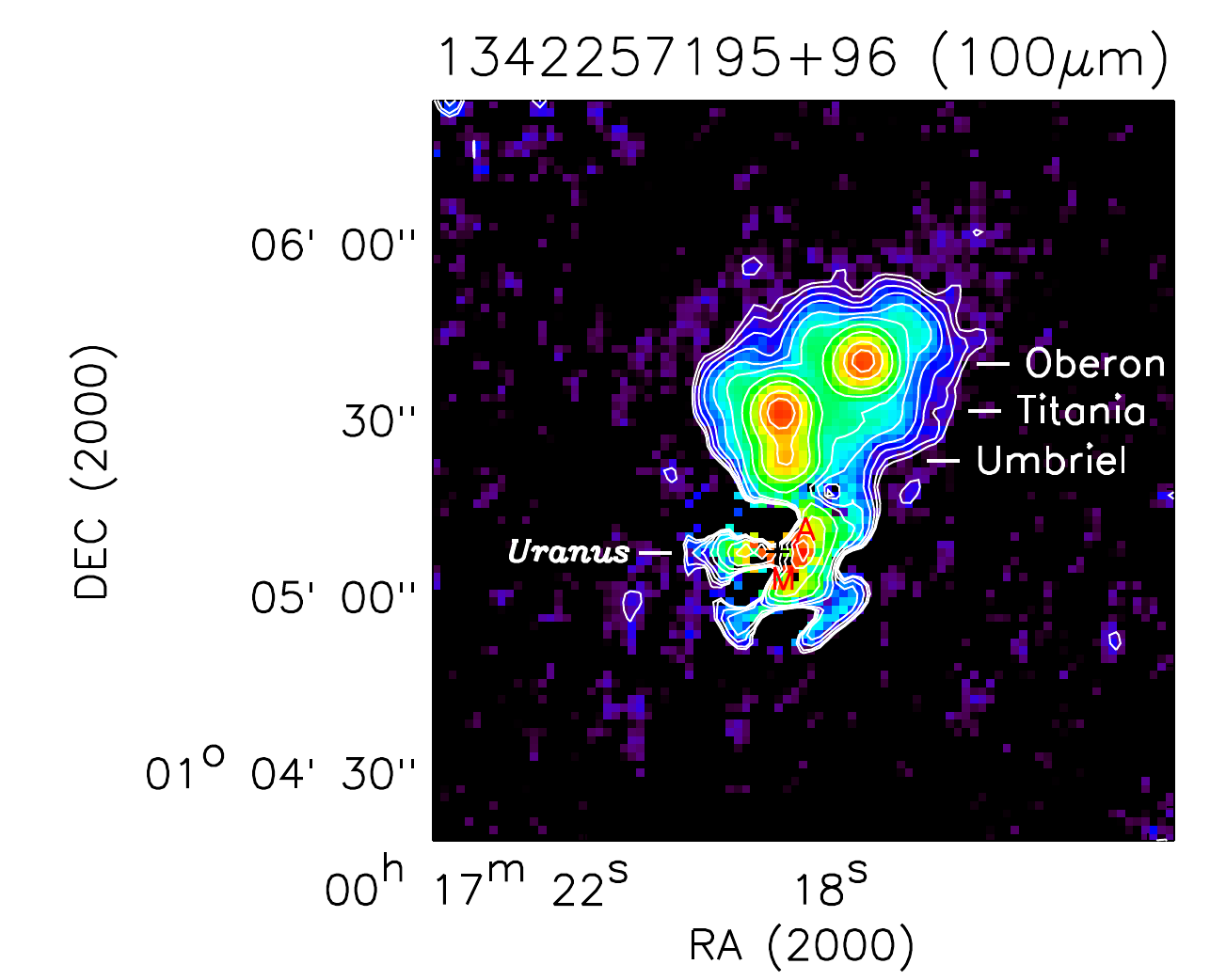}
   \includegraphics[height=0.37\linewidth]{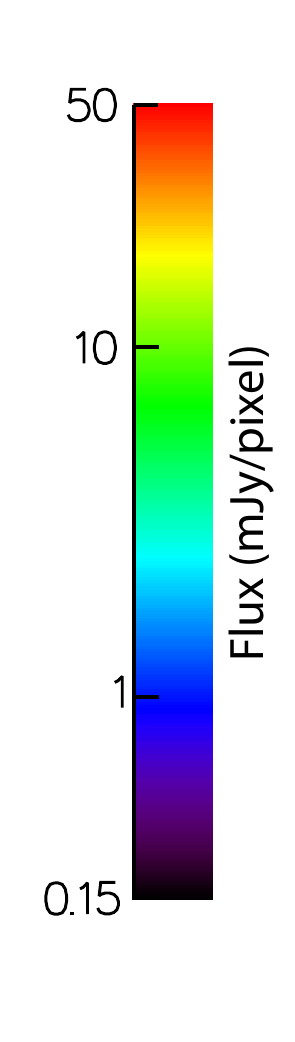}
      \caption{100\,$\mu$m maps of the Uranus moons after subtraction of the Uranus 
               PSF reference. Pixel scale is 1\farcs4. Moons for which the PSF is not or 
               only slightly affected by PSF subtraction residuals are labelled in white. 
               The central positions of moons for which the PSF is more significantly
               affected by PSF subtraction residuals or which are located closely together
               are marked by a red cross. If only the initial of a moon is
               labelled, then its PSF peak is located inside the critical
               residual area. The central position of Uranus is marked by a
               black or white cross and labelled in italics. 
%               The positions of the moons
%               relative to Uranus within plus minus one day of the observation are
%               indicated by small red lines (note these refer to the Uranus position
%               at the time of observation). Since Miranda has a orbital period P = 1.413\,d,
%               a closed orbit line is seen.
              }
         \label{fig:moonmaps100}
   \end{figure*}

\clearpage

\subsection{160\,$\mu$m maps of Uranian moons}
\label{sect:maps160obsid}

%
%                                                Two column figure
%-----------------------------------------------------------
   \begin{figure*}[ht!]
   \centering
   \includegraphics[width=0.47\textwidth]{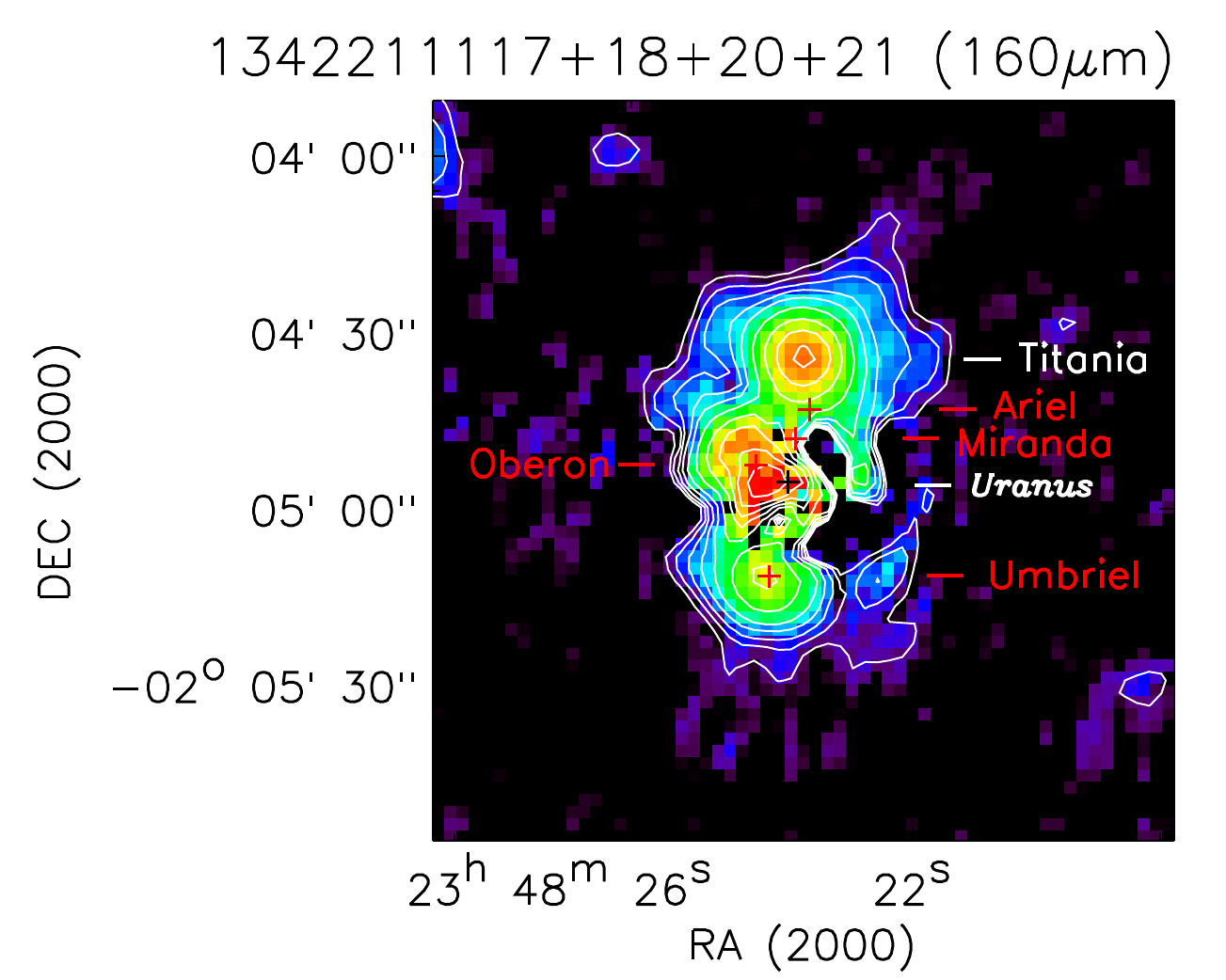}
   \includegraphics[width=0.47\textwidth]{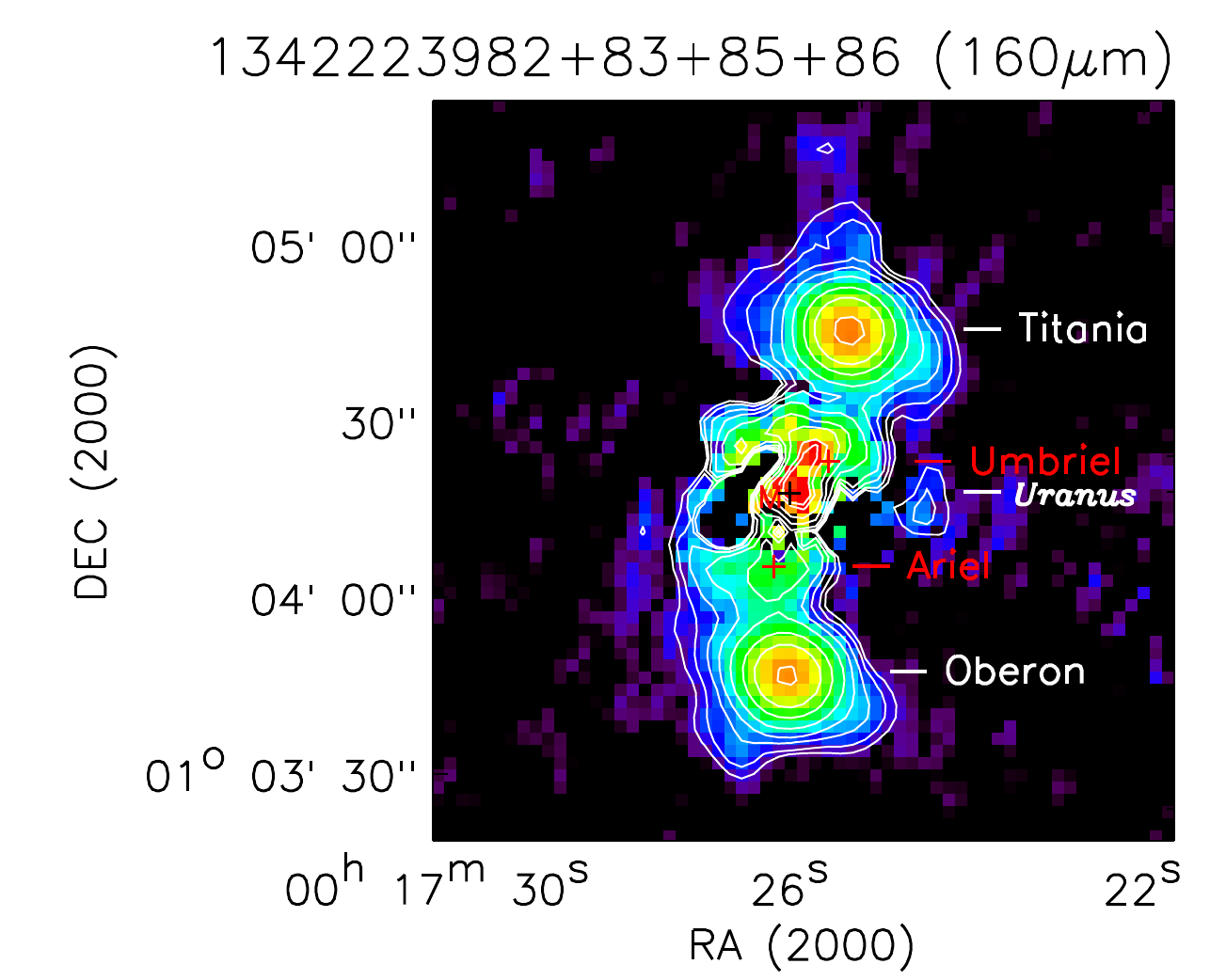}
   \includegraphics[width=0.47\textwidth]{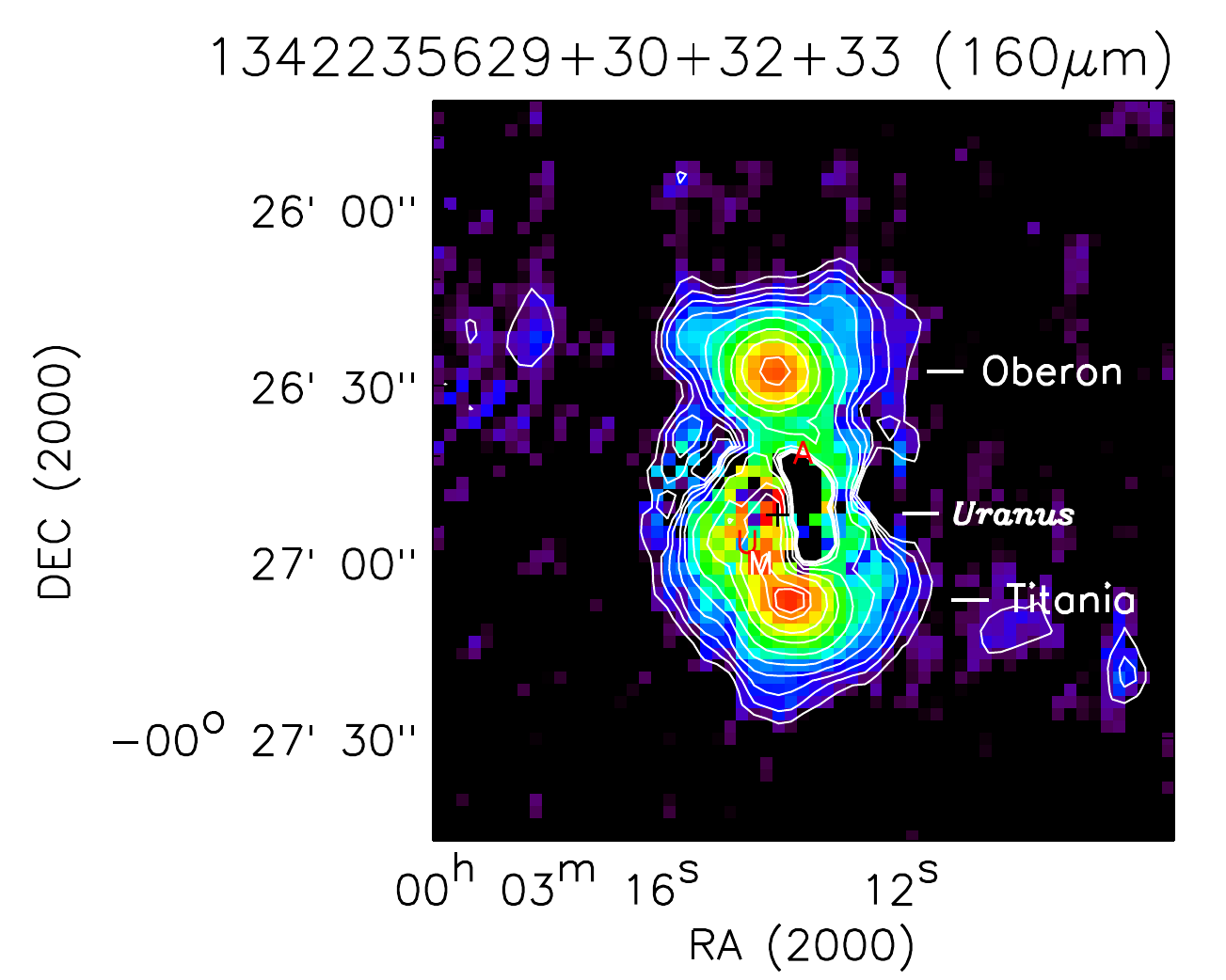}
   \includegraphics[width=0.47\textwidth]{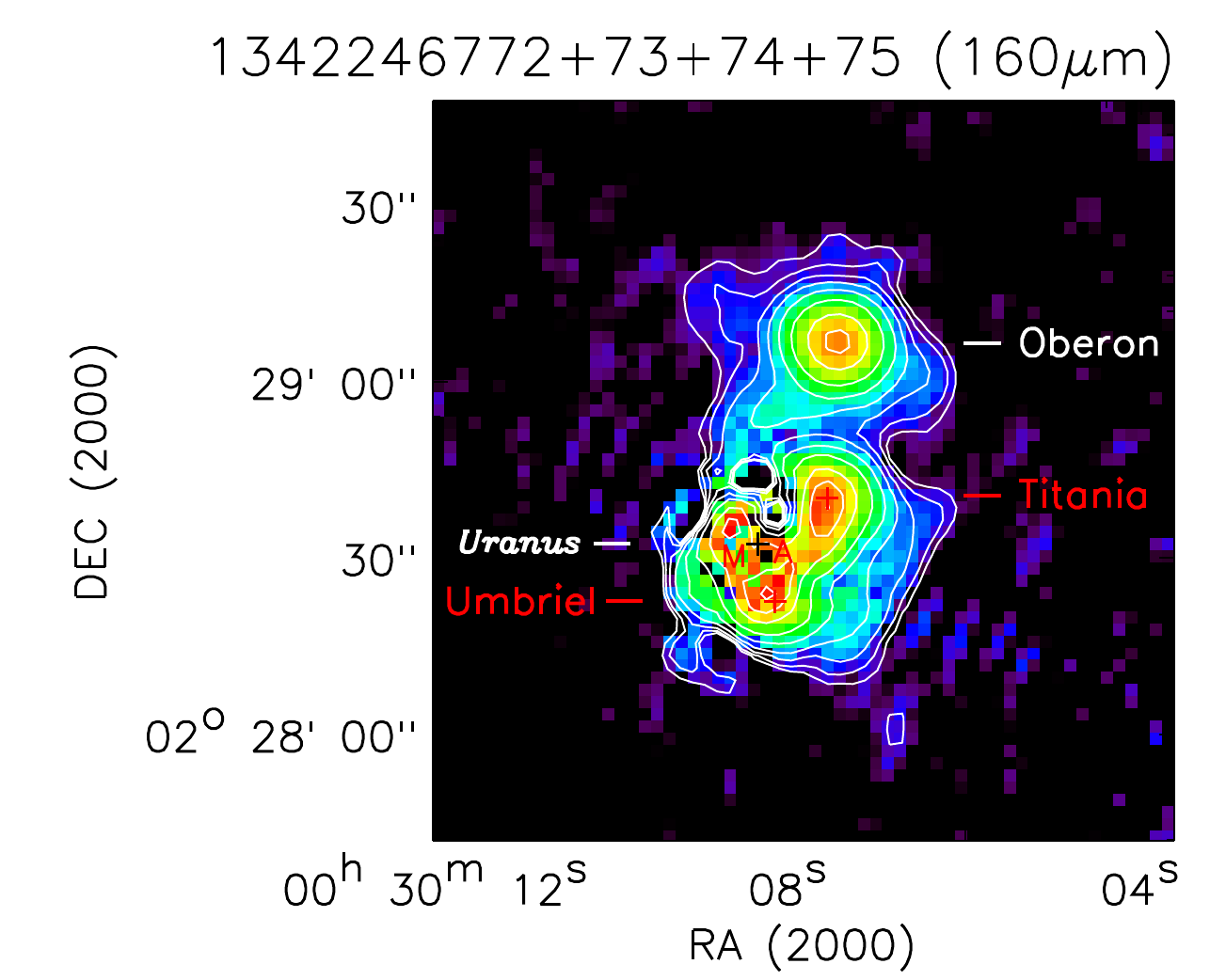}
   \includegraphics[width=0.47\textwidth]{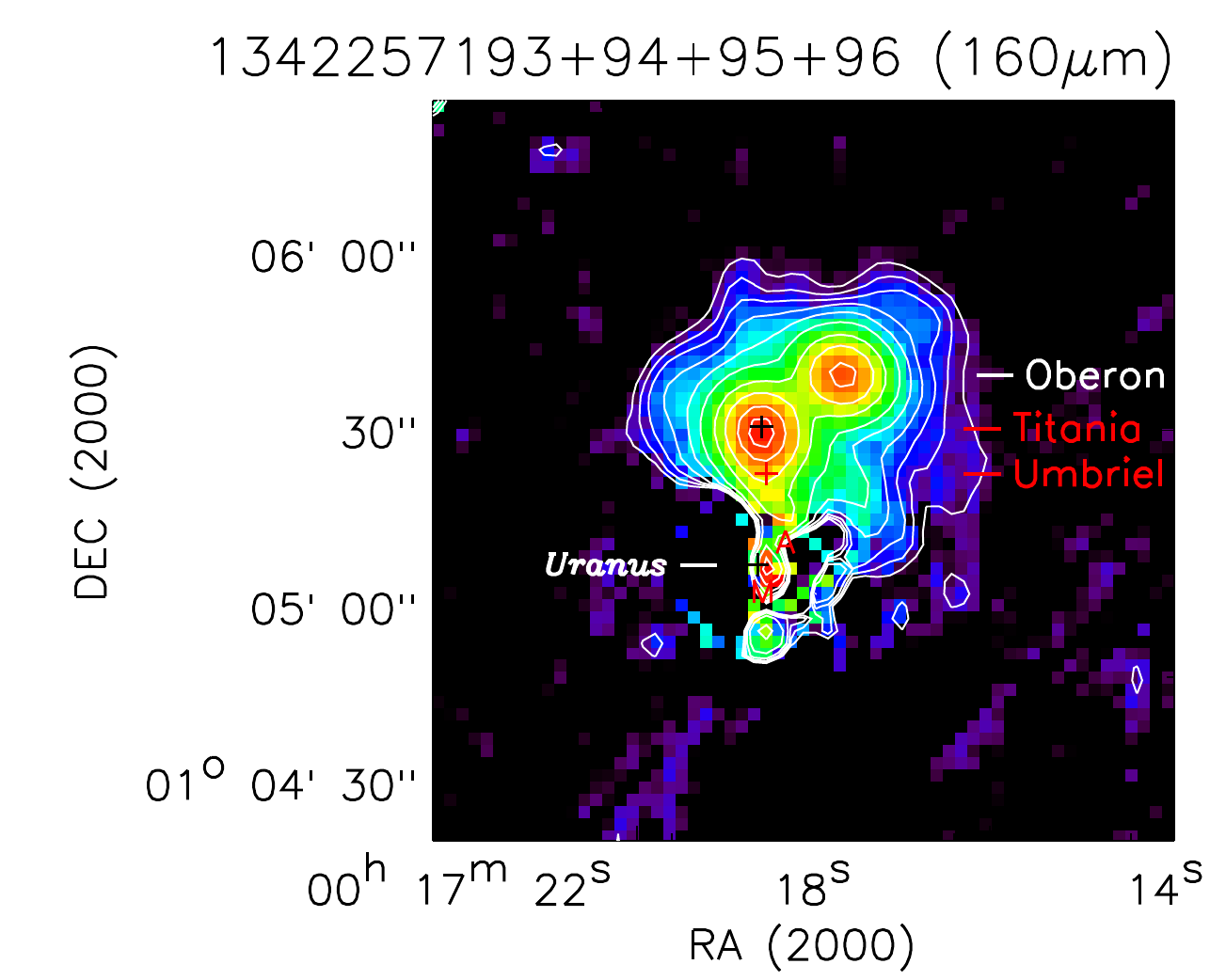}
   \includegraphics[height=0.37\linewidth]{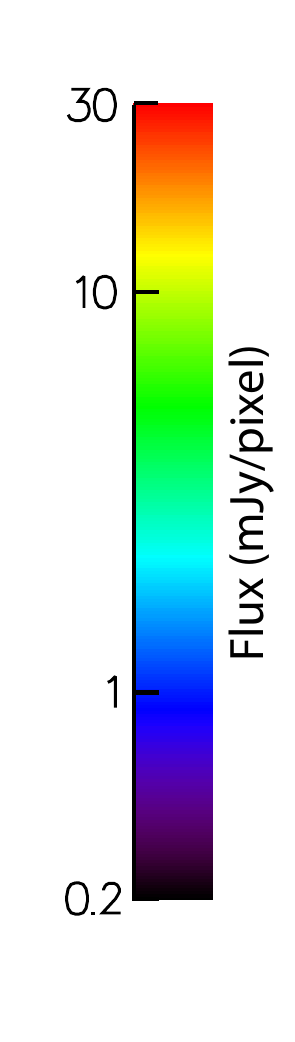}
      \caption{160\,$\mu$m maps of the Uranus moons after subtraction of the Uranus 
               PSF reference. Pixel scale is 2\farcs1. Moons for which the PSF is not or 
               only slightly affected by PSF subtraction residuals are labelled in white. 
               The red cross marks the central positions of moons for which the PSF is more significantly
               affected by PSF subtraction residuals or which are located closely together
               are marked. If only the initial of a moon is
               labelled, then its PSF peak is located inside the critical
               residual area. The central position of Uranus is marked by a
               black or white cross and labelled in italics. 
%               The positions of the moons
%               relative to Uranus within plus minus one day of the observation are
%               indicated by small red lines (note these refer to the Uranus position
%               at the time of observation). Since Miranda has a orbital period P = 1.413\,d,
%               a closed orbit line is seen.
              }
         \label{fig:moonmaps160}
   \end{figure*}

\clearpage

\subsection{Photometry of Uranus from individual maps}
\label{sect:taburanusphot}

\begin{table*}[ht!]
\caption{Photometry of Uranus. f$_{\rm total}$ gives the total flux of the Uranus system (Uranus plus
the moons). f$_{\rm Uranus}^{\rm measured}$ gives the flux at the actual distance of Uranus to {\it Herschel}. 
f$_{\rm Uranus}^{\rm distance corrected}$ gives the flux corrected to the same mean distance  
($\Delta_{\rm obs,mean}$ = 20.024\,AU) by scaling f$_{\rm Uranus}^{\rm distance corrected} = 
\left( \frac{\Delta_{\rm obs}}{\Delta_{\rm obs,mean}} \right)^2\,\times$\,f$_{\rm Uranus}^{\rm measured}$. 
$\Delta_{\rm obs}$ is the range of target centre wrt.\ the observer, i.e.\ {\it Herschel},
r$_{\rm helio}$ is the light-time corrected heliocentric range.
$\sigma_{\rm par}$ is the uncertainty in the PSF fitting parameters, $\sigma_{\rm red}$ is the uncertainty 
due to the reduction method (dependence on the map parameter selection), $\sigma_{tot}$ is the geometrical 
mean of the latter two uncertainties. The five 
epochs with 4 observations each are separated by horizontal lines.
%The colour correction factors cc are 0.984, 0.995 and 1.018 for 70, 100, and 160\,$\mu$m respectively.
}             % title of Table
\label{table:psfphotUranus}      % is used to refer this table in the text
\centering                          % used for centering table
\begin{tabular}{l r c r c c c c c c c c}        % centered columns (4 columns)
\hline\hline                 % inserts double horizontal lines
            \noalign{\smallskip}
OBSID      &  OD  &   MJD         &$\lambda_{\rm ref}$ & f$_{\rm total}$ & f$_{\rm Uranus}^{\rm measured}$ & $\sigma_{\rm par}$ & $\sigma_{\rm red}$& $\sigma_{tot}$ & f$_{\rm Uranus}^{\rm distance corrected}$ & r$_{\rm helio}$ & $\Delta_{\rm obs}$\\
           &      & mid-time obs. & $\mu$m&  (Jy)   &   (Jy)  &   (Jy) & (Jy) & (Jy) & (Jy) & (AU) & (AU)   \\   
            \noalign{\smallskip}
\hline
            \noalign{\smallskip}
1342211117 &  579 & 55543.72060 &  70.0 & 871.595 & 866.586 & 0.336 & 0.535 & 0.632 & 862.636 & 20.090 & 19.975 \\ 
           &      &             & 160.0 & 667.551 & 665.241 & 0.420 & 0.453 & 0.617 & 662.207 &        &        \\
1342211118 &  579 & 55543.72464 &  70.0 & 869.527 & 864.426 & 0.321 & 1.072 & 1.119 & 860.492 & 20.090 & 19.975 \\
           &      &             & 160.0 & 670.427 & 668.005 & 0.467 & 0.505 & 0.688 & 664.963 &        &        \\
1342211120 &  579 & 55543.73128 & 100.0 & 888.362 & 884.066 & 0.317 & 0.354 & 0.475 & 880.049 & 20.090 & 19.975 \\
           &      &             & 160.0 & 669.061 & 666.696 & 0.415 & 0.411 & 0.584 & 663.668 &        &        \\
1342211121 &  579 & 55543.73532 & 100.0 & 887.348 & 882.873 & 0.322 & 0.302 & 0.441 & 878.867 & 20.090 & 19.975 \\
           &      &             & 160.0 & 670.997 & 668.587 & 0.475 & 0.742 & 0.881 & 665.555 &        &        \\
            \noalign{\smallskip}
\hline
            \noalign{\smallskip}
1342223982 &  789 & 55754.05794 &  70.0 & 886.482 & 881.336 & 0.366 & 0.360 & 0.513 & 862.139 & 20.083 & 19.801 \\
           &      &             & 160.0 & 678.418 & 675.964 & 0.397 & 0.364 & 0.539 & 661.239 &        &        \\
1342223983 &  789 & 55754.06198 &  70.0 & 882.336 & 877.252 & 0.377 & 0.338 & 0.540 & 858.138 & 20.083 & 19.801 \\
           &      &             & 160.0 & 681.133 & 678.773 & 0.432 & 0.380 & 0.575 & 663.982 &        &        \\
1342223985 &  789 & 55754.06862 & 100.0 & 905.048 & 900.614 & 0.321 & 0.433 & 0.539 & 880.977 & 20.083 & 19.801 \\
           &      &             & 160.0 & 677.957 & 675.510 & 0.416 & 0.109 & 0.430 & 660.783 &        &        \\
1342223986 &  789 & 55754.07266 & 100.0 & 901.061 & 896.569 & 0.299 & 0.588 & 0.656 & 877.015 & 20.083 & 19.801 \\
           &      &             & 160.0 & 683.285 & 680.924 & 0.442 & 0.291 & 0.529 & 666.075 &        &        \\
            \noalign{\smallskip}
\hline
            \noalign{\smallskip}
1342235629 &  957 & 55921.94924 &  70.0 & 858.994 & 853.920 & 0.381 & 1.098 & 1.162 & 862.251 & 20.076 & 20.118 \\
           &      &             & 160.0 & 656.857 & 654.588 & 0.404 & 0.491 & 0.636 & 660.973 &        &        \\
1342235630 &  957 & 55921.95328 &  70.0 & 856.286 & 851.288 & 0.371 & 1.424 & 1.472 & 859.599 & 20.076 & 20.118 \\
           &      &             & 160.0 & 659.070 & 656.766 & 0.441 & 0.322 & 0.546 & 663.176 &        &        \\
1342235632 &  957 & 55921.95992 & 100.0 & 874.231 & 870.139 & 0.307 & 0.405 & 0.509 & 878.640 & 20.076 & 20.118 \\
           &      &             & 160.0 & 657.920 & 655.652 & 0.428 & 0.672 & 0.796 & 662.060 &        &        \\
1342235633 &  957 & 55921.96396 & 100.0 & 872.182 & 867.978 & 0.318 & 0.512 & 0.603 & 876.465 & 20.076 & 20.118 \\
           &      &             & 160.0 & 661.010 & 658.770 & 0.470 & 0.366 & 0.596 & 665.212 &        &        \\
            \noalign{\smallskip}
\hline
            \noalign{\smallskip}
1342246772 & 1121 & 56086.18500 &  70.0 & 835.110 & 830.155 & 0.384 & 0.910 & 0.988 & 862.297 & 20.069 & 20.404 \\
           &      &             & 160.0 & 638.110 & 635.998 & 0.414 & 0.569 & 0.704 & 660.620 &        &        \\
1342246773 & 1121 & 56086.18904 &  70.0 & 833.527 & 828.737 & 0.392 & 1.504 & 1.554 & 860.819 & 20.069 & 20.404 \\
           &      &             & 160.0 & 640.467 & 638.325 & 0.426 & 0.900 & 0.996 & 663.034 &        &        \\
1342246774 & 1121 & 56086.19308 & 100.0 & 851.492 & 847.114 & 0.337 & 0.207 & 0.395 & 879.898 & 20.069 & 20.404 \\
           &      &             & 160.0 & 639.878 & 637.673 & 0.413 & 0.709 & 0.820 & 662.352 &        &        \\
1342246775 & 1121 & 56086.19712 & 100.0 & 849.254 & 844.839 & 0.332 & 0.554 & 0.646 & 877.528 & 20.069 & 20.404 \\
           &      &             & 160.0 & 640.252 & 637.971 & 0.482 & 0.340 & 0.590 & 662.657 &        &        \\
            \noalign{\smallskip}
\hline
            \noalign{\smallskip}
1342257193 & 1310 & 56275.07361 &  70.0 & 883.668 & 878.500 & 0.361 & 0.910 & 0.979 & 861.196 & 20.059 & 19.823 \\
           &      &             & 160.0 & 677.071 & 674.596 & 0.417 & 0.445 & 0.610 & 661.307 &        &        \\
1342257194 & 1310 & 56275.07765 &  70.0 & 880.011 & 874.862 & 0.360 & 1.234 & 1.286 & 857.635 & 20.059 & 19.823 \\
           &      &             & 160.0 & 681.238 & 678.930 & 0.427 & 0.644 & 0.773 & 665.561 &        &        \\
1342257195 & 1310 & 56275.08169 & 100.0 & 902.984 & 898.793 & 0.296 & 0.459 & 0.546 & 881.098 & 20.059 & 19.823 \\
           &      &             & 160.0 & 678.415 & 676.120 & 0.410 & 0.463 & 0.619 & 662.810 &        &        \\
1342257196 & 1310 & 56275.08573 & 100.0 & 899.319 & 894.907 & 0.321 & 0.478 & 0.576 & 877.294 & 20.059 & 19.823 \\
           &      &             & 160.0 & 680.801 & 678.403 & 0.458 & 0.556 & 0.720 & 665.053 &        &        \\
            \noalign{\smallskip}
\hline\hline                 % inserts double horizontal lines
\end{tabular}
\end{table*}

~\\
~\\
\subsection{PSF photometry of moons from individual maps}
\label{sect:tabmoonphot}

\clearpage

\begin{sidewaystable*}[h!]
\small
%\begin{table*}[h!]
\caption{PSF photometry of Titania. Applied colour correction factors cc to derive colour corrected fluxes f$_{\rm moon,cc}$, are listed in Table~\ref{table:derivedmodelparams}. r$_{\rm helio}$ is the light-time corrected heliocentric range, $\Delta_{\rm obs}$ is the range of target centre wrt.\ the observer, i.e.\ {\it Herschel}, $\alpha$ is the phase angle with indication of L(eading) or T(railing) the Sun and $\theta_{\rm U-O}$ is the angular separation from Uranus. $\dot{r}_{\rm helio}$ is the heliocentric range rate and $\frac{\dot{r}_{\rm helio}}{|{\dot{r}_{\rm helio}^{\rm max}}|}$ indicates L(eading) H(emisphere) or T(railing) H(emisphere), if the absolute value of the ratio is greater than $\frac{2}{3}$, or B(oth) H(emispheres) otherwise. All five figures have been computed with the JPL Horizons On-Line Ephemeris System. The five epochs with 4 observations each are separated by horizontal lines.
}             % title of Table
\label{table:psfphotTitania}      % is used to refer this table in the text
\centering                          % used for centering table
\begin{tabular}{l r c r c c c c c c c c c c c c c}        % centered columns (4 columns)
\hline\hline                 % inserts double horizontal lines
            \noalign{\smallskip}
OBSID      &  OD  & MJD   &$\lambda_{\rm ref}$ &$\frac{f_{\rm moon}}{f_{\rm Uranus}}$ & f$_{\rm moon}$ & $\sigma_{\rm par}$ & $\sigma_{\rm red}$& $\sigma_{tot}$ & f$_{\rm moon,cc}$ & f$_{\rm model}$ & $\frac{f_{\rm moon,cc}}{f_{\rm model}}$ & r$_{\rm helio}$ & $\Delta_{\rm obs}$& $\alpha$ & $\theta_{\rm U-O}$ & $\frac{\dot{r}_{\rm helio}}{|{\dot{r}_{\rm helio}^{\rm max}}|}$ \\ %& l$_{\rm SSP}$\\
            \noalign{\smallskip}
           &      & mid-time obs. &$\mu$m&(10$^{-3}$)& (Jy)& (Jy) & (Jy) & (Jy) &   (Jy)  &  (Jy)   &   & (AU) & (AU) & (deg) & (\arcsec) &   \\ %(deg) \\      
            \noalign{\smallskip}
\hline
            \noalign{\smallskip}
1342211117 &  579 & 55543.72060 &  70.0 & 1.926 & 1.669 & 0.0011 & 0.0164 & 0.0165 & 1.696 & 1.650 & 1.028 & 20.088 & 19.973 & L~2.826 & 21.86 &$+$0.72\,(TH) \\ %11.72 \\
           &      &             & 160.0 & 1.315 & 0.875 & 0.0061 & 0.0094 & 0.0112 & 0.848 & 0.839 & 1.010 &        &        &         &       &      \\
1342211118 &  579 & 55543.72464 &  70.0 & 1.943 & 1.680 & 0.0011 & 0.0181 & 0.0181 & 1.707 & 1.650 & 1.034 & 20.088 & 19.973 & L~2.826 & 21.92 &$+$0.72\,(TH) \\ %11.72 \\
           &      &             & 160.0 & 1.282 & 0.856 & 0.0061 & 0.0149 & 0.0161 & 0.830 & 0.839 & 0.989 &        &        &         &       &      \\
1342211120 &  579 & 55543.73128 & 100.0 & 1.597 & 1.412 & 0.0018 & 0.0058 & 0.0061 & 1.413 & 1.385 & 1.020 & 20.088 & 19.973 & L~2.826 & 22.01 &$+$0.72\,(TH) \\ %11.72 \\
           &      &             & 160.0 & 1.296 & 0.864 & 0.0060 & 0.0082 & 0.0102 & 0.837 & 0.839 & 0.998 &        &        &         &       &       \\
1342211121 &  579 & 55543.73532 & 100.0 & 1.583 & 1.398 & 0.0019 & 0.0056 & 0.0059 & 1.399 & 1.385 & 1.010 & 20.088 & 19.973 & L~2.826 & 22.07 &$+$0.72\,(TH) \\ %11.72 \\
           &      &             & 160.0 & 1.277 & 0.854 & 0.0058 & 0.0125 & 0.0138 & 0.827 & 0.839 & 0.986 &        &        &         &       &       \\
            \noalign{\smallskip}
\hline
            \noalign{\smallskip}
1342223982 &  789 & 55754.05794 &  70.0 & 1.934 & 1.705 & 0.0012 & 0.0151 & 0.0152 & 1.732 & 1.697 & 1.021 & 20.084 & 19.802 & T~2.837 & 29.52 &$+$0.97\,(TH) \\ %13.95 \\
           &      &             & 160.0 & 1.307 & 0.884 & 0.0031 & 0.0087 & 0.0093 & 0.856 & 0.860 & 0.996 &        &        &         &       &       \\
1342223983 &  789 & 55754.06198 &  70.0 & 1.949 & 1.710 & 0.0012 & 0.0163 & 0.0164 & 1.737 & 1.697 & 1.024 & 20.084 & 19.802 & T~2.837 & 29.50 &$+$0.97\,(TH) \\ %13.95 \\
           &      &             & 160.0 & 1.336 & 0.907 & 0.0033 & 0.0064 & 0.0072 & 0.878 & 0.860 & 1.022 &        &        &         &       &       \\
1342223985 &  789 & 55754.06862 & 100.0 & 1.631 & 1.469 & 0.0015 & 0.0059 & 0.0061 & 1.471 & 1.422 & 1.034 & 20.084 & 19.802 & T~2.837 & 29.47 &$+$0.97\,(TH) \\ %13.95 \\
           &      &             & 160.0 & 1.285 & 0.868 & 0.0032 & 0.0152 & 0.0156 & 0.841 & 0.859 & 0.979 &        &        &         &       &       \\
1342223986 &  789 & 55754.07266 & 100.0 & 1.623 & 1.455 & 0.0014 & 0.0051 & 0.0053 & 1.457 & 1.422 & 1.024 & 20.084 & 19.802 & T~2.837 & 29.45 &$+$0.97\,(TH) \\ %13.95 \\
           &      &             & 160.0 & 1.297 & 0.883 & 0.0033 & 0.0086 & 0.0092 & 0.856 & 0.859 & 0.996 &        &        &         &       &       \\
            \noalign{\smallskip}
\hline
            \noalign{\smallskip}
1342235629 &  957 & 55921.94924 &  70.0 & 1.909 & 1.630 & 0.0022 & 0.0083 & 0.0986 & 1.656 & 1.635 & 1.013 & 20.079 & 20.121 & L~2.828 & 14.56 &$-$0.46\,(BH) \\ % 15.73 \\
           &      &             & 160.0 & 1.326 & 0.868 & 0.0243 & 0.0151 & 0.0286 & 0.841 & 0.829 & 1.015 &        &        &         &       &       \\
1342235630 &  957 & 55921.95328 &  70.0 & 1.978 & 1.684 & 0.0022 & 0.0207 & 0.0209 & 1.711 & 1.635 & 1.047 & 20.079 & 20.121 & L~2.828 & 14.63 &$-$0.47\,(BH) \\ % 15.73 \\
           &      &             & 160.0 & 1.276 & 0.838 & 0.0234 & 0.0449 & 0.0507 & 0.812 & 0.829 & 0.980 &        &        &         &       &       \\
1342235632 &  957 & 55921.95992 & 100.0 & 1.666 & 1.449 & 0.0057 & 0.0058 & 0.0082 & 1.451 & 1.371 & 1.058 & 20.079 & 20.121 & L~2.828 & 14.74 &$-$0.47\,(BH) \\ % 15.73 \\
           &      &             & 160.0 & 1.269 & 0.832 & 0.0246 & 0.0155 & 0.0291 & 0.806 & 0.829 & 0.972 &        &        &         &       &       \\
1342235633 &  957 & 55921.96396 & 100.0 & 1.689 & 1.466 & 0.0067 & 0.0093 & 0.0114 & 1.468 & 1.371 & 1.070 & 20.079 & 20.121 & L~2.828 & 14.81 &$-$0.47\,(BH) \\ % 15.73 \\
           &      &             & 160.0 & 1.340 & 0.883 & 0.0233 & 0.0113 & 0.0259 & 0.855 & 0.829 & 1.032 &        &        &         &       &       \\
            \noalign{\smallskip}
\hline
            \noalign{\smallskip}
1342246772 & 1121 & 56086.18500 &  70.0 & 1.895 & 1.573 & 0.0017 & 0.0115 & 0.0116 & 1.599 & 1.608 & 0.994 & 20.071 & 20.407 & T~2.744 & 14.34 &$+$0.36\,(BH) \\ %17.48 \\
           &      &             & 160.0 & 1.349 & 0.858 & 0.0134 & 0.0075 & 0.0154 & 0.831 & 0.813 & 1.023 &        &        &         &       &       \\
1342246773 & 1121 & 56086.18904 &  70.0 & 1.967 & 1.630 & 0.0014 & 0.0382 & 0.0382 & 1.657 & 1.608 & 1.030 & 20.071 & 20.407 & T~2.744 & 14.28 &$+$0.35\,(BH) \\ %17.48 \\
           &      &             & 160.0 & 1.276 & 0.815 & 0.0138 & 0.0293 & 0.0324 & 0.789 & 0.813 & 0.971 &        &        &         &       &       \\
1342246774 & 1121 & 56086.19308 & 100.0 & 1.556 & 1.318 & 0.0031 & 0.0094 & 0.0099 & 1.319 & 1.346 & 0.980 & 20.071 & 20.407 & T~2.744 & 14.23 &$+$0.35\,(BH) \\ %17.48 \\
           &      &             & 160.0 & 1.327 & 0.847 & 0.0143 & 0.0262 & 0.0299 & 0.820 & 0.813 & 1.009 &        &        &         &       &       \\
1342246775 & 1121 & 56086.19712 & 100.0 & 1.590 & 1.343 & 0.0026 & 0.0072 & 0.0077 & 1.344 & 1.346 & 0.999 & 20.071 & 20.407 & T~2.744 & 14.18 &$+$0.35\,(BH) \\ %17.48 \\
           &      &             & 160.0 & 1.274 & 0.813 & 0.0139 & 0.0205 & 0.0248 & 0.788 & 0.813 & 0.969 &        &        &         &       &       \\
            \noalign{\smallskip}
\hline
            \noalign{\smallskip}
1342257193 & 1310 & 56275.07361 &  70.0 & 1.963 & 1.724 & 0.0011 & 0.0174 & 0.0175 & 1.752 & 1.695 & 1.034 & 20.057 & 19.821 & L~2.774 & 23.80 &$+$0.76\,(TH) \\ %19.49 \\
           &      &             & 160.0 & 1.379 & 0.931 & 0.0047 & 0.0161 & 0.0168 & 0.902 & 0.857 & 1.052 &        &        &         &       &       \\
1342257194 & 1310 & 56275.07765 &  70.0 & 1.977 & 1.729 & 0.0012 & 0.0137 & 0.0138 & 1.757 & 1.695 & 1.037 & 20.057 & 19.821 & L~2.774 & 23.85 &$+$0.76\,(TH) \\ %19.49 \\
           &      &             & 160.0 & 1.374 & 0.933 & 0.0048 & 0.0157 & 0.0165 & 0.904 & 0.857 & 1.055 &        &        &         &       &       \\
1342257195 & 1310 & 56275.08169 & 100.0 & 1.654 & 1.487 & 0.0014 & 0.0144 & 0.0145 & 1.488 & 1.419 & 1.048 & 20.057 & 19.821 & L~2.774 & 23.90 &$+$0.76\,(TH) \\ %19.49 \\
           &      &             & 160.0 & 1.412 & 0.955 & 0.0047 & 0.0243 & 0.0248 & 0.925 & 0.857 & 1.079 &        &        &         &       &       \\
1342257196 & 1310 & 56275.08573 & 100.0 & 1.642 & 1.470 & 0.0017 & 0.0046 & 0.0049 & 1.471 & 1.419 & 1.037 & 20.057 & 19.821 & L~2.774 & 23.95 &$+$0.76\,(TH) \\ %19.49 \\
           &      &             & 160.0 & 1.404 & 0.952 & 0.0047 & 0.0265 & 0.0269 & 0.923 & 0.857 & 1.077 &        &        &         &       &       \\
            \noalign{\smallskip}
\hline\hline                 % inserts double horizontal lines
\end{tabular}
\end{sidewaystable*}

%\addtocounter{figure}{-1}

\clearpage

% mean ratio 70um:  0.00182820 +-   6.84409e-06
% mean ratio 100um: 0.00151792 +-   6.09095e-06
% mean ratio 160um: 0.00115738 +-   0.000132394

\begin{sidewaystable*}[h!]
\small
%\begin{table*}[h!]
\caption{PSF photometry of Oberon. Applied colour correction factors cc to derive colour corrected fluxes f$_{\rm moon,cc}$, are listed in Table~\ref{table:derivedmodelparams}. r$_{\rm helio}$ is the light-time corrected heliocentric range, $\Delta_{\rm obs}$ is the range of target centre wrt.\ the observer, i.e.\ {\it Herschel}, $\alpha$ is the Phase angle with indication of L(eading) or T(railing) the Sun, and $\theta_{\rm U-O}$ is the angular separation from Uranus. $\dot{r}_{\rm helio}$ is the heliocentric range rate and $\frac{\dot{r}_{\rm helio}}{|{\dot{r}_{\rm helio}^{\rm max}}|}$ indicates L(eading) H(emisphere) or T(railing) H(emisphere), if the absolute value of the ratio is greater than $\frac{2}{3}$, or B(oth) H(emispheres) otherwise. All five figures have been computed with the JPL Horizons On-Line Ephemeris System. The five epochs with 4 observations each are separated by horizontal lines.
% l$_{\rm SSP}$ is the latitude of the sub-solar point. 
}             % title of Table
\label{table:psfphotOberon}      % is used to refer this table in the text
\centering                          % used for centering table
\begin{tabular}{l r c r c c c c c c c c c c c c c}        % centered columns (4 columns)
\hline\hline                 % inserts double horizontal lines
            \noalign{\smallskip}
OBSID      &  OD  & MJD   &$\lambda_{\rm ref}$ &$\frac{f_{\rm moon}}{f_{\rm Uranus}}$ & f$_{\rm moon}$ & $\sigma_{\rm par}$ & $\sigma_{\rm red}$& $\sigma_{tot}$ & f$_{\rm moon,cc}$ & f$_{\rm model}$ & $\frac{f_{\rm moon,cc}}{f_{\rm model}}$ & r$_{\rm helio}$ & $\Delta_{\rm obs}$& $\alpha$ & $\theta_{\rm U-O}$ & $\frac{\dot{r}_{\rm helio}}{|{\dot{r}_{\rm helio}^{\rm max}}|}$ \\ %& l$_{\rm SSP}$\\
            \noalign{\smallskip}
           &      & mid-time obs. &$\mu$m&(10$^{-3}$)& (Jy)& (Jy) & (Jy) & (Jy) &   (Jy)  &  (Jy)   &   & (AU) & (AU) & (deg) & (\arcsec) &   \\ %(deg) \\
            \noalign{\smallskip}
\hline
            \noalign{\smallskip}
1342211117 &  579 & 55543.72060 &  70.0 & 1.892 & 1.639 & 0.0229 & 0.1034 & 0.1059 & 1.666 & 1.529 & 1.090 & 20.086 & 19.971 & L~2.826 &  6.31 &$+$0.03\,(BH) \\ %11.63 \\
           &      &             & 160.0 & 1.212 & 0.806 & 0.1168 & 0.0594 & 0.1310 & 0.781 & 0.782 & 0.999 &        &        &         &       &       \\
1342211118 &  579 & 55543.72464 &  70.0 & 1.796 & 1.553 & 0.0241 & 0.0561 & 0.0610 & 1.578 & 1.529 & 1.032 & 20.086 & 19.971 & L~2.826 &  6.32 &$+$0.03\,(BH) \\ %11.63 \\
           &      &             & 160.0 & 1.266 & 0.846 & 0.1205 & 0.0469 & 0.1293 & 0.819 & 0.782 & 1.048 &        &        &         &       &       \\
1342211120 &  579 & 55543.73128 & 100.0 & 1.612 & 1.425 & 0.0316 & 0.1059 & 0.1105 & 1.426 & 1.288 & 1.107 & 20.086 & 19.971 & L~2.826 &  6.35 &$+$0.03\,(BH) \\ %11.63 \\
           &      &             & 160.0 & 1.226 & 0.817 & 0.1143 & 0.0902 & 0.1456 & 0.792 & 0.782 & 1.013 &        &        &         &       &       \\
1342211121 &  579 & 55543.73532 & 100.0 & 1.776 & 1.568 & 0.0319 & 0.0324 & 0.0454 & 1.569 & 1.288 & 1.218 & 20.086 & 19.971 & L~2.826 &  6.37 &$+$0.03\,(BH) \\ %11.63 \\
           &      &             & 160.0 & 1.216 & 0.813 & 0.1161 & 0.0586 & 0.1300 & 0.788 & 0.782 & 1.007 &        &        &         &       &       \\
            \noalign{\smallskip}
\hline
            \noalign{\smallskip}
1342223982 &  789 & 55754.05794 &  70.0 & 1.828 & 1.611 & 0.0010 & 0.0077 & 0.0078 & 1.637 & 1.578 & 1.038 & 20.086 & 19.804 & T~2.837 & 30.90 &$-$0.74\,(LH) \\ % 13.85 \\
           &      &             & 160.0 & 1.193 & 0.807 & 0.0029 & 0.0069 & 0.0075 & 0.781 & 0.803 & 0.973 &        &        &         &       &       \\
1342223983 &  789 & 55754.06198 &  70.0 & 1.867 & 1.638 & 0.0011 & 0.0116 & 0.0117 & 1.664 & 1.578 & 1.055 & 20.086 & 19.804 & T~2.837 & 30.95 &$-$0.74\,(LH) \\ % 13.85 \\
           &      &             & 160.0 & 1.232 & 0.836 & 0.0028 & 0.0113 & 0.0116 & 0.810 & 0.803 & 1.009 &        &        &         &       &       \\
1342223985 &  789 & 55754.06862 & 100.0 & 1.526 & 1.374 & 0.0012 & 0.0114 & 0.0115 & 1.376 & 1.326 & 1.037 & 20.086 & 19.804 & T~2.837 & 31.02 &$-$0.74\,(LH) \\ % 13.85 \\
           &      &             & 160.0 & 1.223 & 0.826 & 0.0029 & 0.0044 & 0.0052 & 0.801 & 0.803 & 0.997 &        &        &         &       &       \\
1342223986 &  789 & 55754.07266 & 100.0 & 1.559 & 1.398 & 0.0015 & 0.0127 & 0.0128 & 1.400 & 1.326 & 1.055 & 20.086 & 19.804 & T~2.837 & 31.07 &$-$0.74\,(LH) \\ % 13.85 \\
           &      &             & 160.0 & 1.232 & 0.839 & 0.0028 & 0.0076 & 0.0081 & 0.813 & 0.803 & 1.012 &        &        &         &       &       \\
            \noalign{\smallskip}
\hline
            \noalign{\smallskip}
1342235629 &  957 & 55921.94924 &  70.0 & 1.842 & 1.573 & 0.0011 & 0.0138 & 0.0138 & 1.599 & 1.519 & 1.053 & 20.073 & 20.115 & L~2.829 & 24.69 &$+$0.58\,(BH) \\ %15.63 \\
           &      &             & 160.0 & 1.251 & 0.819 & 0.0038 & 0.0106 & 0.0113 & 0.794 & 0.774 & 1.025 &        &        &         &       &       \\
1342235630 &  957 & 55921.95328 &  70.0 & 1.851 & 1.576 & 0.0012 & 0.0154 & 0.0154 & 1.601 & 1.519 & 1.054 & 20.073 & 20.115 & L~2.829 & 24.75 &$+$0.58\,(BH) \\ %15.63 \\
           &      &             & 160.0 & 1.219 & 0.800 & 0.0038 & 0.0043 & 0.0058 & 0.776 & 0.774 & 1.002 &        &        &         &       &       \\
1342235632 &  957 & 55921.95992 & 100.0 & 1.517 & 1.320 & 0.0014 & 0.0089 & 0.0090 & 1.321 & 1.277 & 1.035 & 20.073 & 20.115 & L~2.829 & 24.84 &$+$0.59\,(BH) \\ %15.63 \\
           &      &             & 160.0 & 1.241 & 0.814 & 0.0037 & 0.0082 & 0.0090 & 0.789 & 0.774 & 1.019 &        &        &         &       &       \\
1342235633 &  957 & 55921.96396 & 100.0 & 1.558 & 1.353 & 0.0019 & 0.0086 & 0.0088 & 1.354 & 1.277 & 1.060 & 20.073 & 20.115 & L~2.829 & 24.89 &$+$0.59\,(BH) \\ %15.63 \\
           &      &             & 160.0 & 1.244 & 0.819 & 0.0038 & 0.0125 & 0.0130 & 0.794 & 0.774 & 1.026 &        &        &         &       &       \\
            \noalign{\smallskip}
\hline
            \noalign{\smallskip}
1342246772 & 1121 & 56086.18500 &  70.0 & 1.923 & 1.597 & 0.0012 & 0.0334 & 0.0335 & 1.623 & 1.500 & 1.082 & 20.070 & 20.406 & T~2.744 & 37.58 &$+$0.95\,(TH) \\ %17.36 \\
           &      &             & 160.0 & 1.246 & 0.792 & 0.0031 & 0.0089 & 0.0095 & 0.768 & 0.761 & 1.009 &        &        &         &       &       \\
1342246773 & 1121 & 56086.18904 &  70.0 & 1.890 & 1.566 & 0.0012 & 0.0328 & 0.0328 & 1.591 & 1.500 & 1.061 & 20.070 & 20.406 & T~2.744 & 37.56 &$+$0.95\,(TH) \\ %17.36 \\
           &      &             & 160.0 & 1.215 & 0.776 & 0.0034 & 0.0081 & 0.0088 & 0.752 & 0.761 & 0.988 &        &        &         &       &       \\
1342246774 & 1121 & 56086.19308 & 100.0 & 1.535 & 1.300 & 0.0014 & 0.0057 & 0.0059 & 1.301 & 1.258 & 1.034 & 20.070 & 20.406 & T~2.744 & 37.54 &$+$0.95\,(TH) \\ %17.36 \\
           &      &             & 160.0 & 1.225 & 0.781 & 0.0031 & 0.0043 & 0.0053 & 0.757 & 0.761 & 0.995 &        &        &         &       &       \\
1342246775 & 1121 & 56086.19712 & 100.0 & 1.525 & 1.289 & 0.0015 & 0.0074 & 0.0075 & 1.290 & 1.258 & 1.025 & 20.070 & 20.405 & T~2.744 & 37.52 &$+$0.95\,(TH) \\ %17.36 \\
           &      &             & 160.0 & 1.195 & 0.762 & 0.0034 & 0.0070 & 0.0077 & 0.739 & 0.761 & 0.971 &        &        &         &       &       \\
            \noalign{\smallskip}
\hline
            \noalign{\smallskip}
1342257193 & 1310 & 56275.07361 &  70.0 & 1.850 & 1.625 & 0.0011 & 0.0129 & 0.0129 & 1.651 & 1.579 & 1.046 & 20.060 & 19.824 & L~2.773 & 35.88 &$+$0.86\,(TH) \\ %19.36 \\
           &      &             & 160.0 & 1.201 & 0.811 & 0.0032 & 0.0054 & 0.0063 & 0.785 & 0.802 & 0.979 &        &        &         &       &       \\
1342257194 & 1310 & 56275.07765 &  70.0 & 1.866 & 1.632 & 0.0011 & 0.0176 & 0.0177 & 1.659 & 1.579 & 1.051 & 20.060 & 19.824 & L~2.773 & 35.84 &$+$0.86\,(TH) \\ %19.36 \\
           &      &             & 160.0 & 1.202 & 0.816 & 0.0033 & 0.0036 & 0.0049 & 0.791 & 0.802 & 0.986 &        &        &         &       &       \\
1342257195 & 1310 & 56275.08169 & 100.0 & 1.553 & 1.396 & 0.0013 & 0.0148 & 0.0149 & 1.397 & 1.325 & 1.055 & 20.060 & 19.824 & L~2.773 & 35.81 &$+$0.86\,(TH) \\ %19.36 \\
           &      &             & 160.0 & 1.224 & 0.828 & 0.0031 & 0.0044 & 0.0054 & 0.802 & 0.802 & 1.000 &        &        &         &       &       \\
1342257196 & 1310 & 56275.08573 & 100.0 & 1.544 & 1.382 & 0.0013 & 0.0128 & 0.0128 & 1.383 & 1.325 & 1.044 & 20.061 & 19.825 & L~2.773 & 35.78 &$+$0.86\,(TH) \\ %19.36 \\
           &      &             & 160.0 & 1.178 & 0.799 & 0.0033 & 0.0097 & 0.0103 & 0.775 & 0.802 & 0.966 &        &        &         &       &       \\
            \noalign{\smallskip}
\hline\hline                 % inserts double horizontal lines
\end{tabular}
%\end{table*}
\end{sidewaystable*}

\clearpage

\begin{sidewaystable*}[h!]
\small
%\begin{table*}[h!]
\caption{PSF photometry of Umbriel. Applied colour correction factors cc to derive colour corrected fluxes f$_{\rm moon,cc}$, are listed in Table~\ref{table:derivedmodelparams}. r$_{\rm helio}$ is the light-time corrected heliocentric range, $\Delta_{\rm obs}$ is the range of target centre wrt.\ the observer, i.e.\ {\it Herschel}, $\alpha$ is the Phase angle with indication of L(eading) or T(railing) the Sun, and $\theta_{\rm U-O}$ is the angular separation from Uranus. $\dot{r}_{\rm helio}$ is the heliocentric range rate and $\frac{\dot{r}_{\rm helio}}{|{\dot{r}_{\rm helio}^{\rm max}}|}$ indicates L(eading) H(emisphere) or T(railing) H(emisphere), if the absolute value of the ratio is greater than $\frac{2}{3}$, or B(oth) H(emispheres) otherwise. All five figures have been computed with the JPL Horizons On-Line Ephemeris System. The five epochs with 4 observations each are separated by horizontal lines.
% l$_{\rm SSP}$ is the latitude of the sub-solar point. 
}             % title of Table
\label{table:psfphotUmbriel}      % is used to refer this table in the text
\centering                          % used for centering table
\begin{tabular}{l r c r c c c c c c c c c c c c c}        % centered columns (4 columns)
\hline\hline                 % inserts double horizontal lines
            \noalign{\smallskip}
OBSID      &  OD  & MJD   &$\lambda_{\rm ref}$ &$\frac{f_{\rm moon}}{f_{\rm Uranus}}$ & f$_{\rm moon}$ & $\sigma_{\rm par}$ & $\sigma_{\rm red}$& $\sigma_{tot}$ & f$_{\rm moon,cc}$ & f$_{\rm model}$ & $\frac{f_{\rm moon,cc}}{f_{\rm model}}$ & r$_{\rm helio}$ & $\Delta_{\rm obs}$& $\alpha$ & $\theta_{\rm U-O}$ & $\frac{\dot{r}_{\rm helio}}{|{\dot{r}_{\rm helio}^{\rm max}}|}$ \\ %& l$_{\rm SSP}$\\
            \noalign{\smallskip}
           &      & mid-time obs. &$\mu$m&(10$^{-3}$)& (Jy)& (Jy) & (Jy) & (Jy) &   (Jy)  &  (Jy)   &   & (AU) & (AU) & (deg) & (\arcsec) &   \\ %(deg) \\
            \noalign{\smallskip}
\hline
            \noalign{\smallskip}
1342211117 &  579 & 55543.72060 &  70.0 & 0.962 & 0.833 & 0.0035 & 0.0182 & 0.0185 & 0.847 & 0.915 & 0.926 & 20.091 & 19.976 & L~2.825 & 16.59 &$-$0.90\,(LH) \\ %11.76 \\
           &      &             & 160.0 & 0.662 & 0.440 & 0.0380 & 0.0315 & 0.0494 & 0.427 & 0.463 & 0.921 &        &        &         &       &       \\
1342211118 &  579 & 55543.72464 &  70.0 & 1.017 & 0.880 & 0.0034 & 0.0116 & 0.0121 & 0.894 & 0.915 & 0.977 & 20.091 & 19.976 & L~2.825 & 16.64 &$-$0.90\,(LH) \\ %11.76 \\
           &      &             & 160.0 & 0.666 & 0.445 & 0.0374 & 0.0304 & 0.0482 & 0.431 & 0.463 & 0.931 &        &        &         &       &       \\
1342211120 &  579 & 55543.73128 & 100.0 & 0.812 & 0.718 & 0.0059 & 0.0076 & 0.0096 & 0.719 & 0.766 & 0.938 & 20.091 & 19.976 & L~2.825 & 16.71 &$-$0.91\,(LH) \\ %11.76 \\
           &      &             & 160.0 & 0.657 & 0.438 & 0.0380 & 0.0235 & 0.0446 & 0.425 & 0.463 & 0.917 &        &        &         &       &       \\
1342211121 &  579 & 55543.73532 & 100.0 & 0.863 & 0.762 & 0.0059 & 0.0051 & 0.0078 & 0.763 & 0.766 & 0.996 & 20.091 & 19.976 & L~2.826 & 16.76 &$-$0.91\,(LH) \\ %11.76 \\
           &      &             & 160.0 & 0.700 & 0.468 & 0.0378 & 0.0311 & 0.0489 & 0.453 & 0.463 & 0.979 &        &        &         &       &       \\
            \noalign{\smallskip}
\hline
            \noalign{\smallskip}
1342223982 &  789 & 55754.05794 &  70.0 & 1.043 & 0.919 & 0.0107 & 0.0157 & 0.0189 & 0.934 & 0.941 & 0.992 & 20.085 & 19.803 & T~2.837 &  8.62 &$+$0.38\,(BH) \\ %13.99 \\
           &      &             & 160.0 & 0.707 & 0.478 & 0.0942 & 0.0687 & 0.1166 & 0.463 & 0.474 & 0.977 &        &        &         &       &       \\
1342223983 &  789 & 55754.06198 &  70.0 & 1.062 & 0.932 & 0.0080 & 0.0200 & 0.0215 & 0.947 & 0.941 & 1.007 & 20.085 & 19.803 & T~2.837 &  8.54 &$+$0.37\,(BH) \\ %13.99 \\
           &      &             & 160.0 & 0.638 & 0.433 & 0.0911 & 0.0345 & 0.0974 & 0.420 & 0.474 & 0.885 &        &        &         &       &       \\
1342223985 &  789 & 55754.06862 & 100.0 & 0.898 & 0.809 & 0.0235 & 0.0144 & 0.0275 & 0.810 & 0.786 & 1.030 & 20.085 & 19.803 & T~2.836 &  8.42 &$+$0.36\,(BH) \\ %13.99 \\
           &      &             & 160.0 & 0.744 & 0.503 & 0.1010 & 0.0956 & 0.1391 & 0.487 & 0.474 & 1.027 &        &        &         &       &       \\
1342223986 &  789 & 55754.07266 & 100.0 & 0.884 & 0.793 & 0.0196 & 0.0125 & 0.0232 & 0.794 & 0.786 & 1.010 & 20.085 & 19.803 & T~2.836 &  8.34 &$+$0.36\,(BH) \\ %13.99 \\
           &      &             & 160.0 & 0.715 & 0.487 & 0.0954 & 0.0522 & 0.1088 & 0.472 & 0.474 & 0.995 &        &        &         &       &       \\
            \noalign{\smallskip}
\hline
            \noalign{\smallskip}
1342235629 &  957 & 55921.94924 &  70.0 & 0.905 & 0.773 & 0.0252 & 0.0502 & 0.0561 & 0.785 & 0.907 & 0.866 & 20.075 & 20.116 & L~2.828 &  7.09 &$-$0.33\,(BH) \\ %15.77 \\
           &      &             & 160.0 & 0.637 & 0.417 & 0.1653 & 0.0313 & 0.1683 & 0.404 & 0.457 & 0.884 &        &        &         &       &       \\
1342235630 &  957 & 55921.95328 &  70.0 & 0.991 & 0.843 & 0.0199 & 0.0250 & 0.0320 & 0.857 & 0.907 & 0.945 & 20.075 & 20.116 & L~2.828 &  7.01 &$-$0.33\,(BH) \\ %15.77 \\
           &      &             & 160.0 & 0.646 & 0.425 & 0.1646 & 0.1205 & 0.2040 & 0.411 & 0.457 & 0.900 &        &        &         &       &       \\
1342235632 &  957 & 55921.95992 & 100.0 & 0.747 & 0.650 & 0.0463 & 0.0437 & 0.0636 & 0.651 & 0.758 & 0.858 & 20.075 & 20.117 & L~2.828 &  6.87 &$-$0.32\,(BH) \\ %15.77 \\
           &      &             & 160.0 & 0.742 & 0.486 & 0.1785 & 0.0488 & 0.1850 & 0.471 & 0.458 & 1.029 &        &        &         &       &       \\
1342235633 &  957 & 55921.96396 & 100.0 & 0.815 & 0.708 & 0.0425 & 0.0293 & 0.0516 & 0.708 & 0.758 & 0.935 & 20.075 & 20.117 & L~2.828 &  6.78 &$-$0.31\,(BH) \\ %15.77 \\
           &      &             & 160.0 & 0.584 & 0.385 & 0.1688 & 0.0519 & 0.1767 & 0.373 & 0.458 & 0.813 &        &        &         &       &       \\
            \noalign{\smallskip}
\hline
            \noalign{\smallskip}
1342246772 & 1121 & 56086.18500 &  70.0 & 1.022 & 0.848 & 0.0063 & 0.0173 & 0.0184 & 0.862 & 0.892 & 0.966 & 20.070 & 20.406 & T~2.744 & 10.30 &$-$0.49\,(BH) \\%17.51 \\
           &      &             & 160.0 & 0.616 & 0.392 & 0.0724 & 0.0611 & 0.0947 & 0.379 & 0.448 & 0.847 &        &        &         &       &       \\
1342246773 & 1121 & 56086.18904 &  70.0 & 1.072 & 0.889 & 0.0082 & 0.0275 & 0.0287 & 0.903 & 0.892 & 1.013 & 20.070 & 20.406 & T~2.744 & 10.37 &$-$0.49\,(BH) \\%17.51 \\
           &      &             & 160.0 & 0.646 & 0.412 & 0.0787 & 0.0508 & 0.0936 & 0.399 & 0.448 & 0.892 &        &        &         &       &       \\
1342246774 & 1121 & 56086.19308 & 100.0 & 0.778 & 0.659 & 0.0132 & 0.0086 & 0.0157 & 0.660 & 0.744 & 0.887 & 20.070 & 20.406 & T~2.744 & 10.44 &$-$0.50\,(BH) \\%17.51 \\
           &      &             & 160.0 & 0.570 & 0.363 & 0.0773 & 0.0651 & 0.1011 & 0.352 & 0.448 & 0.786 &        &        &         &       &       \\
1342246775 & 1121 & 56086.19712 & 100.0 & 0.736 & 0.622 & 0.0162 & 0.0308 & 0.0348 & 0.623 & 0.744 & 0.837 & 20.070 & 20.406 & T~2.744 & 10.51 &$-$0.50\,(BH) \\%17.51 \\
           &      &             & 160.0 & 0.442 & 0.282 & 0.0734 & 0.0477 & 0.0876 & 0.273 & 0.448 & 0.610 &        &        &         &       &       \\
            \noalign{\smallskip}
\hline
            \noalign{\smallskip}
1342257193 & 1310 & 56275.07361 &  70.0 & 1.078 & 0.947 & 0.0029 & 0.0075 & 0.0081 & 0.963 & 0.940 & 1.024 & 20.058 & 19.822 & L~2.774 & 15.83 &$+$0.84\,(TH) \\ %19.52 \\
           &      &             & 160.0 & 0.510 & 0.344 & 0.0344 & 0.0396 & 0.0524 & 0.333 & 0.473 & 0.705 &        &        &         &       &       \\
1342257194 & 1310 & 56275.07765 &  70.0 & 1.102 & 0.964 & 0.0034 & 0.0104 & 0.0110 & 0.980 & 0.940 & 1.042 & 20.058 & 19.822 & L~2.774 & 15.88 &$+$0.85\,(TH) \\ %19.52 \\
           &      &             & 160.0 & 0.573 & 0.389 & 0.0314 & 0.0435 & 0.0536 & 0.377 & 0.473 & 0.797 &        &        &         &       &       \\
1342257195 & 1310 & 56275.08169 & 100.0 & 0.905 & 0.814 & 0.0064 & 0.0042 & 0.0077 & 0.815 & 0.785 & 1.038 & 20.058 & 19.822 & L~2.774 & 15.94 &$+$0.85\,(TH) \\ %19.52 \\
           &      &             & 160.0 & 0.523 & 0.354 & 0.0373 & 0.0469 & 0.0599 & 0.343 & 0.473 & 0.725 &        &        &         &       &       \\
1342257196 & 1310 & 56275.08573 & 100.0 & 0.896 & 0.802 & 0.0056 & 0.0038 & 0.0068 & 0.803 & 0.785 & 1.023 & 20.058 & 19.822 & L~2.774 & 15.99 &$+$0.86\,(TH) \\ %19.52 \\
           &      &             & 160.0 & 0.445 & 0.302 & 0.0380 & 0.0625 & 0.0731 & 0.293 & 0.473 & 0.618 &        &        &         &       &       \\
            \noalign{\smallskip}
\hline\hline                 % inserts double horizontal lines
\end{tabular}
%\end{table*}
\end{sidewaystable*}

\clearpage

\begin{sidewaystable*}[h!]
\small
%\begin{table*}[h!]
\caption{PSF photometry of Ariel. Applied colour correction factors cc to derive colour corrected fluxes f$_{\rm moon,cc}$, are listed in Table~\ref{table:derivedmodelparams}. r$_{\rm helio}$ is the light-time corrected heliocentric range, $\Delta_{\rm obs}$ is the range of target centre wrt.\ the observer, i.e.\ {\it Herschel}, $\alpha$ is the Phase angle with indication of L(eading) or T(railing) the Sun, and $\theta_{\rm U-O}$ is the angular separation from Uranus. $\dot{r}_{\rm helio}$ is the heliocentric range rate and $\frac{\dot{r}_{\rm helio}}{|{\dot{r}_{\rm helio}^{\rm max}}|}$ indicates L(eading) H(emisphere) or T(railing) H(emisphere), if the absolute value of the ratio is greater than $\frac{2}{3}$, or B(oth) H(emispheres) otherwise. All five figures have been computed with the JPL Horizons On-Line Ephemeris System. The five epochs with 4 observations each are separated by horizontal lines. 
% l$_{\rm SSP}$ is the latitude of the sub-solar point. 
}             % title of Table
\label{table:psfphotAriel}      % is used to refer this table in the text
\centering                          % used for centering table
\begin{tabular}{l r c r c c c c c c c c c c c c c}        % centered columns (4 columns)
\hline\hline                 % inserts double horizontal lines
            \noalign{\smallskip}
OBSID      &  OD  & MJD   &$\lambda_{\rm ref}$ &$\frac{f_{\rm moon}}{f_{\rm Uranus}}$ & f$_{\rm moon}$ & $\sigma_{\rm par}$ & $\sigma_{\rm red}$& $\sigma_{tot}$ & f$_{\rm moon,cc}$ & f$_{\rm model}$ & $\frac{f_{\rm moon,cc}}{f_{\rm model}}$ & r$_{\rm helio}$ & $\Delta_{\rm obs}$& $\alpha$ & $\theta_{\rm U-O}$ & $\frac{\dot{r}_{\rm helio}}{|{\dot{r}_{\rm helio}^{\rm max}}|}$ \\ %& l$_{\rm SSP}$\\
            \noalign{\smallskip}
           &      & mid-time obs. &$\mu$m&(10$^{-3}$)& (Jy)& (Jy) & (Jy) & (Jy) &   (Jy)  &  (Jy)   &   & (AU) & (AU) & (deg) & (\arcsec) &   \\ %(deg) \\
            \noalign{\smallskip}
\hline
            \noalign{\smallskip}
1342211117 &  579 & 55543.72060 &  70.0 & 0.859 & 0.745 & 0.0063 & 0.0093 & 0.0112 & 0.758 & 0.751 & 1.009 & 20.090 & 19.975 & L~2.826 & 13.15 &$+$1.00\,(TH) \\ %11.96 \\
           &      &             & 160.0 & 0.278 & 0.185 & 0.1387 & 0.0797 & 0.1600 & 0.179 & 0.410 & 0.438 &        &        &         &       &       \\
1342211118 &  579 & 55543.72464 &  70.0 & 0.845 & 0.730 & 0.0069 & 0.0236 & 0.0246 & 0.743 & 0.751 & 0.989 & 20.090 & 19.975 & L~2.826 & 13.14 &$+$1.00\,(TH) \\ %11.96 \\
           &      &             & 160.0 & 0.364 & 0.243 & 0.1345 & 0.0906 & 0.1621 & 0.236 & 0.410 & 0.576 &        &        &         &       &       \\
1342211120 &  579 & 55543.73128 & 100.0 & 0.728 & 0.644 & 0.0130 & 0.0285 & 0.0313 & 0.645 & 0.656 & 0.984 & 20.090 & 19.975 & L~2.826 & 13.12 &$+$1.00\,(TH) \\ %11.97 \\
           &      &             & 160.0 & 0.333 & 0.222 & 0.1316 & 0.0938 & 0.1616 & 0.216 & 0.410 & 0.526 &        &        &         &       &       \\
1342211121 &  579 & 55543.73532 & 100.0 & 0.660 & 0.582 & 0.0148 & 0.0201 & 0.0249 & 0.584 & 0.656 & 0.890 & 20.090 & 19.975 & L~2.826 & 13.11 &$+$0.99\,(TH) \\ %11.97 \\
           &      &             & 160.0 & 0.413 & 0.276 & 0.1188 & 0.0816 & 0.1442 & 0.268 & 0.410 & 0.654 &        &        &         &       &       \\
            \noalign{\smallskip}
\hline
            \noalign{\smallskip}
1342223982 &  789 & 55754.05794 &  70.0 & 0.932 & 0.821 & 0.0052 & 0.0107 & 0.0119 & 0.835 & 0.780 & 1.071 & 20.083 & 19.802 & T~2.837 & 13.05 &$-$0.98\,(LH) \\ % 14.22 \\
           &      &             & 160.0 & 0.345 & 0.233 & 0.0974 & 0.0681 & 0.1188 & 0.226 & 0.422 & 0.537 &        &        &         &       &       \\
1342223983 &  789 & 55754.06198 &  70.0 & 0.910 & 0.798 & 0.0056 & 0.0225 & 0.0232 & 0.812 & 0.779 & 1.042 & 20.083 & 19.802 & T~2.837 & 13.07 &$-$0.98\,(LH) \\ % 14.22 \\
           &      &             & 160.0 & 0.239 & 0.162 & 0.0914 & 0.0904 & 0.1286 & 0.157 & 0.422 & 0.372 &        &        &         &       &       \\
1342223985 &  789 & 55754.06862 & 100.0 & 0.868 & 0.782 & 0.0103 & 0.0161 & 0.0191 & 0.784 & 0.678 & 1.157 & 20.083 & 19.801 & T~2.837 & 13.11 &$-$0.98\,(LH) \\ % 14.22 \\
           &      &             & 160.0 & 0.362 & 0.244 & 0.0963 & 0.0438 & 0.1058 & 0.237 & 0.422 & 0.562 &        &        &         &       &       \\
1342223986 &  789 & 55754.07266 & 100.0 & 0.943 & 0.846 & 0.0122 & 0.0153 & 0.0196 & 0.848 & 0.678 & 1.251 & 20.083 & 19.801 & T~2.837 & 13.13 &$-$0.99\,(LH) \\ % 14.22 \\
           &      &             & 160.0 & 0.224 & 0.153 & 0.0886 & 0.0798 & 0.1193 & 0.148 & 0.422 & 0.351 &        &        &         &       &       \\
            \noalign{\smallskip}
\hline
            \noalign{\smallskip}
1342235629 &  957 & 55921.94924 &  70.0 & 0.850 & 0.726 & 0.0084 & 0.0332 & 0.0343 & 0.739 & 0.748 & 0.987 & 20.077 & 20.119 & L~2.829 & 11.59 &$+$0.88\,(TH) \\ %16.02 \\
           &      &             & 160.0 & 0.250 & 0.164 & 0.1338 & 0.0517 & 0.1434 & 0.159 & 0.406 & 0.391 &        &        &         &       &       \\
1342235630 &  957 & 55921.95328 &  70.0 & 0.824 & 0.702 & 0.0093 & 0.0146 & 0.0173 & 0.714 & 0.748 & 0.954 & 20.077 & 20.119 & L~2.829 & 11.53 &$+$0.87\,(TH) \\ %16.02 \\
           &      &             & 160.0 & 0.226 & 0.149 & 0.1264 & 0.0721 & 0.1455 & 0.144 & 0.406 & 0.355 &        &        &         &       &       \\
1342235632 &  957 & 55921.95992 & 100.0 & 0.615 & 0.536 & 0.0208 & 0.0358 & 0.0414 & 0.537 & 0.651 & 0.825 & 20.077 & 20.119 & L~2.828 & 11.43 &$+$0.86\,(TH) \\ %16.02 \\
           &      &             & 160.0 & 0.201 & 0.132 & 0.1345 & 0.0725 & 0.1528 & 0.128 & 0.406 & 0.316 &        &        &         &       &       \\
1342235633 &  957 & 55921.96396 & 100.0 & 0.683 & 0.593 & 0.0231 & 0.0206 & 0.0310 & 0.594 & 0.651 & 0.913 & 20.077 & 20.119 & L~2.828 & 11.37 &$+$0.86\,(TH) \\ %16.02 \\
           &      &             & 160.0 & 0.232 & 0.153 & 0.1330 & 0.0450 & 0.1404 & 0.149 & 0.406 & 0.366 &        &        &         &       &       \\
            \noalign{\smallskip}
\hline
            \noalign{\smallskip}
1342246772 & 1121 & 56086.18500 &  70.0 & 0.852 & 0.708 & 0.0811 & 0.1371 & 0.1593 & 0.720 & 0.743 & 0.969 & 20.070 & 20.406 & T~2.744 &  4.47 &$+$0.03\,(BH) \\ %17.78 \\
           &      &             & 160.0 & 0.072 & 0.046 & 1.6721 & 0.0569 & 1.6731 & 0.045 & 0.401 & 0.111 &        &        &         &       &       \\
1342246773 & 1121 & 56086.18904 &  70.0 & 0.633 & 0.525 & 0.0778 & 0.1864 & 0.2020 & 0.534 & 0.743 & 0.718 & 20.070 & 20.406 & T~2.744 &  4.46 &$+$0.02\,(BH) \\ %17.78 \\
           &      &             & 160.0 & 0.200 & 0.128 & 0.6343 & 0.1172 & 0.6450 & 0.124 & 0.401 & 0.309 &        &        &         &       &       \\
1342246774 & 1121 & 56086.19308 & 100.0 & 1.077 & 0.912 & 0.1160 & 0.0960 & 0.1506 & 0.915 & 0.645 & 1.418 & 20.070 & 20.406 & T~2.744 &  4.45 &$+$0.01\,(BH) \\ %17.78 \\
           &      &             & 160.0 & 0.303 & 0.193 & 0.6355 & 0.0828 & 0.6409 & 0.188 & 0.401 & 0.468 &        &        &         &       &       \\
1342246775 & 1121 & 56086.19712 & 100.0 & 1.215 & 1.027 & 0.1131 & 0.0851 & 0.1415 & 1.030 & 0.645 & 1.597 & 20.070 & 20.405 & T~2.744 &  4.45 &$-$0.00\,(BH) \\ % 17.78 \\
           &      &             & 160.0 & 0.663 & 0.423 & 0.7415 & 0.3194 & 0.8073 & 0.410 & 0.401 & 1.024 &        &        &         &       &       \\
            \noalign{\smallskip}
\hline
            \noalign{\smallskip}
1342257193 & 1310 & 56275.07361 &  70.0 & 0.663 & 0.582 & 0.0510 & 0.1044 & 0.1162 & 0.592 & 0.781 & 0.758 & 20.060 & 19.824 & L~2.774 &  6.13 &$+$0.36\,(BH) \\ %19.80 \\
           &      &             & 160.0 & 0.499 & 0.337 & 0.3590 & 0.1382 & 0.3847 & 0.327 & 0.422 & 0.775 &        &        &         &       &       \\
1342257194 & 1310 & 56275.07765 &  70.0 & 0.751 & 0.657 & 0.0442 & 0.0321 & 0.0546 & 0.668 & 0.781 & 0.856 & 20.060 & 19.824 & L~2.774 &  6.03 &$+$0.35\,(BH) \\ %19.80 \\
           &      &             & 160.0 & 0.246 & 0.167 & 0.3202 & 0.1194 & 0.3418 & 0.162 & 0.422 & 0.384 &        &        &         &       &       \\
1342257195 & 1310 & 56275.08169 & 100.0 & 0.503 & 0.452 & 0.0673 & 0.0493 & 0.0834 & 0.453 & 0.678 & 0.668 & 20.060 & 19.824 & L~2.774 &  5.94 &$+$0.34\,(BH) \\ %19.80 \\
           &      &             & 160.0 & 0.214 & 0.144 & 0.3606 & 0.0476 & 0.3637 & 0.140 & 0.422 & 0.332 &        &        &         &       &       \\
1342257196 & 1310 & 56275.08573 & 100.0 & 0.678 & 0.607 & 0.0604 & 0.0215 & 0.0641 & 0.609 & 0.678 & 0.898 & 20.060 & 19.824 & L~2.774 &  5.85 &$+$0.33\,(BH) \\ %19.80 \\
           &      &             & 160.0 & 0.473 & 0.321 & 0.3281 & 0.1268 & 0.3517 & 0.312 & 0.422 & 0.739 &        &        &         &       &       \\
            \noalign{\smallskip}
\hline\hline                 % inserts double horizontal lines
\end{tabular}
%\end{table*}
\end{sidewaystable*}

\clearpage

\begin{sidewaystable*}[ht!]
\small
%\begin{table*}[h!]
\caption{PSF photometry of Miranda. Applied colour correction factors cc to derive colour corrected fluxes f$_{\rm moon,cc}$, are listed in Table~\ref{table:derivedmodelparams}. r$_{\rm helio}$ is the light-time corrected heliocentric range, $\Delta_{\rm obs}$ is the range of target centre wrt.\ the observer, i.e.\ {\it Herschel}, $\alpha$ is the Phase angle with indication of L(eading) or T(railing) the Sun, and $\theta_{\rm U-O}$ is the angular separation from Uranus. $\dot{r}_{\rm helio}$ is the heliocentric range rate and $\frac{\dot{r}_{\rm helio}}{|{\dot{r}_{\rm helio}^{\rm max}}|}$ indicates L(eading) H(emisphere) or T(railing) H(emisphere), if the absolute value of the ratio is greater than $\frac{2}{3}$, or B(oth) H(emispheres) otherwise. All five figures have been computed with the JPL Horizons On-Line Ephemeris System. 
% l$_{\rm SSP}$ is the latitude of the sub-solar point. 
}             % title of Table
\label{table:psfphotMiranda}      % is used to refer this table in the text
\centering                          % used for centering table
\begin{tabular}{l r c r c c c c c c c c c c c c c}        % centered columns (4 columns)
\hline\hline                 % inserts double horizontal lines
            \noalign{\smallskip}
OBSID      &  OD  & MJD   &$\lambda_{\rm ref}$ &$\frac{f_{\rm moon}}{f_{\rm Uranus}}$ & f$_{\rm moon}$ & $\sigma_{\rm par}$ & $\sigma_{\rm red}$& $\sigma_{tot}$ & f$_{\rm moon,cc}$ & f$_{\rm model}$ & $\frac{f_{\rm moon,cc}}{f_{\rm model}}$ & r$_{\rm helio}$ & $\Delta_{\rm obs}$& $\alpha$ & $\theta_{\rm U-O}$ & $\frac{\dot{r}_{\rm helio}}{|{\dot{r}_{\rm helio}^{\rm max}}|}$ \\ %& l$_{\rm SSP}$\\
            \noalign{\smallskip}
           &      & mid-time obs. &$\mu$m&(10$^{-3}$)& (Jy)& (Jy) & (Jy) & (Jy) &   (Jy)  &  (Jy)   &   & (AU) & (AU) & (deg) & (\arcsec) &   \\ %(deg) \\
            \noalign{\smallskip}
\hline
            \noalign{\smallskip}
1342211117 &  579 & 55543.72060 &  70.0 & 0.141 & 0.122 & 0.0996 & 0.0650 & 0.1189 & 0.124 & 0.129 & 0.964 & 20.089 & 19.974 & L~2.826 &  7.56  &$+$0.84\,(TH) \\ %16.04 \\
%           &      &               & 160.0 &       &       &        &        &        &       &       &       &        &       \\
1342211118 &  579 & 55543.72464 &  70.0 & 0.299 & 0.258 & 0.1219 & 0.0253 & 0.1244 & 0.263 & 0.129 & 2.038 & 20.089 & 19.975 & L~2.826 &  7.65  &$+$0.85\,(TH) \\ %16.04 \\
%           &      &               & 160.0 &       &        &      &     &     &      &       &       &       &        &       \\
1342211120 &  579 & 55543.73128 & 100.0 & 0.110 & 0.097 & 0.2645 & 0.0254 & 0.2657 & 0.097 & 0.112 & 0.870 & 20.090 & 19.975 & L~2.826 &  7.78  &$+$0.86\,(TH) \\ %16.04 \\
%           &      &               & 160.0 &       &        &      &     &     &      &      &        &       &        &       \\
1342211121 &  579 & 55543.73532 & 100.0 & 0.187 & 0.165 & 0.2922 & 0.0810 & 0.3032 & 0.166 & 0.112 & 1.480 & 20.090 & 19.975 & L~2.826 &  7.85  &$+$0.87\,(TH) \\ %16.04 \\
%           &      &               & 160.0 &       &        &      &     &     &      &       &       &       &        &       \\
            \noalign{\smallskip}
\hline
            \noalign{\smallskip}
1342223982 &  789 & 55754.05794 &  70.0 & 0.102 & 0.090 & 0.3742 & 0.0720 & 0.3810 & 0.092 & 0.134 & 0.684 & 20.082 & 19.801 & T~2.837 &  3.50  &$-$0.21\,(BH) \\ %17.82 \\
%           &      &               & 160.0 &       &        &      &     &     &      &       &       &       &        &       \\
1342223983 &  789 & 55754.06198 &  70.0 & 0.042 & 0.037 & 0.6195 & 0.0482 & 0.6213 & 0.037 & 0.134 & 0.279 & 20.082 & 19.800 & T~2.837 &  3.43  &$-$0.19\,(BH) \\ %17.82 \\
%           &      &               & 160.0 &       &        &      &     &     &      &       &       &       &        &       \\
1342223985 &  789 & 55754.06862 & 100.0 &   --  &   --  &   --   &   --   &   --   &  --   & 0.115 &  --   & 20.082 & 19.800 & T~2.837 &  3.33  &$-$0.17\,(BH) \\ %17.82 \\
%           &      &               & 160.0 &       &        &      &     &     &      &       &       &       &        &       \\
1342223986 &  789 & 55754.07266 & 100.0 &   --  &   --  &   --   &   --   &   --   &  --   & 0.116 &  --   & 20.082 & 19.800 & T~2.837 &  3.27  &$-$0.15\,(BH) \\ %17.82 \\
%           &      &               & 160.0 &       &        &      &     &     &      &       &       &       &        &       \\
            \noalign{\smallskip}
\hline
            \noalign{\smallskip}
1342235629 &  957 & 55921.94924 &  70.0 & 0.435 & 0.371 & 0.1109 & 0.0656 & 0.1289 & 0.378 & 0.128 & 2.954 & 20.077 & 20.118 & L~2.828 &  8.91  &$-$1.00\,(LH) \\ %19.15 \\
%           &      &               & 160.0 &       &        &      &     &     &      &       &       &       &        &       \\
1342235630 &  957 & 55921.95328 &  70.0 & 0.228 & 0.194 & 0.0926 & 0.0744 & 0.1188 & 0.197 & 0.128 & 1.542 & 20.076 & 20.118 & L~2.828 &  8.91  &$-$1.00\,(LH) \\ %19.15 \\
%           &      &               & 160.0 &       &        &      &     &     &      &       &       &       &        &       \\
1342235632 &  957 & 55921.95992 & 100.0 & 0.158 & 0.137 & 0.2457 & 0.0765 & 0.2573 & 0.137 & 0.111 & 1.238 & 20.076 & 20.118 & L~2.828 &  8.91  &$-$1.00\,(LH) \\ %19.15 \\
%           &      &               & 160.0 &       &        &      &     &     &      &       &       &       &        &       \\
1342235633 &  957 & 55921.96396 & 100.0 & 0.098 & 0.085 & 0.2618 & 0.0242 & 0.2629 & 0.085 & 0.111 & 0.767 & 20.076 & 20.118 & L~2.828 &  8.91  &$-$1.00\,(LH) \\ %19.15 \\
%           &      &               & 160.0 &       &        &      &     &     &      &       &       &       &        &       \\
            \noalign{\smallskip}
\hline
            \noalign{\smallskip}
1342246772 & 1121 & 56086.18500 &  70.0 & 0.276 & 0.230 & 0.3037 & 0.1285 & 0.3298 & 0.233 & 0.127 & 1.838 & 20.068 & 20.404 & T~2.744 &  4.53  &$-$0.39\,(BH) \\ %20.39 \\
%           &      &               & 160.0 &       &        &      &     &     &      &       &       &       &        &       \\
1342246773 & 1121 & 56086.18904 &  70.0 & 0.217 & 0.180 & 0.4240 & 0.0450 & 0.4264 & 0.183 & 0.127 & 1.441 & 20.068 & 20.404 & T~2.745 &  4.44  &$-$0.38\,(BH) \\ %20.39 \\
%           &      &               & 160.0 &       &        &      &     &     &      &       &       &       &        &       \\
1342246774 & 1121 & 56086.19308 & 100.0 & 0.222 & 0.188 & 0.6144 & 0.0148 & 0.6146 & 0.189 & 0.110 & 1.714 & 20.068 & 20.404 & T~2.745 &  4.35  &$-$0.36\,(BH) \\ %20.39 \\
%           &      &               & 160.0 &       &        &      &     &     &      &       &       &       &        &       \\
1342246775 & 1121 & 56086.19712 & 100.0 & 0.160 & 0.135 & 0.5923 & 0.0373 & 0.5934 & 0.135 & 0.110 & 1.231 & 20.068 & 20.404 & T~2.745 &  4.26  &$-$0.34\,(BH) \\ %20.39 \\
%           &      &               & 160.0 &       &        &      &     &     &      &       &       &       &        &       \\
            \noalign{\smallskip}
\hline
            \noalign{\smallskip}
1342257193 & 1310 & 56275.07361 &  70.0 & 0.329 & 0.289 & 0.2983 & 0.0739 & 0.3073 & 0.294 & 0.134 & 2.193 & 20.059 & 19.823 & L~2.773 &  4.74  &$-$0.46\,(BH) \\ %21.75 \\
%           &      &               & 160.0 &       &        &      &     &     &      &       &       &       &        &       \\
1342257194 & 1310 & 56275.07765 &  70.0 & 0.191 & 0.167 & 0.2981 & 0.0610 & 0.3042 & 0.170 & 0.134 & 1.266 & 20.059 & 19.823 & L~2.773 &  4.85  &$-$0.48\,(BH) \\ %21.75 \\
%           &      &               & 160.0 &       &        &      &     &     &      &       &       &       &        &       \\
1342257195 & 1310 & 56275.08169 & 100.0 & 0.048 & 0.044 & 0.4470 & 0.0228 & 0.4476 & 0.044 & 0.116 & 0.376 & 20.059 & 19.823 & L~2.773 &  4.97  &$-$0.49\,(BH) \\ %21.75 \\
%           &      &               & 160.0 &       &        &      &     &     &      &       &       &       &        &       \\
1342257196 & 1310 & 56275.08573 & 100.0 & 0.170 & 0.152 & 0.4948 & 0.0481 & 0.4971 & 0.153 & 0.116 & 1.315 & 20.059 & 19.823 & L~2.773 &  5.08  &$-$0.51\,(BH) \\ %21.75 \\
%           &      &               & 160.0 &       &        &      &     &     &      &       &       &       &        &       \\
            \noalign{\smallskip}
\hline\hline                 % inserts double horizontal lines
\end{tabular}
%\end{table*}
\end{sidewaystable*}

\clearpage

\subsection{Comparison of PSF photometry with aperture photometry for selected measurements}

%\begin{sidewaystable*}[h!]
\begin{table*}[h!]
\caption{Aperture photometry of Uranian moons for selected measurements when the moon image was well
separated from Uranus (minimum aperture edge distance from Uranus was
8\farcs77, 10\farcs79, and 16\farcs57 at 70, 100, and 
160\,$\mu$m, respectively) and any PSF residuals and comparison with the corresponding PSF photometry.
The aperture photometry has been corrected for the finite aperture size according to the description
in the PACS Handbook~\citep{exter18}, sect.\ 7.5.2.
}             % title of Table
\label{table:aperphotcomp_1}      % is used to refer this table in the text
\centering                          % used for centering table
\begin{tabular}{l r c l r c c c c c c c}        % centered columns (4 columns)
\hline\hline                 % inserts double horizontal lines
            \noalign{\smallskip}
OBSID      &  OD  & MJD   & object &$\lambda_{\rm ref}$ & r$^{~aperture}$ & f$_{\rm moon}^{~aperture}$& $\sigma_{\rm aper}$ & f$_{\rm moon}^{~PSF}$ & $\sigma_{tot}$ & $\frac{f^{\rm PSF}}{f^{\rm aper}}$ & $\sigma_{\rm ratio}$ \\
            \noalign{\smallskip}
           &      & mid-time obs. &        &    $\mu$m        &  (\arcsec)   &  (Jy)                 &        (Jy)        &       (Jy)          &     (Jy)      &    &  \\  
            \noalign{\smallskip}
\hline
            \noalign{\smallskip}
1342211117 &  579 & 55543.72060 & Titania &  70.0 &  5.6 & 1.635 & 0.012 & 1.669 & 0.017 & 1.021 & 0.013 \\
           &      &             &         & 160.0 & 10.7 & 0.827 & 0.030 & 0.875 & 0.011 & 1.058 & 0.041 \\
           &      &             & Umbriel &  70.0 &  5.6 & 0.807 & 0.017 & 0.833 & 0.019 & 1.032 & 0.032 \\
1342211118 &  579 & 55543.72464 & Titania &  70.0 &  5.6 & 1.622 & 0.012 & 1.680 & 0.018 & 1.036 & 0.013 \\
           &      &             &         & 160.0 & 10.7 & 0.833 & 0.029 & 0.856 & 0.016 & 1.028 & 0.041 \\
           &      &             & Umbriel &  70.0 &  5.6 & 0.844 & 0.017 & 0.880 & 0.012 & 1.043 & 0.025 \\
1342211120 &  579 & 55543.73128 & Titania & 100.0 &  6.8 & 1.397 & 0.009 & 1.412 & 0.006 & 1.011 & 0.008 \\
           &      &             &         & 160.0 & 10.7 & 0.813 & 0.045 & 0.864 & 0.010 & 1.063 & 0.060 \\
           &      &             & Umbriel & 100.0 &  6.8 & 0.592 & 0.043 & 0.718 & 0.010 & 1.213 & 0.090 \\
1342211121 &  579 & 55543.73532 & Titania & 100.0 &  6.8 & 1.370 & 0.013 & 1.398 & 0.006 & 1.020 & 0.011 \\
           &      &             &         & 160.0 & 10.7 & 0.838 & 0.031 & 0.854 & 0.014 & 1.019 & 0.041 \\
           &      &             & Umbriel & 100.0 &  6.8 & 0.704 & 0.028 & 0.762 & 0.008 & 1.082 & 0.045 \\
1342223982 &  789 & 55754.05794 & Titania &  70.0 &  5.6 & 1.655 & 0.025 & 1.705 & 0.015 & 1.030 & 0.018 \\
           &      &             &         & 160.0 & 10.7 & 0.838 & 0.012 & 0.884 & 0.009 & 1.055 & 0.019 \\
           &      &             & Oberon  &  70.0 &  5.6 & 1.549 & 0.017 & 1.611 & 0.008 & 1.040 & 0.013 \\
           &      &             &         & 160.0 & 10.7 & 0.779 & 0.016 & 0.807 & 0.008 & 1.036 & 0.024 \\
           &      &             & Ariel   &  70.0 &  5.6 & 0.833 & 0.045 & 0.821 & 0.012 & 0.986 & 0.055 \\
1342223983 &  789 & 55754.06198 & Titania &  70.0 &  5.6 & 1.641 & 0.023 & 1.710 & 0.016 & 1.042 & 0.018 \\
           &      &             &         & 160.0 & 10.7 & 0.869 & 0.011 & 0.907 & 0.007 & 1.044 & 0.015 \\
           &      &             & Oberon  &  70.0 &  5.6 & 1.581 & 0.014 & 1.638 & 0.012 & 1.036 & 0.012 \\
           &      &             &         & 160.0 & 10.7 & 0.836 & 0.014 & 0.836 & 0.012 & 1.000 & 0.022 \\
           &      &             & Ariel   &  70.0 &  5.6 & 0.681 & 0.061 & 0.798 & 0.023 & 1.172 & 0.110 \\
1342223985 &  789 & 55754.06862 & Titania & 100.0 &  6.8 & 1.418 & 0.010 & 1.455 & 0.005 & 1.026 & 0.008 \\
           &      &             &         & 160.0 & 10.7 & 0.862 & 0.015 & 0.868 & 0.016 & 1.007 & 0.026 \\
           &      &             & Oberon  & 100.0 &  6.8 & 1.331 & 0.011 & 1.374 & 0.012 & 1.032 & 0.012 \\
           &      &             &         & 160.0 & 10.7 & 0.810 & 0.011 & 0.826 & 0.005 & 1.020 & 0.015 \\
1342223986 &  789 & 55754.07266 & Titania & 100.0 &  6.8 & 1.434 & 0.013 & 1.455 & 0.005 & 1.015 & 0.010 \\
           &      &             &         & 160.0 & 10.7 & 0.833 & 0.015 & 0.883 & 0.009 & 1.060 & 0.022 \\
           &      &             & Oberon  & 100.0 &  6.8 & 1.347 & 0.016 & 1.398 & 0.013 & 1.038 & 0.016 \\
           &      &             &         & 160.0 & 10.7 & 0.838 & 0.015 & 0.839 & 0.008 & 1.001 & 0.020 \\
1342235629 &  957 & 55921.94924 & Titania &  70.0 &  5.6 & 1.592 & 0.044 & 1.630 & 0.097 & 1.024 & 0.067 \\
           &      &             & Oberon  &  70.0 &  5.6 & 1.557 & 0.013 & 1.573 & 0.014 & 1.010 & 0.012 \\
           &      &             &         & 160.0 & 10.7 & 0.785 & 0.012 & 0.819 & 0.011 & 1.043 & 0.021 \\
           &      &             & Ariel   &  70.0 &  5.6 & 0.644 & 0.129 & 0.726 & 0.034 & 1.127 & 0.232 \\
1342235630 &  957 & 55921.95328 & Titania &  70.0 &  5.6 & 1.550 & 0.041 & 1.684 & 0.021 & 1.086 & 0.032 \\
           &      &             & Oberon  &  70.0 &  5.6 & 1.556 & 0.023 & 1.576 & 0.015 & 1.013 & 0.018 \\
           &      &             &         & 160.0 & 10.7 & 0.758 & 0.011 & 0.800 & 0.006 & 1.055 & 0.017 \\
           &      &             & Ariel   &  70.0 &  5.6 & 0.651 & 0.062 & 0.702 & 0.017 & 1.078 & 0.106 \\
1342235632 &  957 & 55921.95992 & Titania & 100.0 &  6.8 & 1.388 & 0.019 & 1.449 & 0.008 & 1.083 & 0.016 \\
           &      &             & Oberon  & 100.0 &  6.8 & 1.296 & 0.012 & 1.320 & 0.009 & 1.019 & 0.012 \\
           &      &             &         & 160.0 & 10.7 & 0.762 & 0.011 & 0.814 & 0.009 & 1.068 & 0.019 \\
1342235633 &  957 & 55921.96396 & Titania & 100.0 &  6.8 & 1.398 & 0.040 & 1.466 & 0.011 & 1.049 & 0.031 \\
           &      &             & Oberon  & 100.0 &  6.8 & 1.306 & 0.019 & 1.353 & 0.009 & 1.036 & 0.017 \\
           &      &             &         & 160.0 & 10.7 & 0.785 & 0.019 & 0.819 & 0.013 & 1.043 & 0.030 \\
1342246772 & 1121 & 56086.18500 & Titania &  70.0 &  5.6 & 1.554 & 0.017 & 1.573 & 0.012 & 1.012 & 0.013 \\
           &      &             & Oberon  &  70.0 &  5.6 & 1.558 & 0.028 & 1.597 & 0.034 & 1.025 & 0.029 \\
           &      &             &         & 160.0 & 10.7 & 0.768 & 0.014 & 0.792 & 0.010 & 1.031 & 0.023 \\
1342246773 & 1121 & 56086.18904 & Titania &  70.0 &  5.6 & 1.581 & 0.030 & 1.630 & 0.038 & 1.031 & 0.031 \\
           &      &             & Oberon  &  70.0 &  5.6 & 1.517 & 0.019 & 1.566 & 0.033 & 1.032 & 0.025 \\
           &      &             &         & 160.0 & 10.7 & 0.768 & 0.007 & 0.776 & 0.009 & 1.010 & 0.015 \\
            \noalign{\smallskip}
\hline\hline                 % inserts double horizontal lines
\end{tabular}
\end{table*}
%\end{sidewaystable*}

\addtocounter{table}{-1}
  
%\begin{sidewaystable*}[h!]
\begin{table*}[h!]
\caption{Aperture photometry of Uranian moons continued.
}             % title of Table
\label{table:aperphotcomp_2}      % is used to refer this table in the text
\centering                          % used for centering table
\begin{tabular}{l r c l r c c c c c c c}        % centered columns (4 columns)
\hline\hline                 % inserts double horizontal lines
            \noalign{\smallskip}
OBSID      &  OD  & MJD   & object &$\lambda_{\rm ref}$ & r$^{~aperture}$ & f$_{\rm moon}^{~aperture}$& $\sigma_{\rm aper}$ & f$_{\rm moon}^{~PSF}$ & $\sigma_{tot}$ & $\frac{f^{\rm PSF}}{f^{\rm aper}}$ & $\sigma_{\rm ratio}$ \\
            \noalign{\smallskip}
           &      & mid-time obs. &        &    $\mu$m        &  (\arcsec)   &  (Jy)                 &        (Jy)        &       (Jy)          &     (Jy)      &    &  \\  
            \noalign{\smallskip}
\hline
            \noalign{\smallskip}
1342246774 & 1121 & 56086.19308 & Titania & 100.0 &  6.8 & 1.283 & 0.011 & 1.318 & 0.010 & 1.027 & 0.012 \\
           &      &             & Oberon  & 100.0 &  6.8 & 1.273 & 0.014 & 1.300 & 0.006 & 1.021 & 0.012 \\
           &      &             &         & 160.0 & 10.7 & 0.751 & 0.018 & 0.781 & 0.005 & 1.040 & 0.026 \\
1342246775 & 1121 & 56086.19712 & Titania & 100.0 &  6.8 & 1.316 & 0.028 & 1.343 & 0.008 & 1.021 & 0.023 \\
           &      &             & Oberon  & 100.0 &  6.8 & 1.242 & 0.008 & 1.289 & 0.008 & 1.038 & 0.009 \\
           &      &             &         & 160.0 & 10.7 & 0.739 & 0.011 & 0.762 & 0.008 & 1.031 & 0.019 \\
1342257193 & 1310 & 56275.07361 & Titania &  70.0 &  5.6 & 1.681 & 0.016 & 1.724 & 0.018 & 1.026 & 0.014 \\
           &      &             &         & 160.0 & 10.7 & 0.894 & 0.027 & 0.931 & 0.017 & 1.041 & 0.037 \\
           &      &             & Oberon  &  70.0 &  5.6 & 1.575 & 0.012 & 1.625 & 0.013 & 1.032 & 0.011 \\
           &      &             &         & 160.0 & 10.7 & 0.791 & 0.011 & 0.811 & 0.006 & 1.025 & 0.016 \\
           &      &             & Umbriel &  70.0 &  5.6 & 0.922 & 0.017 & 0.947 & 0.008 & 1.027 & 0.021 \\
1342257194 & 1310 & 56275.07765 & Titania &  70.0 &  5.6 & 1.656 & 0.019 & 1.729 & 0.014 & 1.044 & 0.015 \\
           &      &             &         & 160.0 & 10.7 & 0.916 & 0.019 & 0.933 & 0.017 & 1.019 & 0.028 \\
           &      &             & Oberon  &  70.0 &  5.6 & 1.556 & 0.022 & 1.632 & 0.018 & 1.049 & 0.019 \\
           &      &             &         & 160.0 & 10.7 & 0.772 & 0.008 & 0.816 & 0.005 & 1.057 & 0.013 \\
           &      &             & Umbriel &  70.0 &  5.6 & 0.920 & 0.012 & 0.964 & 0.011 & 1.048 & 0.018 \\
1342257195 & 1310 & 56275.08169 & Titania & 100.0 &  6.8 & 1.449 & 0.013 & 1.487 & 0.015 & 1.026 & 0.014 \\
           &      &             &         & 160.0 & 10.7 & 0.915 & 0.037 & 0.955 & 0.025 & 1.044 & 0.050 \\
           &      &             & Oberon  & 100.0 &  6.8 & 1.326 & 0.017 & 1.396 & 0.015 & 1.053 & 0.018 \\
           &      &             &         & 160.0 & 10.7 & 0.797 & 0.012 & 0.828 & 0.005 & 1.039 & 0.017 \\
           &      &             & Umbriel & 100.0 &  6.8 & 0.736 & 0.026 & 0.814 & 0.008 & 1.106 & 0.041 \\
1342257196 & 1310 & 56275.08573 & Titania & 100.0 &  6.8 & 1.420 & 0.014 & 1.470 & 0.005 & 1.035 & 0.011 \\
           &      &             &         & 160.0 & 10.7 & 0.958 & 0.031 & 0.952 & 0.027 & 0.994 & 0.043 \\
           &      &             & Oberon  & 100.0 &  6.8 & 1.324 & 0.008 & 1.382 & 0.013 & 1.044 & 0.012 \\
           &      &             &         & 160.0 & 10.7 & 0.766 & 0.013 & 0.799 & 0.010 & 1.043 & 0.022 \\
           &      &             & Umbriel & 100.0 &  6.8 & 0.703 & 0.009 & 0.802 & 0.007 & 1.141 & 0.018 \\
            \noalign{\smallskip}
\hline\hline                 % inserts double horizontal lines
\end{tabular}
\end{table*}
%\end{sidewaystable*}

\clearpage

\end{onecolumn}

\begin{twocolumn}

\section{Relation of PACS photometer detector response with the telescope background power in the 70, 100, and 160\,$\mu$m PACS filters}
\label{sect:appb}

According to \citet{exter18}, Sect.\ 7.4.2, the monochromatic PACS flux density
is inversely proportional to the detector response R:

\begin{equation}
f_{\nu,1}(\lambda_{0})~[Jy] = U_{\rm sig}~\frac{C_{\rm conv}}{R} = \frac{U_{\rm sig}}{R_{\nu,1}}
\end{equation}
with
\begin{equation}
R_{\nu,1}~~[V/Jy] = \frac{R~[V/W]}{C_{\rm conv}~[Jy/pW]}
.\end{equation}

R$_{\nu,1}$ is actually not a constant. It depends on the operational temperature 
of the bolometers and the IR total flux load, hence R$_{\nu,1}$ = f(T, B$_{total flux}$).
B$_{total flux}$ is dominated by the background of the only passively cooled telescope
B$_{telescope}$. A first description of this detector response effect by the telescope
background was given by \citet{balog14}. In that study the telescope background was 
described as flux/per spectrometer pixel. \citet{klaas16} describes a telescope 
background model, from which a telescope background per photometer pixel can be
calculated for each {\it Herschel} Operational Day (OD). In Sect.~6 there detector 
response relations with regard to this calculated telescope background are shown 
which are based on observations of standard stars. In particular at 160\,$\mu$m the 
stars are already quite faint ($<$3\,Jy) and no significant correlation could be derived 
due to the uncertainties of the measured fluxes and hence a large scatter of the 
data points. 

However, the Uranus observations offer high S/N data points for all three filters.
The only prerequisite is to scale all observations to the same distance (dc). 
Fig.~\ref{fig:reldetrespwithtelbgpow} shows the derived relations for the 
correction factors c$_{\rm telbg}$(B$_{\rm telescope}$) = $\left( \frac{f_{meas}^{Uranus}(B_{\rm telescope})}{f_{model}^{Uranus}} \right)_{\rm dc}$. 
These are (from PSF photometry):
\begin{eqnarray}
\label{eqn:corrtelbg1}
~~~70\,{\mu}m: c_{\rm telbg} = 1.2445 - 0.1041\,\times\,B_{\rm telescope}~(pW) \\
\label{eqn:corrtelbg2}
~100\,{\mu}m: c_{\rm telbg} = 1.1859 - 0.1496\,\times\,B_{\rm telescope}~(pW) \\
\label{eqn:corrtelbg3}
~160\,{\mu}m: c_{\rm telbg} = 1.3678 - 0.2015\,\times\,B_{\rm telescope}~(pW)
\end{eqnarray} 

For our Uranus observations the following correction factors in Table~\ref{table:corrtelbg} 
are applied to the fluxes (by division, since R is inversely proportional with 
f$_{\nu,1}(\lambda_{0})$).

%
%                                                One column table
%-----------------------------------------------------------
\begin{table}[h!]
\caption{Telescope background correction factors for the Uranus observations
         derived from Eqns~\ref{eqn:corrtelbg1} to~\ref{eqn:corrtelbg3}.
         The last column gives the distance correction (dc) factor to bring
         the Uranus photometry to a mean distance.  
         }
\label{table:corrtelbg}
\begin{tabular}{r c c c c}
\hline\hline
            \noalign{\smallskip}
OD   & c$_{\rm telbg}^{70}$ &  c$_{\rm telbg}^{100}$ &  c$_{\rm telbg}^{160}$ & c$_{\rm dist}$    \\
     &                   &                     &                     &                  \\
\noalign{\smallskip} \hline \noalign{\smallskip}
 579 & 1.0001 & 0.9989 & 1.0009 & 1.0046 \\
 789 & 1.0020 & 1.0007 & 1.0038 & 1.0223 \\
 957 & 0.9939 & 0.9965 & 0.9979 & 0.9903 \\
1121 & 0.9976 & 0.9990 & 1.0020 & 0.9627 \\
1310 & 0.9892 & 0.9947 & 0.9959 & 1.0201 \\
\noalign{\smallskip} \hline\hline \noalign{\smallskip}
\end{tabular}
\end{table}

While the effect of the distance correction ($\frac{c_{\rm dist}^{\rm max}}{c_{\rm dist}^{\rm min}}$)
is in the order of 6\%, the effect of the detector response change with telescope
background ($\frac{c_{\rm telbg}^{max}}{c_{\rm telbg}^{min}}$) is in the order of
1.3\%, 0.6\%, and 0.8\% at 70, 100, and 160\,$\mu$m, respectively, for the
data set of Uranus and its satellites.

%
%                                                Two column figure
%-----------------------------------------------------------
   \begin{figure}[hb]
   \centering
   \includegraphics[width=0.385\textwidth]{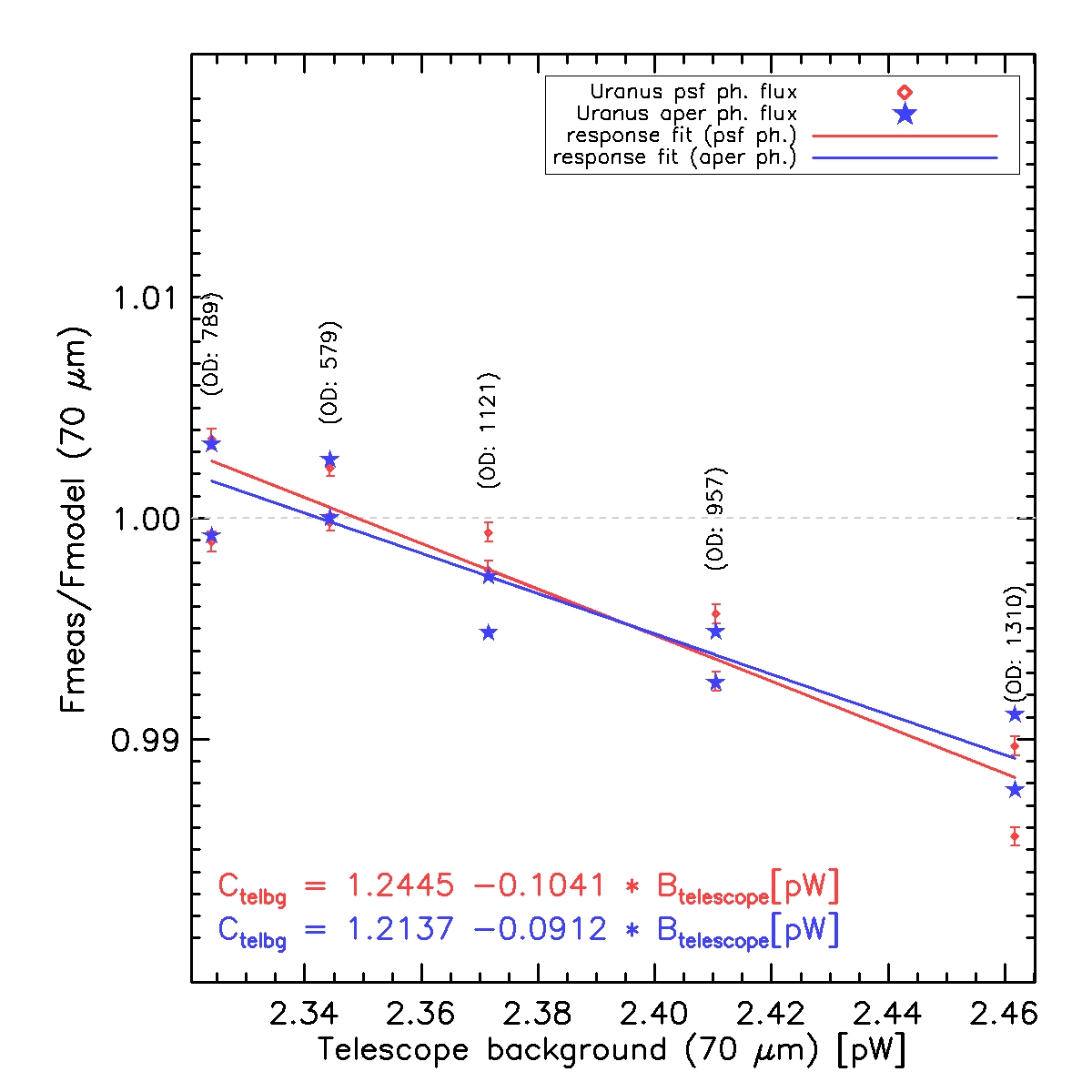}
   \includegraphics[width=0.385\textwidth]{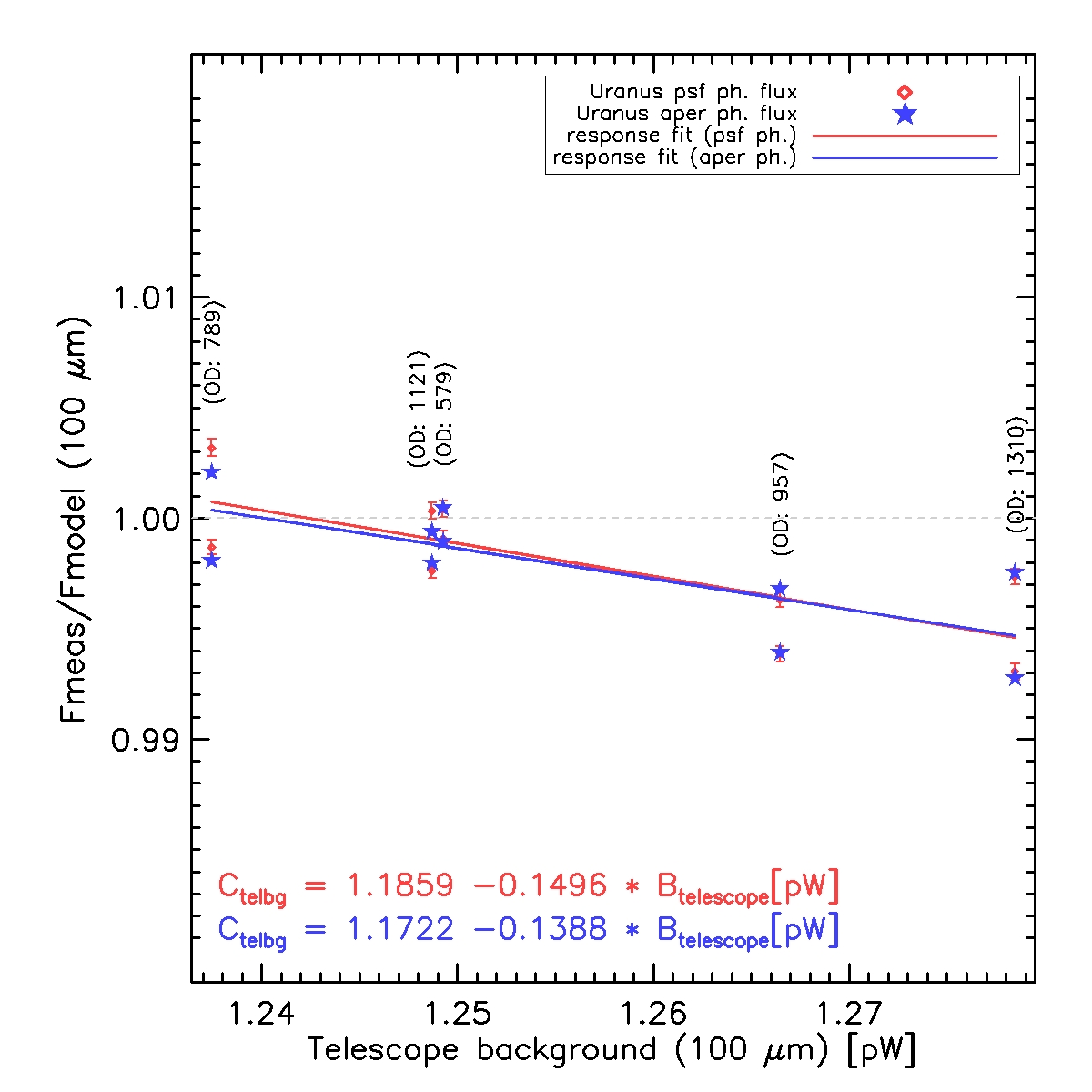}
   \includegraphics[width=0.385\textwidth]{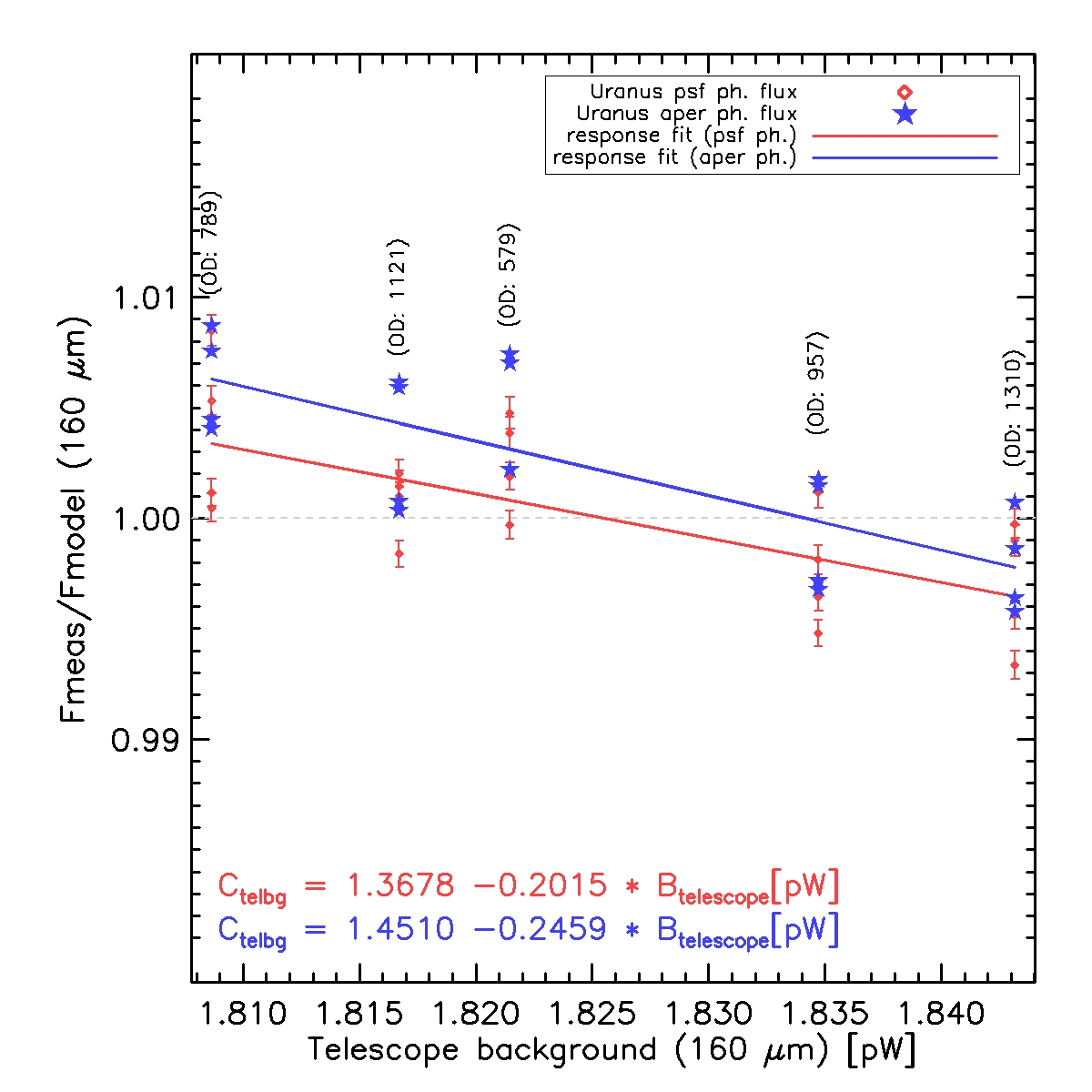}
      \caption{Relation of PACS photometer detector response, as indicated by a
               normalised flux level, with the telescope background power for
               the 70, 100, and 160\,$\mu$m filters. The fits were done both
               for PSF photometry (red) and aperture photometry (blue).
               }
         \label{fig:reldetrespwithtelbgpow}
   \end{figure}

\end{twocolumn}

\end{appendix}

\end{document}